\documentclass[aps,noshowpacs,preprintnumbers,onecolumn]{revtex4}
\usepackage{epsfig}
\usepackage{graphicx}
\usepackage{color}
\usepackage{latexsym,amsmath,amssymb,amstext}
\begin{document}
\title{QCD thermodynamics on the lattice}
\author{Sayantan\ \surname{Sharma}}
\email{sayantan@physik.uni-bielefeld.de}
\affiliation{Fakult\"at f\"ur Physik, Universit\"at Bielefeld, 
           D-33615 Bielefeld, Germany}
\begin{abstract}
A remarkable progress has been made in the understanding of the hot and dense QCD matter using  
lattice gauge theory. The issues which are very well understood as well those which require 
both conceptual as well as algorithmic advances are highlighted.
The recent lattice results on QCD thermodynamics which are important in the context
of the heavy ion experiments are reviewed. 
Instances of greater synergy between the lattice theory and the 
experiments in the recent years are discussed where lattice results could be directly used as benchmarks for experiments 
and results from the experiments would be a crucial input for lattice computations.
\end{abstract}
\maketitle
\section{Introduction}
A large part of the visible matter in our universe is made up of protons and neutrons, collectively 
called hadrons. Hadrons are made up of more fundamental particles called quarks and gluons. The 
quantum theory for these particles is Quantum Chromodynamics(QCD). QCD is a strongly interacting 
theory and the strength of interaction becomes vanishingly small only at asymptotically high energies.
Due to this reason, the quarks and gluons are not visible directly in our world and remain 
confined within the hadrons. Lattice gauge theory has emerged as the most successful non-perturbative 
tool to study QCD, with very precise lattice results available for hadron masses and decay constants which are 
in excellent agreement with the experimental values~\cite{lat1}. 

It is expected that at high enough temperatures that existed in the early universe, the hadrons would melt 
into a quark gluon plasma(QGP) phase. Signatures of such a phase have been seen during the last decade in the 
Heavy ion collision experiments at the Relativistic Heavy Ion Collider(RHIC), in Brookhaven National Laboratory  
This is particularly exciting for the lattice theory community which has been predicting such a phase 
transition since a long time \cite{lat2}. The formation of the QGP phase occurs at temperatures near 
$\Lambda_{QCD}$, where QCD is strongly interacting, which means lattice is the most reliable 
tool to understand the properties of the hot QCD medium. Over the past three decades, the lattice community 
has contributed significantly to the understanding of the physics of heavy ion experiments and strongly interacting 
matter under extreme conditions, in general. Lattice computations are 
entering into the precision regime, where lattice data can be directly used for interpreting the experimental 
results and set benchmarks for the heavy ion experiments at RHIC and at the ALICE facility in CERN.  It is now generally 
believed that the hot and dense matter created due to the collision of two heavy nuclei at RHIC and ALICE, equilibrates 
within 1 fm/c of the initial impact. The equilibrated QGP medium then expands and cools down in the process,  
ultimately forming hadrons at the chemical freezeout. The evolution of the fireball from its equilibration till 
the chemical freezeout  is described by relativistic hydrodynamics~\cite{hydro}. The QCD Equation of State(EoS) is 
an input for the hydrodynamic equations and lattice can provide a non-perturbative estimate of this quantity from first principles. 
The lattice data for the speed of sound in the QCD medium is also an important input for the hydrodynamic 
study, once bulk viscosity is considered.

In this article, I have selected the most recent results form lattice QCD thermodynamics 
that are relevant for the heavy ion phenomenology. I have tried to review the necessary background, but not attempted 
to provide a comprehensive account of the development of the subject throughout these years.
I have divided this article into two major sections. The first section deals with QCD at finite temperature 
and zero baryon density, where lattice methods are very robust. I have given a basic introduction to the lattice techniques, 
and how the continuum limit is taken, which is essential to relate the lattice data with the real world experiments. 
I have discussed the current understanding we have of the nature of QCD phase transition as a function of quark masses, 
inferred from lattice studies. 
Subsequently the different aspects of the hot QCD medium for physical quark masses are discussed; the EoS, the 
nature and the temperature of transition and the behaviour of various thermodynamic observables in the different phases. In the study of thermodynamics, the contribution of the lighter u, d and s quarks are usually considered. 
The effect of heavier charm quarks on QCD thermodynamics is discussed in this section, in view of their relevance for the 
heavy ion experiments at LHC, where hydrodynamic evolution is expected to set in already at temperatures about 500 MeV 
and also for the physics of early 
universe. The relevance of chiral symmetry for the QCD phase diagram and the effects of chiral anomaly are discussed in 
detail. The chiral anomaly is believed to have an important role in shaping the phase diagram and several lattice studies in the 
recent years are trying to understand its effect. It is a difficult problem and I have tried to compile the recent results 
and review the general understanding within the community, about how to improve upon them.
 
The second section is about lattice QCD at finite density, where there is an inherent short-coming of the lattice algorithms 
due to the so-called sign problem. A brief overview of the different methods used and those being developed 
by the lattice practitioners to circumvent this problem, is given. It is an active field of research, with a lot 
of understanding of the origin and the severity of this problem gained in recent years, which is motivating 
the search for its possible cure. 
In the regime where the density of baryons is not too large, which is being probed by the experiments at RHIC, lattice 
techniques have been used successfully to produce some interesting results. One such important proposal in the recent time, is the first principles determination of the chemical freezeout curve using experimental data on the electric charge fluctuations. This and the 
lattice results on the fluctuations of different quantum numbers in the hot medium and the EoS at finite baryon density are discussed in detail. An important feature of the QCD phase diagram is the possible presence of a critical end-point for the chiral first order transition. 
Since critical end-point search is one of the main objectives at RHIC, I have reviewed the current lattice results on 
this topic. The presence of the critical end-point is still not conclusively proven from lattice studies. It is a very challenging 
problem and I mention about the further work in progress to address this problem effectively. Fermions 
with exact chiral symmetry on the lattice are important in this context. I have discussed the recent 
successful development to construct fermion operators that have exact chiral symmetry even at finite density which 
would be important for future studies on the critical end-point.
The signatures of the critical end-point could be detected in the experiments if the critical region is not separated from 
the freezeout curve. It is thus very crucial to estimate the curvature of the critical line from first principles and I 
devote an entire subsection to discuss the lattice results on this topic.

I apologize for my inability to include all the pioneering works that have firmly established this subject and also to review the extensive 
set of interesting contemporary works. For a comprehensive review of the current activity in lattice thermodynamics, at 
finite temperature and density, I refer to the excellent review talks of the Lattice conference, 2012 \cite{maria,gaarts}.

\section{QCD at finite temperature on the lattice}
The starting point of any thermodynamic study is the partition function. The QCD partition function
for $N_f$ flavours of quarks in the canonical ensemble is given as,
\begin{equation}
 \mathcal{Z}_{QCD}(T,V)=\int \mathcal {D}U_\mu(x) \prod_{f=1}^{N_f} det D_f~\rm{e}^{-S_G}
\end{equation}
where $D_f$ is the fermion operator for each flavour of quark $f$. 
$U_\mu$ is the gauge link defined as $U_\mu(x)=exp(-ig\int^x A_\mu (x') d x')$ in terms of 
gauge fields $A_\mu$, which are adjoint representation of the $SU(3)$ color group and 
$g$ is the strength of the gauge coupling.
$S_G$ is the gluon action in Euclidean space of finite temporal extent of size
denoted by the inverse of the temperature of the system, $T$.
Lattice QCD involves discretizing the spacetime into a lattice with a spacing 
denoted by $a$. The volume of the lattice is given as $V=N^3 a^3$, where  $N$ are the number 
of lattice sites along the spatial directions and the temperature being 
$T=1/(N_\tau a)$, where $N_\tau$ are the number of sites along the temporal direction. The 
lattice is usually denoted as $N^3\times N_\tau$. The 
gluon action and the fermion determinant are discretized on the lattice. The simplest gluon action, 
known as Wilson plaquette action is of the form,
\begin{equation}
 S_G=\frac{6}{g^2}\sum_{x,\mu,\nu,\mu<\nu}(1-\frac{1}{3}Tr\text{Re}~U_{\mu,\nu}(x))~,~
 U_{\mu,\nu}(x)=U_{\mu}(x)U_{\nu}(x+\mu)U^\dagger_{\mu}(x+\nu)U^\dagger_{\nu}(x).
\end{equation}
where $U_{\mu,\nu}(x)$ is called a plaquette. The naive discretization of the continuum 
Dirac equation on the lattice results in the fermion operator of the form,
\begin{equation}
 D_f(x,y)=\sum_{x,y}\left[\sum_{\mu=1}^4\frac{1}{2}\gamma_\mu \left(U_{\mu}(x)\delta_{y,x+\mu}-
U^\dagger_{\mu}(y) \delta_{y,x-\mu}\right)+ a m_f \delta_{x,y}\right].
\end{equation}
where in each of the expressions the site index runs from $x=1$ to $N^3\times N_\tau$.
The discretization of the gluon and fermion 
operators  are not unique and there are several choices which give the correct continuum limit.
Usually discretized operators with small finite $a$ corrections are preferred. Reducing $a$-dependent corrections 
by adding suitable ``irrelevant'' terms in the Renormalization Group(RG) sense, is known 
as improvement of the operator. Another issue related to the discretization of the fermion 
operator is called as the ``fermion doubling problem''. It arises because the 
naive discretization of the continuum fermion operator introduces extra unphysical fermion 
species called the doublers. The existence of the doublers can be traced back to a No-Go theorem \cite{nn}
on the lattice which states that fermion actions which are ultra-local, have exact chiral symmetry, and have the 
correct continuum limit, cannot be free from the doublers. Doublers are  problematic
since in the continuum limit we would get a theory with 16 fermion species and QCD with 16 flavours is very close
to the upper bound of the number of flavours beyond which the asymptotic freedom is lost. It is thus
important to ensure that the discrete fermion operator should be free of the doublers. In order to do so 
the chiral symmetry is explicitly broken on the lattice, like for the case of Wilson fermions \cite{wilson}, or only a 
remnant of it is preserved for the staggered fermions \cite{ks}. The staggered fermion discretization retains the 
doubling problem in a milder form. In the continuum limit, the staggered fermion determinant would give contribution 
of four degenerate fermion species or tastes. However on a finite lattice, there is a considerable mixing among the 
tastes so a simple fourth root of the determinant would not yield the contribution of a single fermion flavour. This 
is called the rooting problem. The severity of rooting problem can be minimized by choosing either the stout-smeared staggered quarks~\cite{stoutref} or the Highly Improved Staggered Quarks(HISQ)~\cite{hisqref}. 
Other improved versions of staggered fermions used for QCD thermodynamics are the p4 and asqtad fermions~\cite{p4ref,asqtadref}.  
Only the overlap \cite{neunar} and the 
domain wall fermions \cite{kaplan} have exact chiral symmetry on the lattice at the expense of breaking 
the ultra-locality condition of the Nielsen-Ninomiya No-go theorem. As a result overlap and domain wall 
fermions are much more expensive to simulate compared to the staggered and the Wilson fermions. For QCD thermodynamics, 
the staggered and to some extent the Wilson fermions are favourites, with very high precision data 
available with improved versions of staggered quarks \cite{bf1,bw1}. 
With the advent of faster computing resources and smarter algorithms, even large scale simulations with 
chiral fermions are becoming a reality \cite{cossu1, bwov,dwold,twqcd}. 

With the choice of a suitable gauge and the fermion operators on the lattice,  
different physical observables are measured on statistically independent configurations generated using suitable 
Monte-Carlo algorithms. To make connection with the continuum physics, one needs to take the
$a\rightarrow0$ limit of the observables measured on the lattice. The gauge coupling is related to the lattice 
spacing through the beta-function and the continuum limit, in turn, implies $g\rightarrow 0$. 
In the space of coupling constants and the 
fermion masses, the continuum limit is a second order fixed point and the approach to the fixed point
should be done along the correct RG trajectory or the lines of constant physics. The line of 
constant physics is defined by setting the mass of hadrons on the lattice to the continuum 
values, at each value of the coupling constant. The number of such relations required depends on the number of fermion flavours. 
To relate the lattice hadron masses to their experimental values, one has to define a scale to express the lattice spacing $a$, 
in terms of some physical units. There are two often used methods in QCD to set the scale, using the quantities $r_1$ 
and the kaon decay constant $f_K$. The $r_1$ scale is defined from the quark-antiquark potential $V_{\bar q q}(r)$, as,
\begin{equation}
 \left(r^2\frac{\partial V_{\bar q q}(r)}{\partial r}\right)_{r=r_1}=1.0.
\end{equation}
On the lattice one measures $V_{\bar q q}(r)$ and $r_1$ is extracted from it using a suitable fit ansatz for
the potential. To quantify the value of $r_1$ in physical units, one uses either the pion decay constant or the  
splitting of energy levels of bottom mesons to set the lattice spacing~\cite{milcscale}. Advantages of this scale 
is that it is not sensitive to fermion discretization effects and 
 to the choice of quark masses that defines the line of constant physics. However, the accurate determination 
of the potential requires very good statistics. One can also set the scale by choosing the $f_K$ measured on the
lattice to its physical value. The $f_K$ is known with very high accuracy from the experiments.
Once the line of constant physics is set, one has to take care of the finite size and lattice spacing 
effects such that the continuum extrapolation is correctly performed.
To minimize such corrections, the correlation length which is given by the inverse of the mass of the 
lowest excitation of the system should be much larger that the lattice spacing but sufficiently smaller than the spatial 
size. Also for thermodynamics, it is crucial to minimize finite volume corrections which is ensured for 
the choice $\zeta\geq3$, where $\zeta=N/N_\tau$.

To characterize different phases one needs to define a suitable order parameter which depends 
on the symmetries of the theory. In the limit of infinitely heavy quark masses QCD is just a 
pure gauge theory with an exact order parameter, the expectation value of the Polyakov loop given as,
\begin{equation}
 L(\mathbf x)=\frac{1}{3}Tr P \prod_{x_4=1}^{N_\tau} U_4(\mathbf x,x_4)~,~\text{P}\Rightarrow\text{Path ordering}.
\end{equation}
The phase transition  from a phase of confined colour degrees of freedom to the deconfined regime of 
free gluons, is of first order and is established very firmly from lattice studies \cite{karsch1}. 
The corresponding  transition temperature is $T_c$(pure-gauge)$=276(2)$ MeV \cite{karsch2} using the 
string tension, $\sqrt{\sigma}$, value to be 425  MeV, to set the scale . 
If the quarks are massless, the QCD partition function with $N_f$ quark flavours has an exact 
$SU(N_f)\otimes SU(N_f)$ chiral symmetry. At some temperature, there is a phase transition from a chiral symmetry 
broken phase to the symmetry restored phase, characterized by the order parameter called the chiral condensate,
\begin{equation}
 \langle\bar\psi_f\psi_f\rangle=\lim_{m_f\rightarrow0}\lim_{V\rightarrow\infty}\frac{T}{V}\frac{\partial ln\mathcal{Z}_{QCD}}{\partial m_f}~,~f=1,..,N_f.
\end{equation}
The phase transition in the chiral limit for $N_f=3$, is expected to be of first order and there are 
several lattice results supporting this \cite{nf3}. For $N_f=2$, the lattice results are contradictory with some
claiming a first order transition \cite{nf2fo} whereas recent results showing that the second order transition is 
also a possibility \cite{mageos}. The current status of $N_f=2$ QCD phase transition in the chiral limit would be discussed 
again in a later subsection. For any finite value of quark masses, however there is no unique order parameter and 
 no sharp phase transition is expected but only a gradual crossover.

\begin{figure}[h!]
\begin{center}
\includegraphics[scale=0.3]{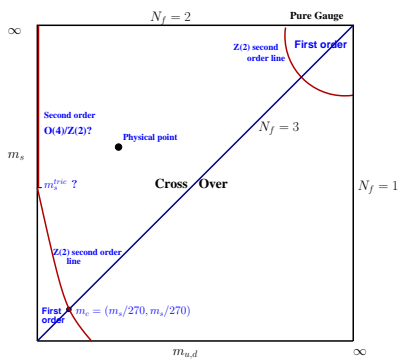}
\caption{The present status of the Columbia plot.}
\label{colplot}
\end{center}
\end{figure}

Based on effective field theories with same symmetries as QCD,  using universality arguments 
and renormalization group inspired techniques, a schematic diagram of different phases of QCD as a function 
of quark mass is summarized in the famous ``Columbia plot'' \cite{colplot}. The first order regions 
in the quenched and the chiral limits are separated from the cross-over region by second order lines
belonging to the $Z(2)$ universality class. These boundaries are schematic, though, and it is important to 
estimate the precise location of the physical point in this diagram. Lattice studies over the years 
have helped to redraw the boundaries more quantitatively. A latest version of the ``Columbia plot'' 
is shown in figure \ref{colplot}. With the high precision lattice data with physical
light and  strange quark masses, it is now known that the QCD transition in our world is a crossover \cite{milc,bf2,yaoki}.
The boundary of the first order region in the upper right hand corner of figure \ref{colplot} is fairly well known \cite{saito}. The extent of the first order region in the bottom  left hand is now believed to be small and much far away from the physical point \cite{ding,bwnf3}.  However the extent of the $Z(2)$ line in the left hand corner is still not well established; it can either continue along the  $m_{u,d}=0$ axis  to the $m_s\rightarrow\infty$ corner or end at a tricritical point. A better understanding of this issue is currently underway. The key to the resolution of this issue is to understand the effects of chiral anomaly through rigorous lattice computations. Since the light u,d-quark masses are much smaller than $\Lambda_{QCD}$, the QCD action has an approximate $SU(2)\times SU(2)\times U_B(1)$ symmetry with an additional classical 
$U_A(1)$ symmetry broken explicitly by quantum effects. This is known as the $U_A(1)$ anomaly \cite{abj,fujikawa}. 
At zero temperature, the magnitude of this anomaly is related to the instanton-density. If the magnitude of this 
anomaly is temperature independent, the phase transition along the $m_{u,d}=0$ axes has to be 
of second order, belonging to the $O(4)$ universality class \cite{piswil}.
This would mean that the $Z(2)$ line has to end at a tri-critical point characterized by the strange quark mass,
$m_s^{tric}$. The difference between the physical and tri-critical mass for the strange quark is not 
yet known with a good precision.

 In the following subsections, the lattice results for the QCD EoS for physical quark 
masses are discussed, which is an input for the hydrodynamics of the QGP medium. 
 The current results on the pseudo-critical temperature, the entropy 
density, the speed of sound are also shown. All the results are for $2+1$ flavour QCD, i.e., two light 
degenerate u and d quarks and a heavier strange quark mass. The effect of the heavy charm quarks on the 
thermodynamic quantities is also highlighted.  At the end of this section, I touch upon the $N_f=2$ QCD 
near the chiral limit and the effects of the $U_A(1)$ anomaly for QCD thermodynamics.

\subsection{Equation of state}
The Equation of State(EoS) is the relation between the pressure and energy density of a system 
in thermal equilibrium.  For estimating the QCD EoS, the most frequently used method by the lattice 
practitioners is the integral method \cite{intmethod}. In this method, one first computes the 
trace anomaly $I(T)$, which is the trace of the energy-momentum tensor. This is 
equal to the quantity $\epsilon-3p$ where $\epsilon$ is the energy density of the 
system and $p$ is the pressure. Moreover it is related to the pressure of the 
system through the following relation
\begin{equation}
 I(T)=T^5\frac{\partial}{\partial T}\frac{p}{T^4}
\end{equation}
So if $I(T)$ is known, the pressure can be computed by integrating $I(T)$ over a range of 
temperature, with the lower value of temperature chosen such that the corresponding value of 
pressure is vanishingly small.
The trace anomaly is related to the chiral condensate and the gluon action as,
\begin{equation}
 \frac{I(T)}{T^4}=-N_\tau^4\left(a\frac{d \beta}{d a}(\langle S_G\rangle-\langle S_G\rangle_0)
+\sum_f a \frac{d (m_f a)}{d a}(\langle \bar\psi_f\psi_f\rangle-\langle \bar\psi_f\psi_f\rangle_0)\right)~,
~\beta=\frac{6}{g^2}~,
\end{equation}
where the subscript zero denotes the vacuum expectation values of the corresponding quantities. The 
subtraction is necessary to remove the  zero temperature ultraviolet divergences and the vacuum expectation 
values are usually computed on a lattice with number of sites $(N_\tau)_0$ in the temporal direction, equal to the 
corresponding spatial number of sites, $N$. The subtraction is an unavoidable expense of this method.
A new idea of deriving thermodynamic observables from cumulants of momentum distribution has emerged, where 
the vacuum subtraction is not required~\cite{giusti} and it would be interesting to check the application of this method 
in QCD. Also one needs to know the functional dependence of the inverse of QCD coupling constant 
$\beta$ and the quark masses with the lattice spacing $a$ along the line of constant physics.
On the lattice, $I(T)$ is known only for a finite number of temperature values. The pressure computed by the numerical integration of 
the $I(T)$ data, has errors both due to statistical fluctuations and systematic uncertainties involved in the numerical interpolation of the data. 

\begin{figure}
\begin{center}
\includegraphics[scale=0.5]{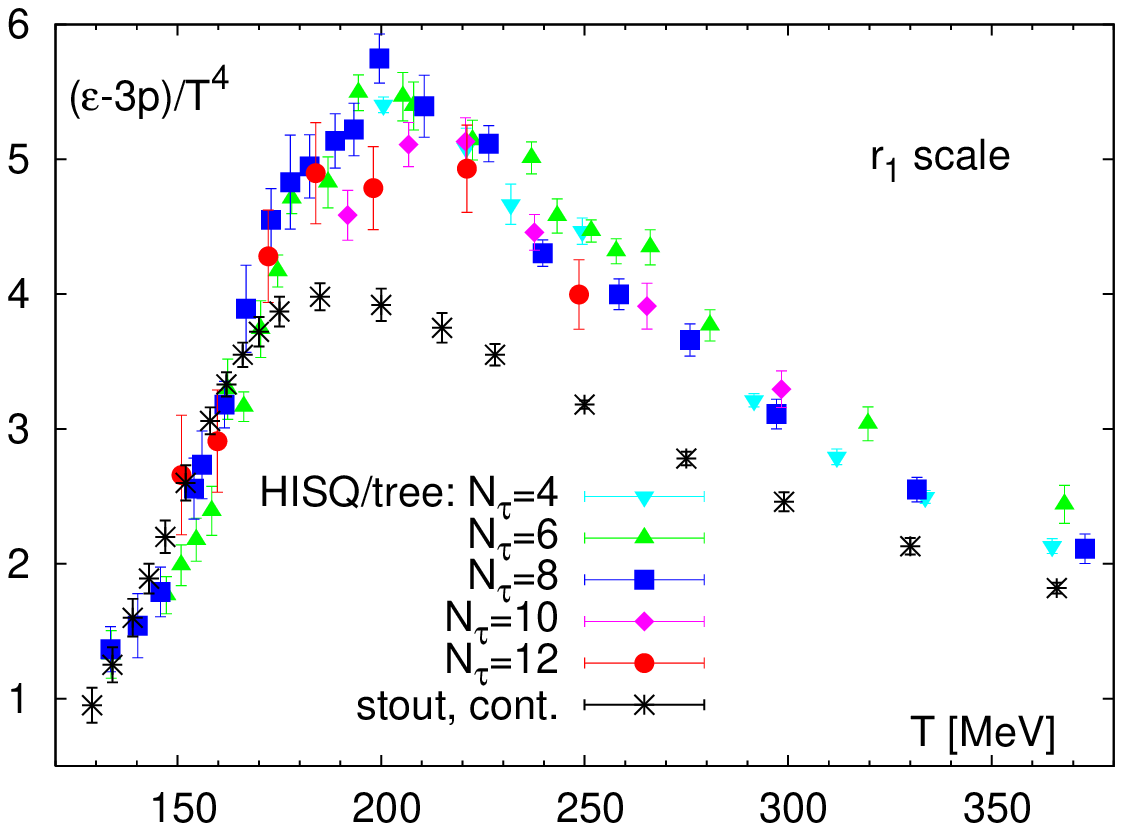}
\includegraphics[scale=0.5]{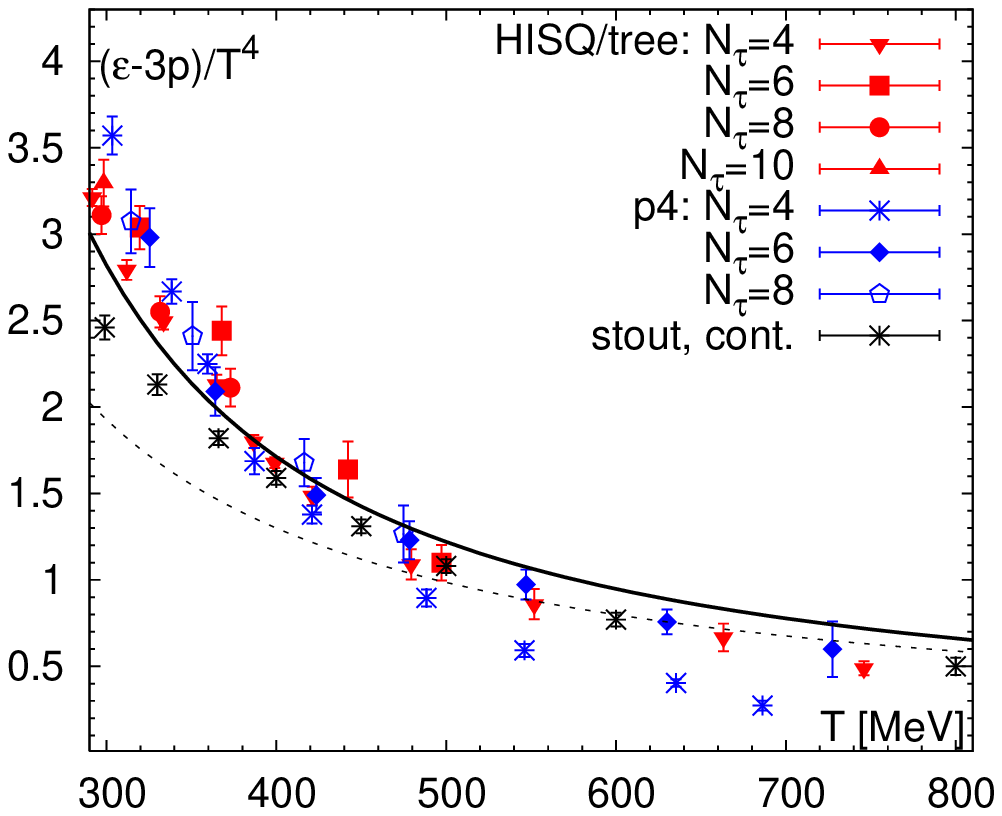}
\caption{The results for the trace anomaly using the HISQ action for low (left panel) and high(right panel) temperatures 
for lattice sizes with temporal extent $N_\tau$ and spatial size $4 N_\tau$, from \cite{petreczky}.  Also in the right panel, the HISQ results are compared to the results using p4 fermions, which has an improved behaviour at high temperatures and to the continuum perturbation theory results at 1-loop(solid line) and 2-loop(dashed line)for the trace anomaly. The stout data are the continuum estimates from the 
$N_\tau=6,8,10$ data in \cite{bw1}.}
\label{intmeashisq}
\end{center}
\end{figure}

\begin{figure}[h!]
\begin{center}
\includegraphics[scale=0.55]{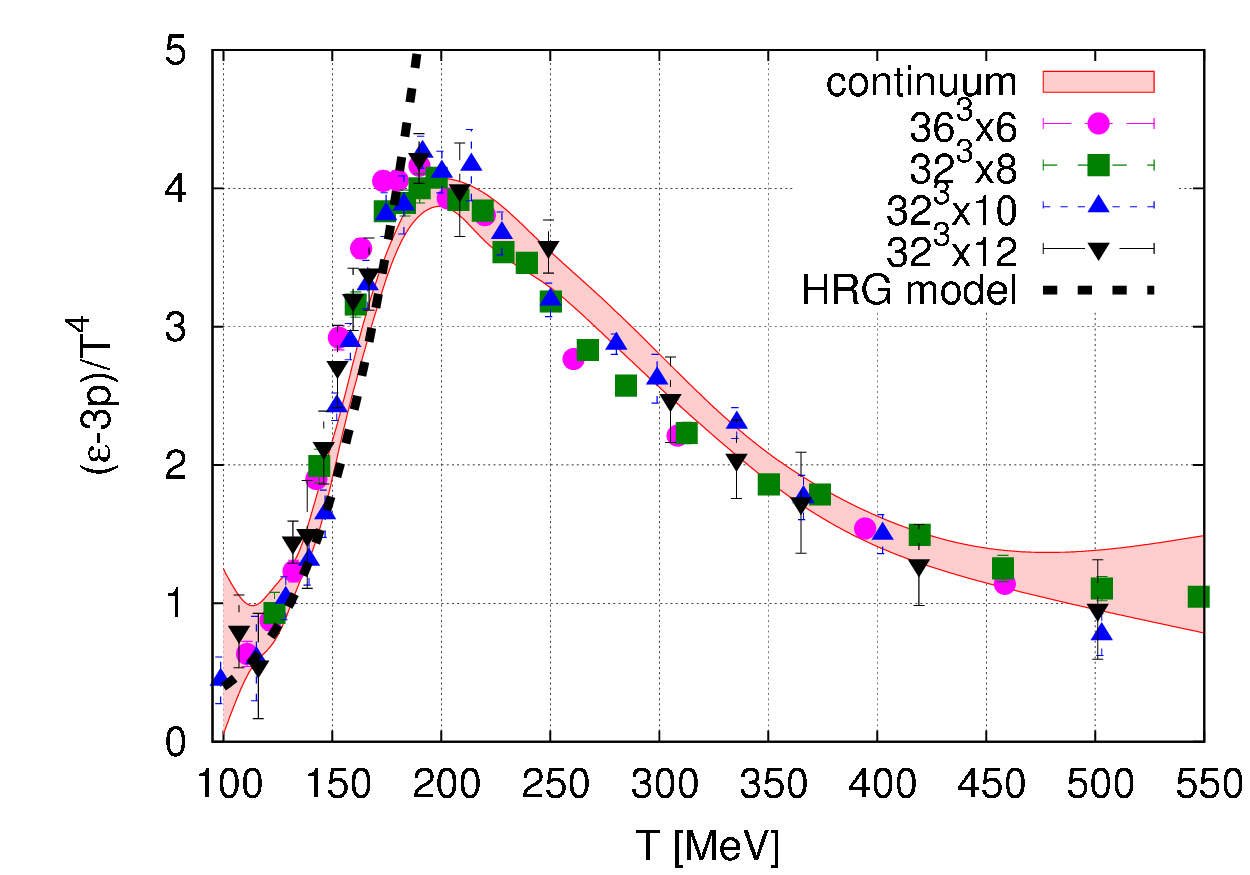}
\includegraphics[scale=0.4]{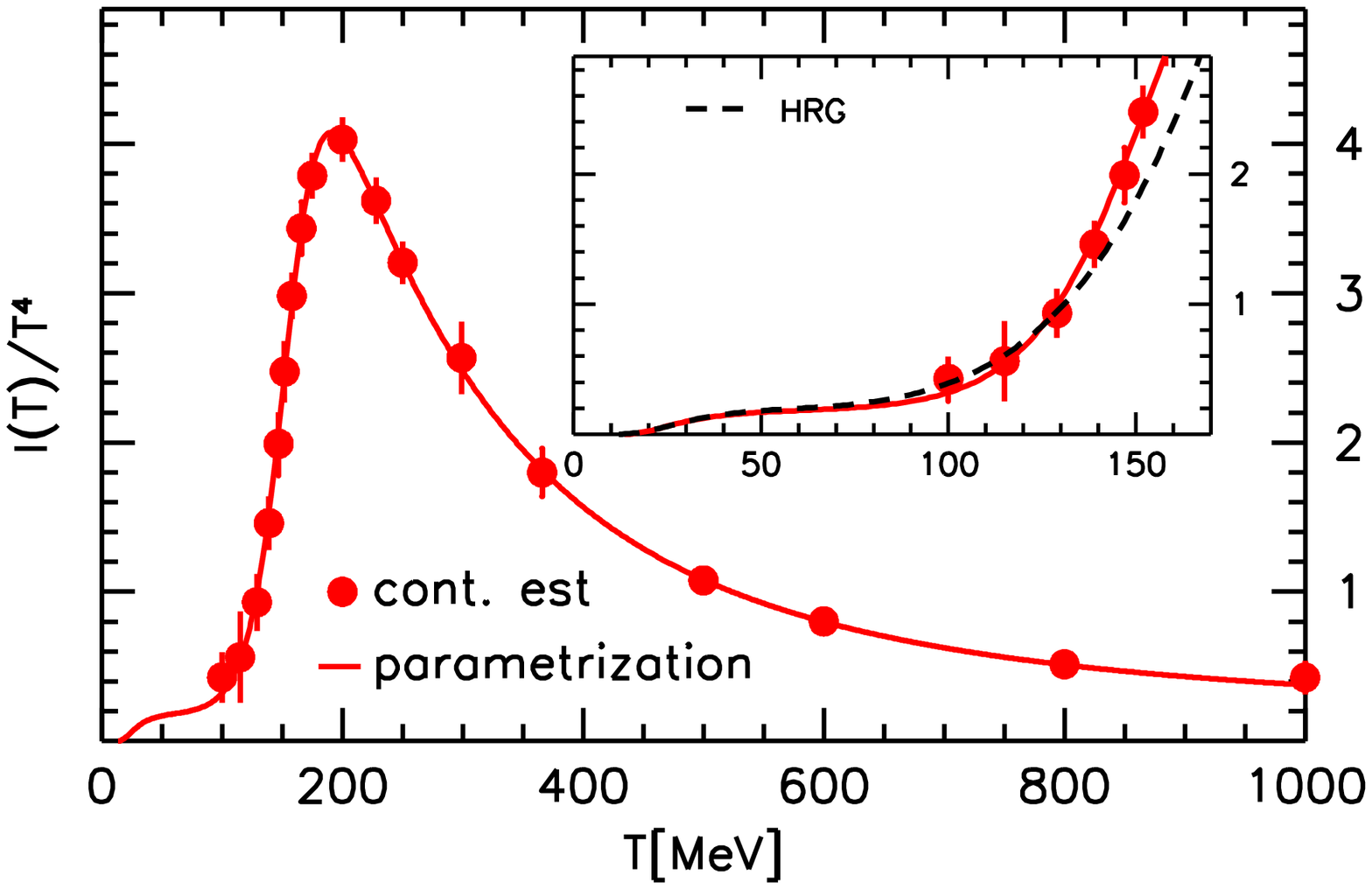}
\caption{The latest data with the stout smeared fermions(left panel), from \cite{bwcharm}. In the right panel, the fit to the 
trace anomaly data from the continuum extrapolation of the $N_\tau=6,8$ results, from \cite{bw1}. The results are in perfect agreement with 
the Hadron resonance gas model calculations for $T<140~$MeV.}
\label{intmeasbw}
\end{center}
\end{figure}

The results for the trace anomaly are available for different lattice discretizations of the fermions. For staggered quarks there are two sets of results, one  from the 
HotQCD collaboration using HISQ discretization \cite{petreczky,hotqcd1} and the other from the Budapest-Wuppertal collaboration using stout smeared staggered quarks \cite{bwcharm, bw1}. These results are compiled in Figures \ref{intmeashisq}, \ref{intmeasbw}. For the HISQ results, the bare lattice parameters are fixed by setting the lowest strange pseudo-scalar meson mass to its physical value at about 686 MeV and $m_\pi=160$ MeV, which  defines the line of constant physics. The kaon decay constant $f_K=156.1$ MeV or alternatively the $r_1=0.3106$ fm from the static quark potential is used to set the scale. The corresponding parameters for the stout smeared quarks are $m_\pi=135$ MeV,  $m_K=498$ MeV and the kaon decay constant.
From figure \ref{intmeashisq}, it is evident that there is a good agreement between the two sets of results for $T<180$MeV and also for high enough temperatures $T>350$ MeV. The stout continuum results in the figure were obtained extrapolation with the $N_\tau=6,8,10$ data from Ref. \cite{bw1}. In the intermediate temperature range there is some discrepancy, specially the peaks of the interaction measure do not coincide for these two different discretization schemes,  which may be due finite lattice spacing effects. However the HISQ $N_\tau=12$ data is inching closer to the stout results in this regime. The recent continuum stout results, obtained from continuum extrapolation of the new $N_\tau=12$ data in addition to the older data are consistent with the HISQ results, with the peak position shifting to 200 MeV(left panel of figure \ref{intmeasbw}). There is also a good agreement of the HISQ and stout data with the trace anomaly obtained from the Hadron Resonance Gas(HRG) model for $T<140$ MeV and with the 
resummed perturbation theory results at high temperatures. Using the $N_\tau=6,8$ data which is available upto temperatures of 1000 MeV, a continuum extrapolation of the stout data
was performed, the result of which is shown in right panel of figure \ref{intmeasbw}. For this entire range of temperature, there is a useful parameterization characterizing the trace anomaly \cite{bw1} with the following parametric form,
\begin{equation}
 \frac{I(T)}{T^4}=\rm{e}^{-h_1/t-h_2/t^2}.\left(h_0+\frac{f_0[tanh(f_1 t+f_2)+1]}{1+g_1 t+g_2 t^2}\right)~,
~t=T/200MeV~.
\end{equation}
where the best fit parameters are
\begin{equation}
 \nonumber
h_0=0.1396~,~h_1=-0.18~,~h_2=0.035~,~f_0=2.76~,~f_1=6.79~,~f_2=-5.29~,~g_1=-0.47~,~g_2=1.04.
\end{equation}
This parametric form could be a useful input for the hydrodynamical simulations, which usually uses the lattice EoS 
before hadronization and that from the HRG after the freezeout of hadrons. 

\begin{figure}[h!]
\begin{center}
\includegraphics[height=4 cm,width=0.3\textwidth]{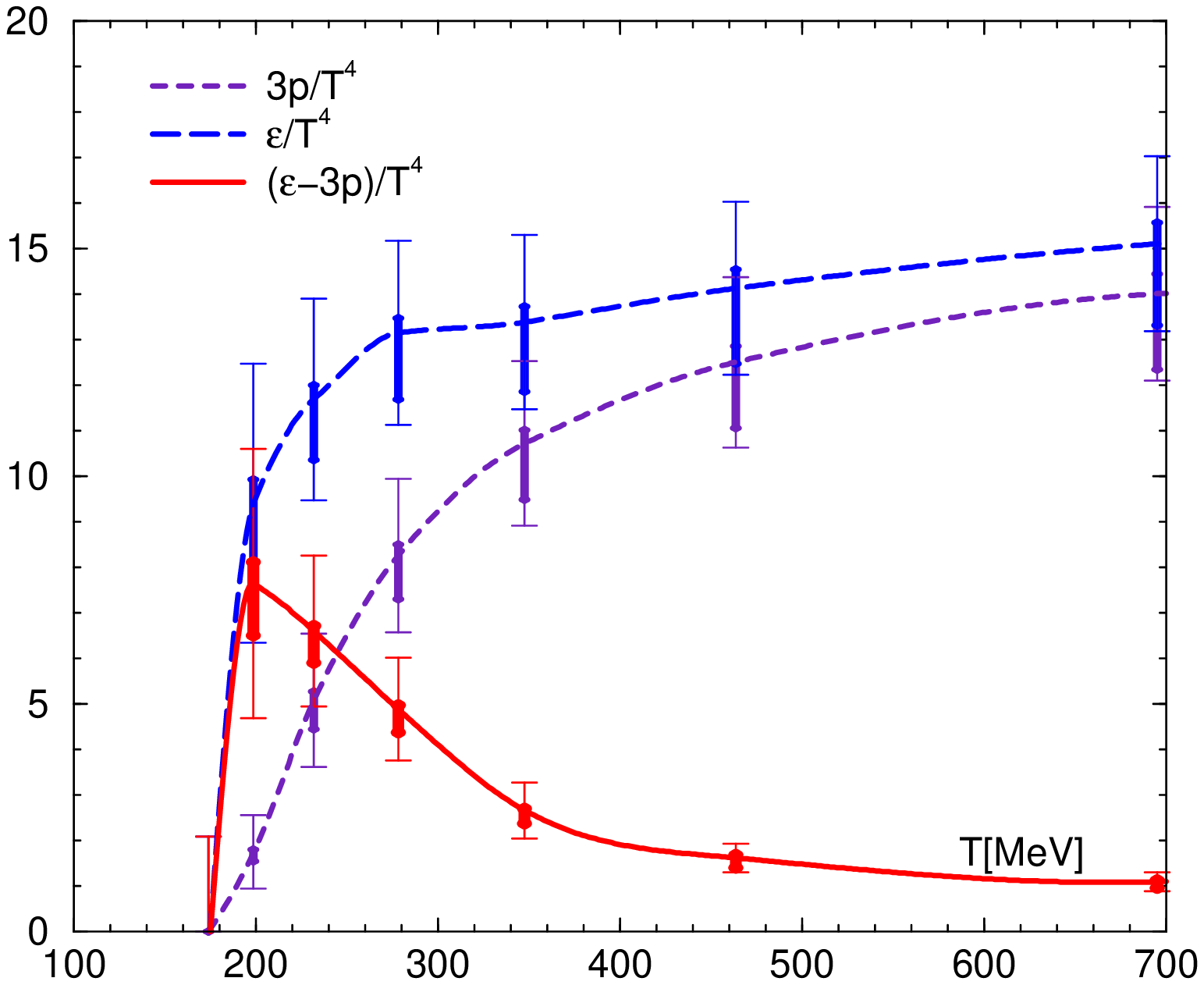}
\includegraphics[height=4 cm,width=0.3\textwidth]{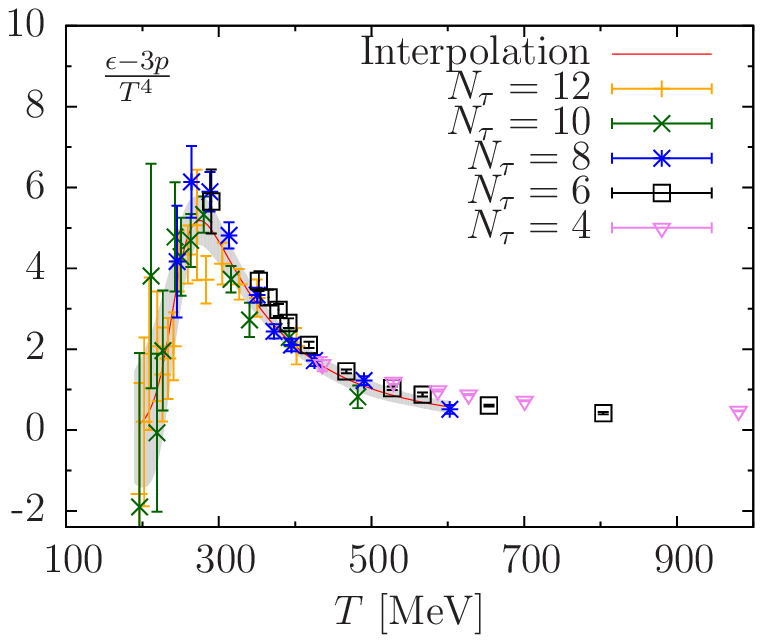}
\caption{The results for the pressure, energy density and the trace anomaly with clover-improved Wilson fermions 
on a $32^3\times8$ lattice, from \cite{whot}(left panel) and the trace anomaly data with the twisted mass Wilson 
fermions, from \cite{tmft}(right panel).}
\label{intmeaswil}
\end{center}
\end{figure}

There are lattice results for the EoS using alternative fermion discretizations, the Wilson fermions. The WHOT-QCD collaboration has results for $2+1$ flavours of improved Wilson fermions \cite{whot} with the physical value of strange quark mass but a large pion mass
equal to $0.63 m_\rho$. The tmfT collaboration has results for 2 flavours of maximally twisted Wilson fermions \cite{tmft} with $m_\pi>400$MeV. Both these results are compiled in figure \ref{intmeaswil}.  These are in rough qualitative agreement with the staggered fermion data,
specially the peak for the WHOT-QCD data occurring at 200 MeV is consistent with the HISQ and stout results. A more quantitative 
agreement at this stage is difficult, since the pion masses for the Wilson fermions are much larger than the physical value.

\subsection{The pseudo-critical temperature}
\begin{figure}
\begin{center}
\includegraphics[scale=0.5]{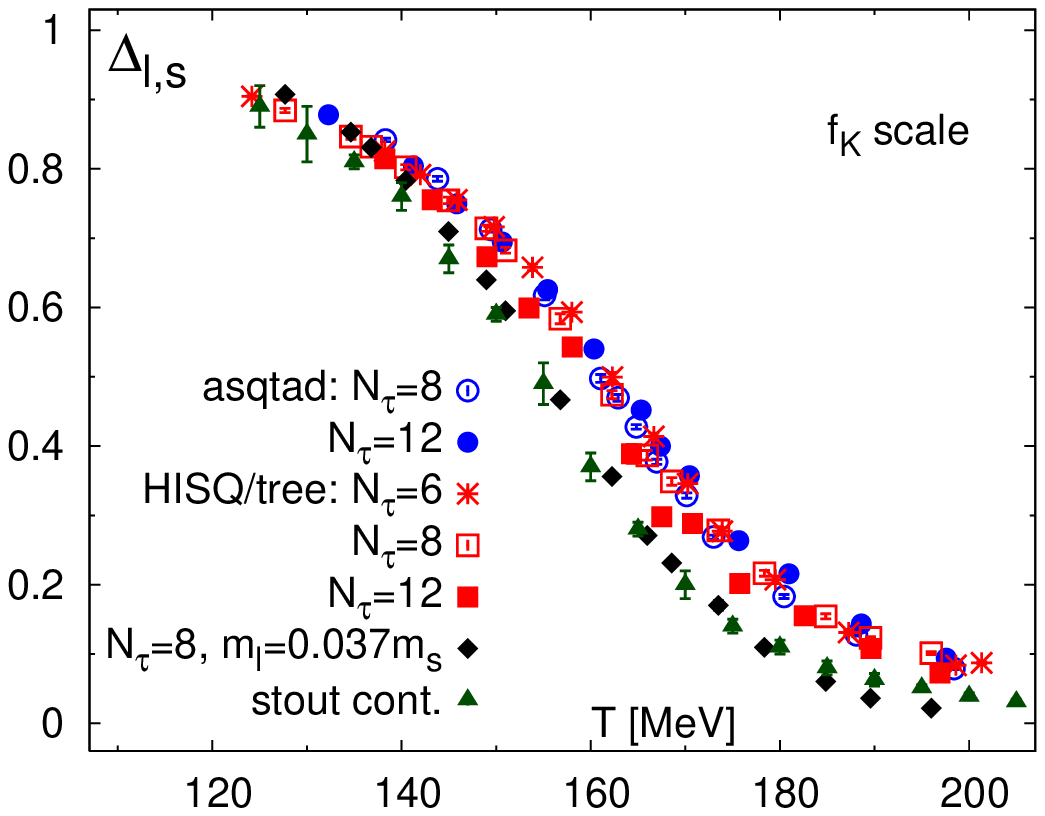}
\includegraphics[scale=0.5]{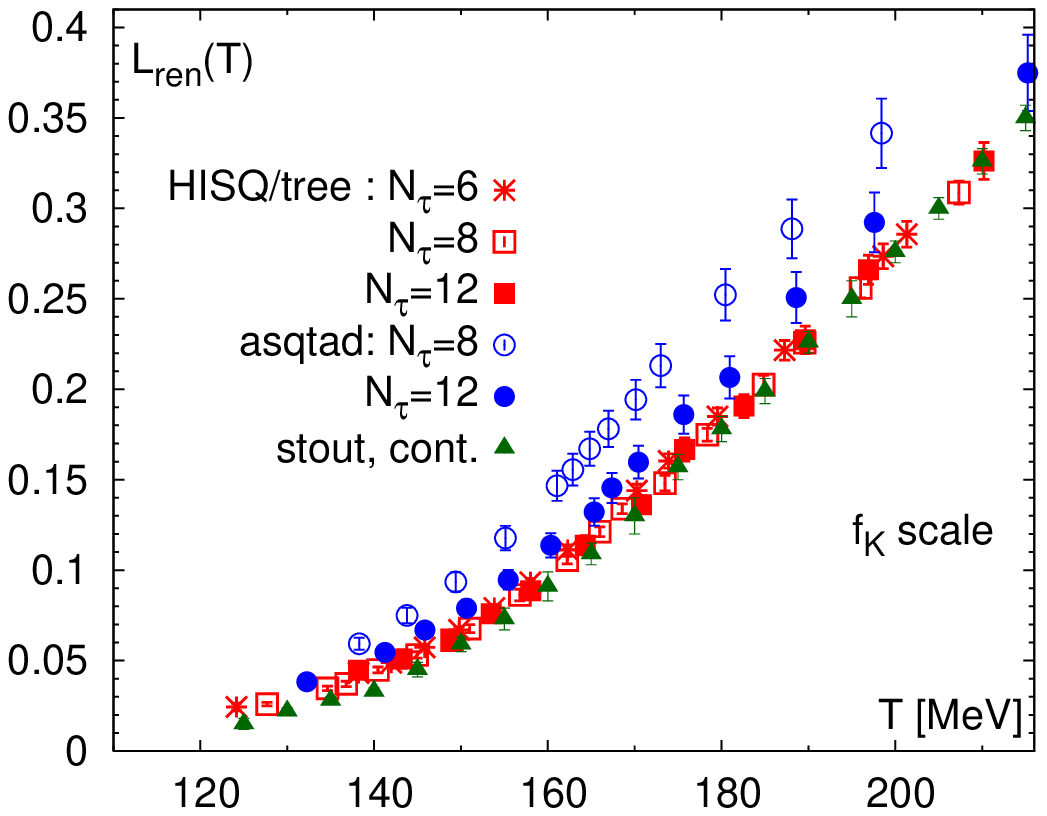}
\caption{The results for the subtracted chiral condensate(left panel) and the renormalized Polyakov loop(right panel)
from the HotQCD collaboration, from \cite{hotqcd1}. These data are compared with the continuum results using stout 
smeared fermions, from \cite{bwtc}.}
\label{deltaLhisq}
\end{center}
\end{figure}

We recall that the QCD transition, from a phase of color singlet states to a phase of colored quantum states 
is an analytic crossover, for physical quark masses. This is fairly well established by now from lattice studies using two different approaches. One approach is to monitor the behaviour of the thermodynamic observables in the transition region for physical values 
of quark masses while the other is to map out the chiral critical line as a function of light quark mass~\cite{fph1}. The absence of 
a sharp phase transition implies that there is no unique transition temperature but only different pseudo-critical temperatures 
corresponding to different observables. There is no order parameter but the observables like the renormalized 
Polyakov loop, $L_R$ has a point of inflexion across the crossover region.  Another observable relevant in the crossover regime is the 
renormalized chiral condensate, which has been defined \cite{hotqcd2} in the following manner to take into account the
multiplicative renormalization as well additive ones due to a finite bare quark mass,
\begin{equation}
\Delta_{l,s}(T)=\frac{\langle \bar\psi\psi\rangle_{l,T}-\frac{m_l}{m_s}\langle \bar\psi\psi\rangle_{s,T}}
{\langle \bar\psi\psi\rangle_{l,0}-\frac{m_l}{m_s}\langle \bar\psi\psi\rangle_{s,0}} ~,~l=u,d.
\end{equation}
The normalized chiral susceptibility $\chi_R$ for the light quarks defined as,
\begin{equation}
 \chi_R=\frac{1}{VT^3}m_l^2\frac{\partial^2 }{\partial m_l^2}(\ln\mathcal{Z}(T)- \ln\mathcal{Z}(0))
\end{equation}
is a good observable as well. Both $L_R$ and $\Delta_{l,s}(T)$ have a point of inflexion at the pseudo-critical temperature and  
$\chi_R$ has a smooth peak. From the continuum extrapolated data of the stout-smeared staggered fermions, 
the pseudo-critical temperatures corresponding to these observables for physical quark masses are,
\[ T_c = \left\{
              \begin{array}{ll}
                   170(4)(3) & \text{for}~ L_R\\
                   157(3)(3) & \Delta_{l,s}\\
	           147(2)(3) & \chi_R
              \end{array}
       \right. 
\] 
The data for $L_R$ and $\Delta_{l,s}$ with the HISQ disretization, is shown in figure \ref{deltaLhisq}. These are for 
lattice of size $N_\tau\times (4 N_\tau)^3$. The HISQ data are in good agreement with the continuum extrapolated stout-smeared 
staggered results from \cite{bwtc}. The fact that the rise of $L_R$ is more gradual than the corresponding rise of $\Delta_{l,s}$ 
signals that the crossover is more likely influenced by the chiral symmetry restoration. 
Previous scaling studies of the renormalized chiral condensate with the p4-staggered quarks 
showed that the physical light quarks already approximate the O(4) critical behaviour of the chiral 
quarks \cite{mageos}. Using the O(4) scaling of the renormalized chiral condensate, the $T_c$ obtained for HISQ quarks 
through chiral and continuum extrapolation is $154\pm 9~$ MeV. This value is in excellent agreement with 
the stout result, implying that the continuum extrapolation done with the staggered fermions is quite robust. 

\subsection{Comparing results for different fermion discretizations}
\begin{figure}[h!]
\begin{center}
\includegraphics[scale=0.85]{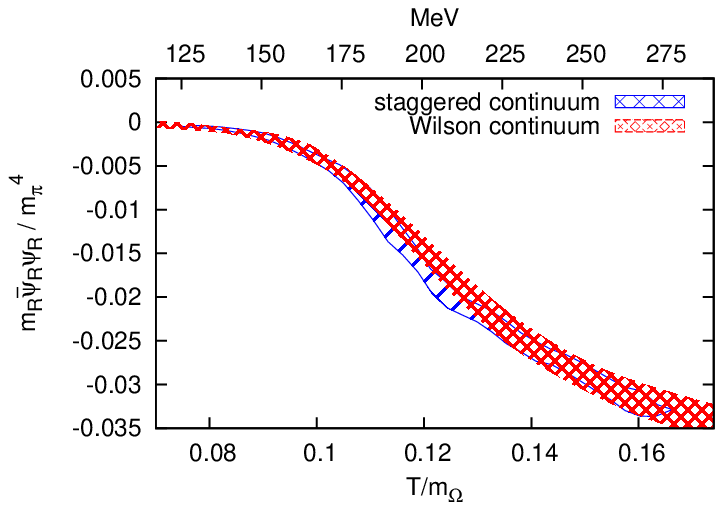}
\includegraphics[scale=0.85]{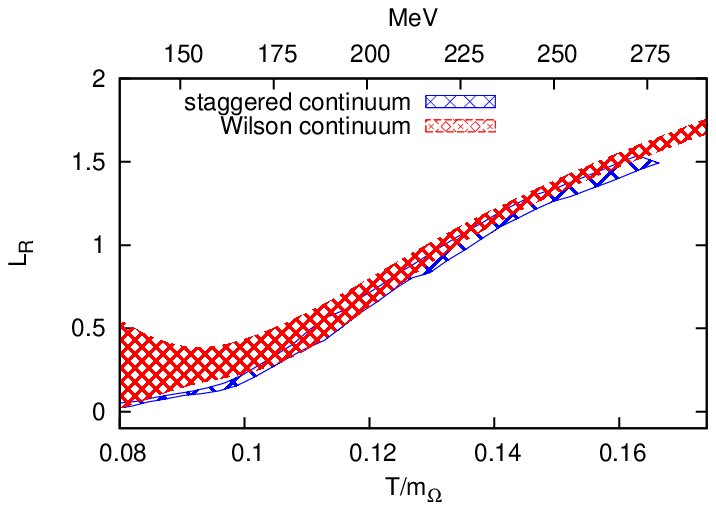}
\caption{The continuum extrapolated renormalized chiral condensate(left panel) and the Polakov loop (right panel) are compared 
for Wilson and stout-smeared staggered fermions, from \cite{bwwil}. }
\label{wil}
\end{center}
\end{figure}

The results for the EoS and the pseudo-critical temperature discussed so far, have been obtained using different improved  
versions of the staggered quarks. For these fermion species, the so called ``rooting'' problem may alter the continuum limit due 
to breaking of the $U_A(1)$ anomaly \cite{creutz} though some other work refutes this claim \cite{gsh}. It is important to
check the effects of the rooting procedure on the continuum 
extrapolation of finite temperature observables. The Budapest-Wuppertal collaboration has recently compared the 
continuum extrapolated results for different observables using the Wilson and staggered fermions \cite{bwwil} 
as the former discretization does not suffer from the rooting problem. 
The scale for the Wilson fermions was determined using $m_{\Omega}=1672~$ MeV and the line of constant physics 
was set using $m_\pi/m_{\Omega}\sim 0.3$ and  $m_K/m_{\Omega}\sim 0.36$. For the staggered quarks, the 
line of constant physics was set such that the ratios $m_\pi/m_{\Omega}$ and $m_K/m_{\Omega}$ are within 
3\% of the corresponding values for the Wilson fermions. This means that the pions are quite heavy 
with $m_\pi\sim 540~$ MeV for both these discretizations. The continuum extrapolated results for $L_R$ and the renormalized chiral condensate 
are shown in figure \ref{wil}. The continuum results for both these quantities are in good agreement for the whole range of temperature,
implying that these two different fermion discretizations indeed have the correct continuum limit.
In all these computations an improved Wilson operator was used, in which the dominant $\mathcal{O}(a)$ 
correction terms due to explicit breaking of chiral symmetry by these fermions were cancelled. It ensured that in both the studies the approach to the continuum limit was chosen to be the same. However at this large value of quark masses, the rooting problem may be mild enough to show any adverse effects and it would be desirable to perform a similar comparison at physical value of the quark masses. 

\begin{figure}[h!]
\begin{center}
\includegraphics[width=0.3\textwidth]{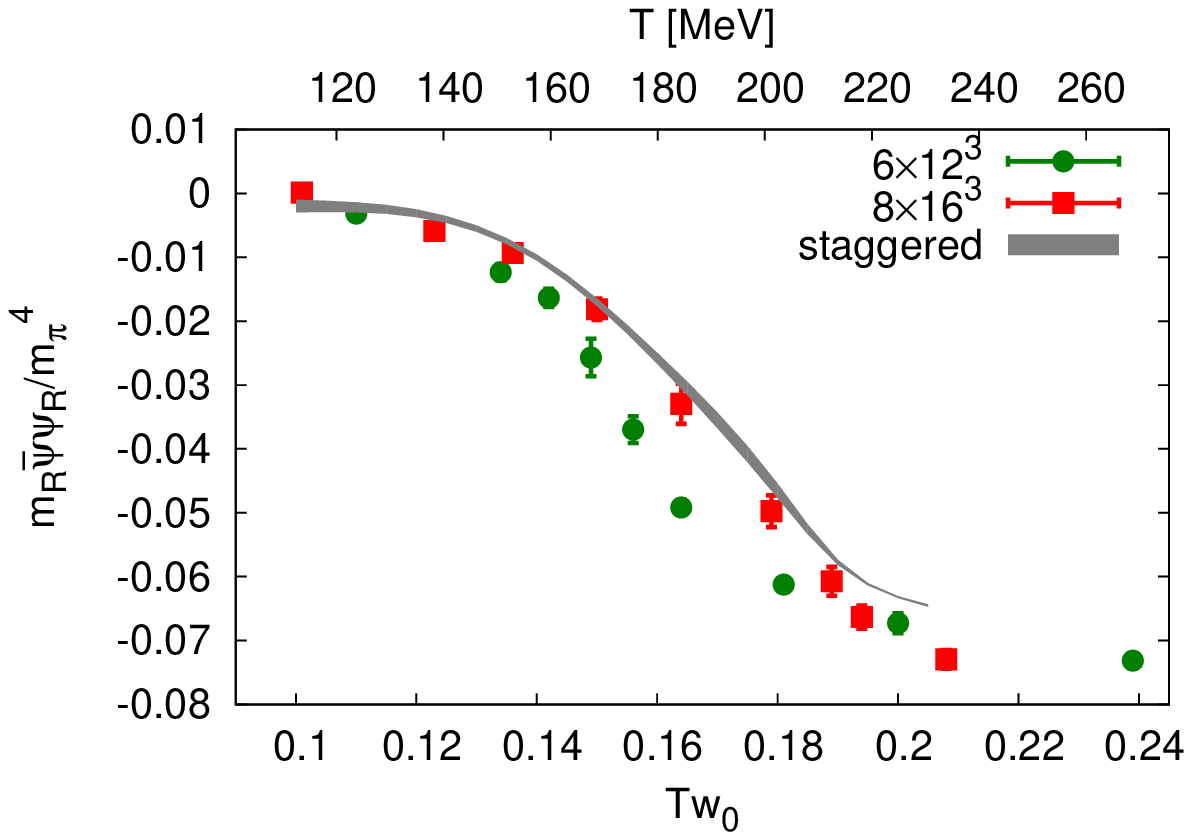}
\includegraphics[width=0.15\textwidth,angle=-90]{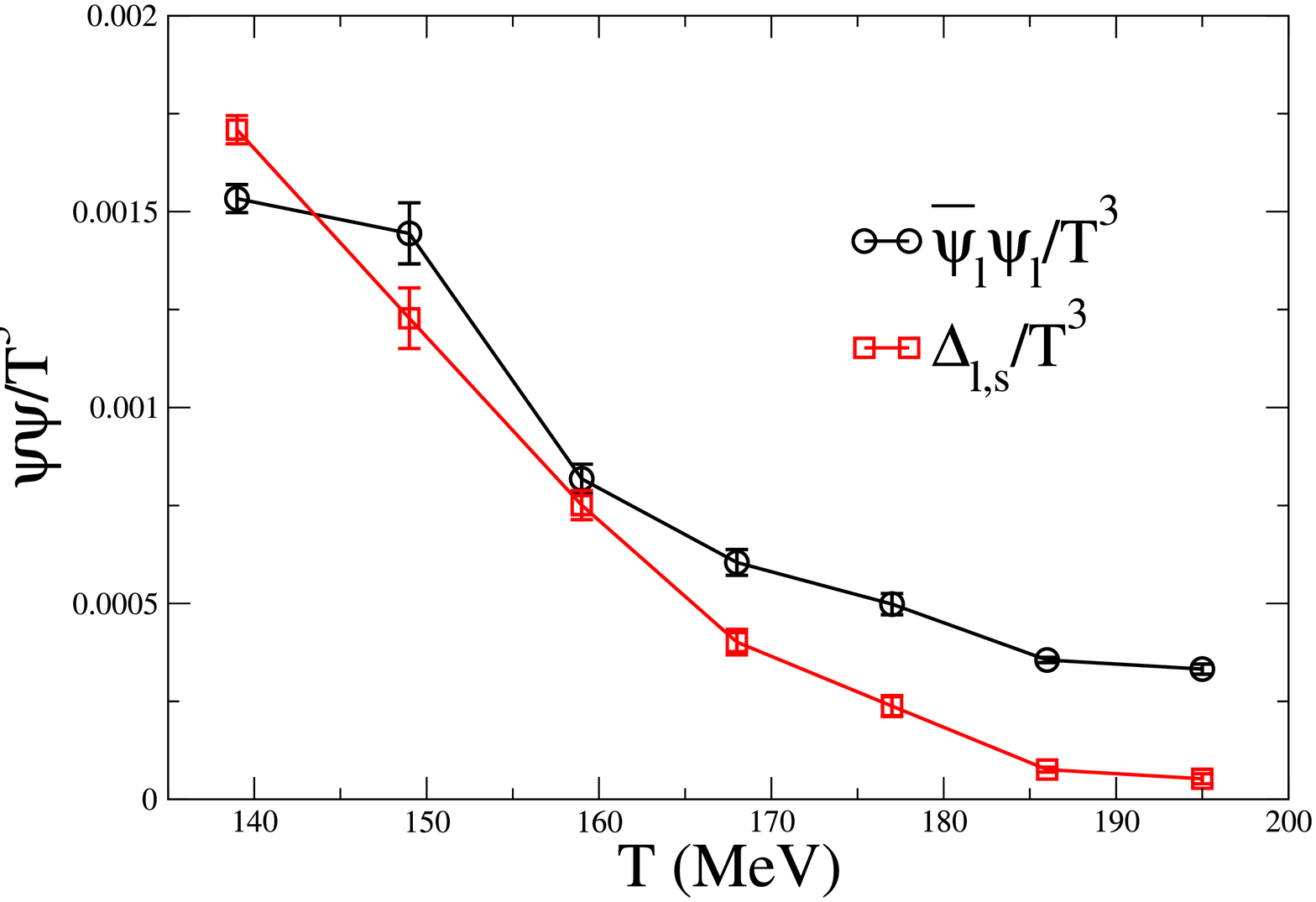}
\caption{The renormalized chiral condensate for the overlap quarks is compared to the continuum extrapolated results using 
the stout smeared staggered quarks in the left panel, from \cite{bwov}. In the right panel, the behaviour of different chiral 
condensates defined using the domain wall fermions are shown in the critical region, from \cite{hotqcddw}.}
\label{ppbchiral}
\end{center}
\end{figure}

Since the effects of chiral symmetry persist in the cross-over region, it is important to compare the existing 
results for $T_c$ with those using fermions with exact chiral symmetry on the lattice. For the Wilson and the 
staggered action, even for massless quarks, the full 
$SU(2)\otimes SU(2)$ chiral symmetry is realized only in the continuum limit. For chiral fermions on the lattice, like the overlap or the domain wall fermions, the chiral and the continuum limits are disentangled, allowing us to understand the remnant effects of chiral symmetry in the cross-over region even on a finite lattice. However, lattice QCD with overlap fermions is computationally prohibitive \cite{fodor} and currently better algorithms are being developed to simulate them with comparatively lesser effort \cite{fixedtop}. The domain wall fermions have exact chiral symmetry only when the extent of the fifth dimension, $N_5$, of the five dimensional lattice on which these fermions are defined, is infinite. For smooth gauge fields, the chiral symmetry violation on a finite lattice is suppressed as an exponential of $N_5$ but the suppression could be much slower, as $1/N_5$ for rough gauge configurations in the crossover region. Better algorithms have been employed to ensure exponential suppression even 
for rough gauge fields \cite{hotqcddw}. The most recent results for the overlap fermions from the Budapest-Wuppertal collaboration \cite{bwov} and the domain wall fermions from the HotQCD collaboration \cite{hotqcddw}, are shown in figure \ref{ppbchiral}. The renormalized chiral condensate for the overlap fermions are qualitatively consistent with the continuum staggered fermion results, even for small volumes and large pion masses of about 350 MeV around the crossover region. The 
lattice cut-off effects seem to be quite small for $N_\tau=8$.
The renormalized chiral condensate and the $\Delta_{l,s}$ for the domain wall fermions are shown in figure \ref{ppbchiral}.
The lattice size is $16^3\times8$ with the number of lattice sites along the fifth dimension taken to be 32 for $T>160~$MeV and 
48 otherwise and the pion mass is about 200 MeV. The lattice volume is comparatively small therefore these results do not show a sharp rise
in the crossover region. With larger volumes, the rise in these thermodynamic quantities is expected to be much steeper. The value of $T_c$ estimated from the peak of the chiral susceptibility i.e the derivative 
of the chiral condensate is between 160-170 MeV which is consistent with the continuum results from the 
HISQ fermions.

\subsection{The thermodynamical observables}
\begin{figure}[h!]
\begin{center}
\includegraphics[scale=0.25]{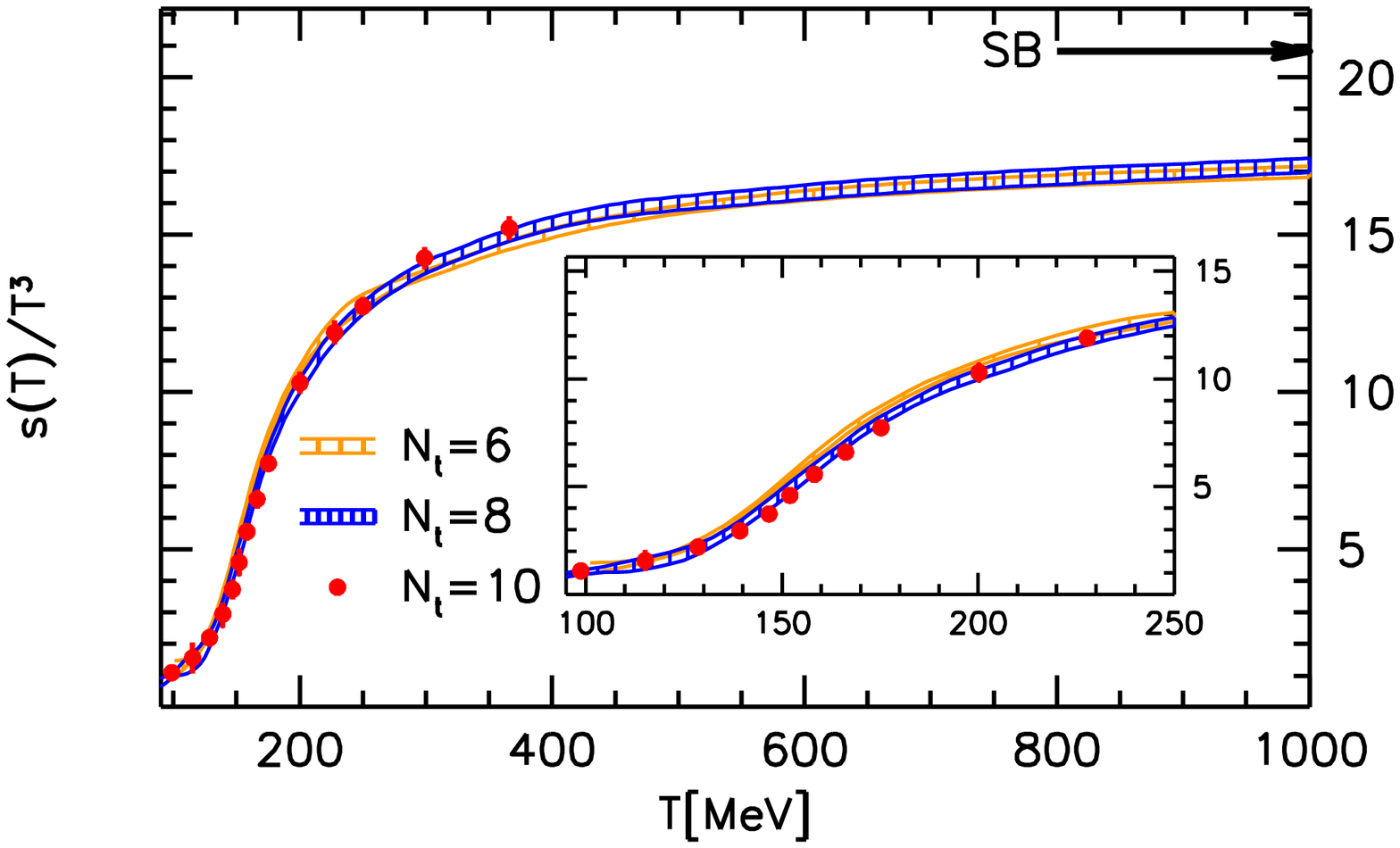}
\includegraphics[scale=0.25]{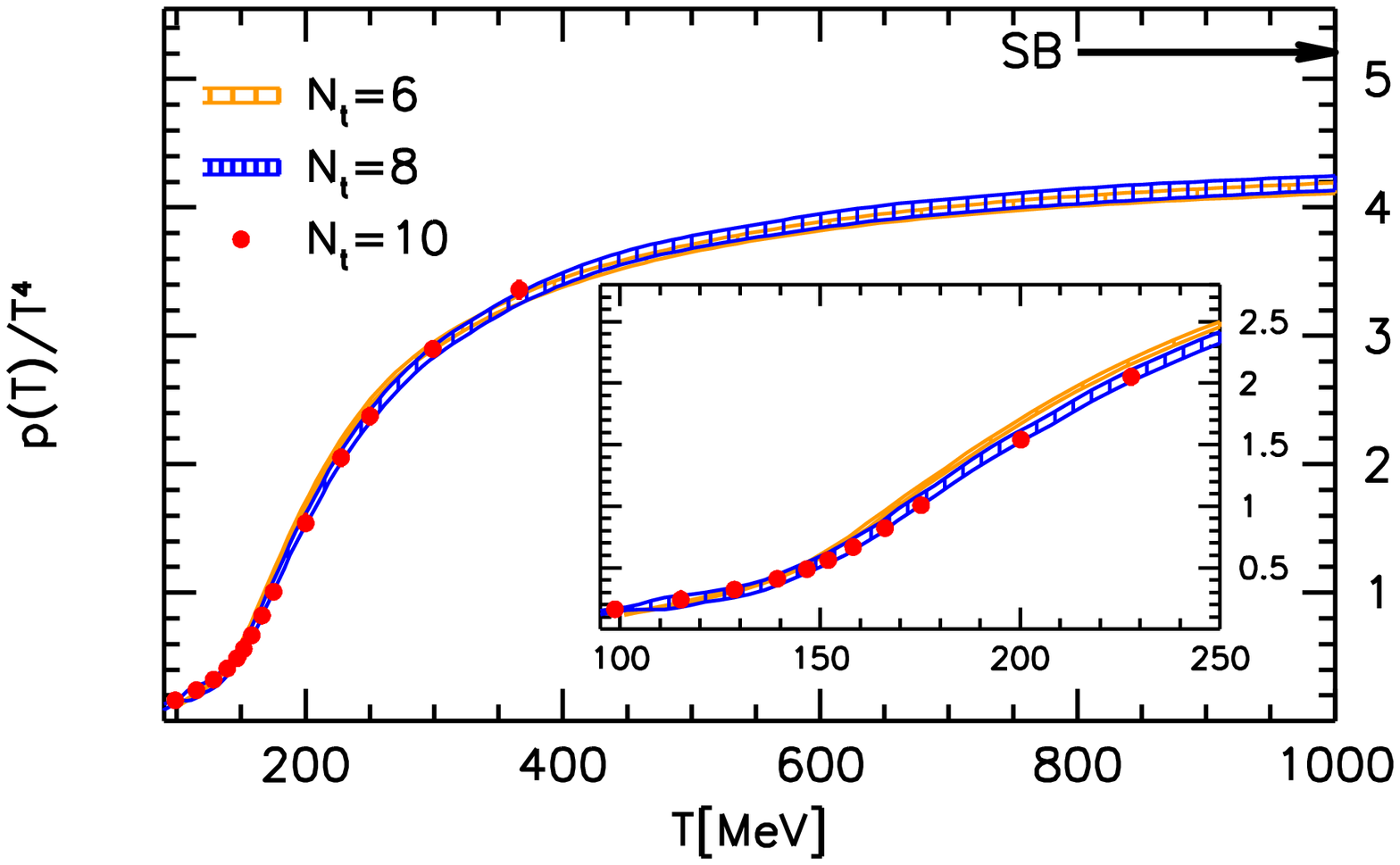}
\includegraphics[scale=0.25]{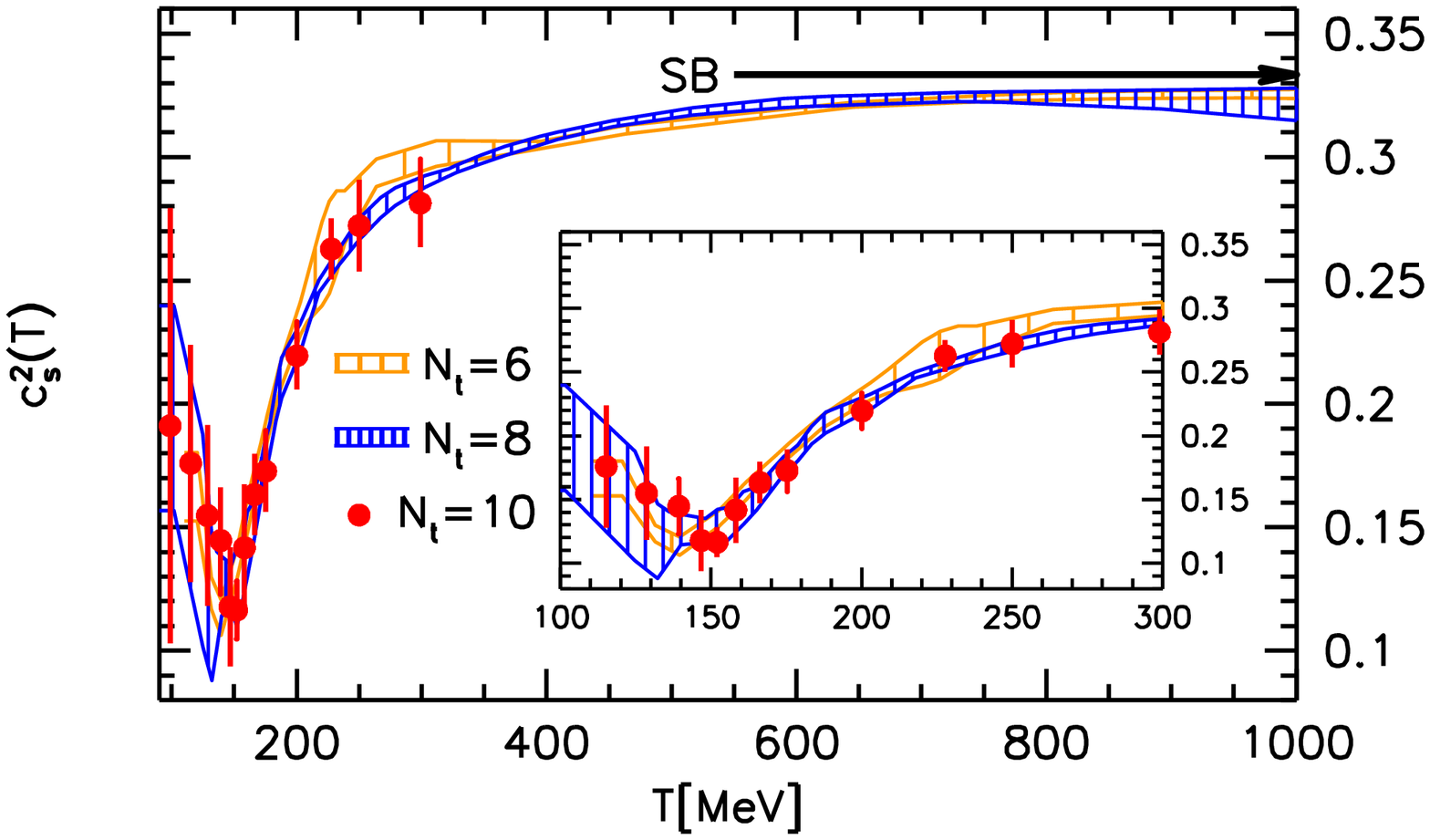}
\vspace{-2 cm}
\caption{The entropy density, pressure and the speed of sound for the stout-smeared fermions as a function of temperature, 
from ~\cite{bw1}. }
\label{obsv}
\end{center}
\end{figure}

Thermodynamic observables characterize the different phases across a phase transition. 
From the behaviour of these observables, one can infer about the degrees of freedom of 
the different phases and the nature of the interactions among the constituents. It was already 
known from an important lattice study that the pressure in high temperature phase of QCD showed a strong 
dependence on the number of quark flavours~\cite{peikert}, signaling deconfinement of the 
quark and gluon degrees of freedom. 
Recent results for the pressure, entropy density and the speed of sound for QCD, using the 
stout-smeared staggered quarks are compiled in figure \ref{obsv}. Though in our 
world there is no real phase transition, the entropy density increases rapidly with temperature, 
again signaling the liberation of a large number of colour degrees of freedom. The entropy density for QCD
is almost 20\% off from the value of a free gas of quarks and gluons, even at temperatures about 1000 MeV. 
The deviation of the pressure of QGP computed at similar temperatures, from its free theory value is even 
more, close to about 25\% of its value. Another observable that characterizes the different phases is 
the speed of sound $c_s$. If QGP at high temperatures was qualitatively close to a strongly interacting 
conformal theory, then the speed of sound would be exactly $1/\sqrt{3}$. However the deviation from conformality is quite 
significant even at temperatures about $T=500 ~$MeV which hints that the AdS-CFT inspired study of the QGP medium should 
be done with more care.
The  values of entropy density computed with different discretizations of staggered fermions like the asqtad or the 
p4 fermions \cite{petreczkymunich} show about $10\%$ deviation from the free theory value at very high temperatures. 
The departure from AdS-CFT values is even more severe using these fermions. The stout results are about $10\%$ lower 
than the corresponding asqtad and p4 results. This deviation is attributed to the fact that the latter discretizations have smaller 
cut-off effects at higher temperatures and would be more closer to the continuum results. The stout continuum values shown in the 
figure were obtained by averaging the $N_\tau=8,10$ data. A proper continuum extrapolation of the results for both the fermion discretizations 
is necessary for resolving the difference and for use of these values for the real world calculations. However, the lattice results 
with at least $10\%$ off from the free theory values even at very high temperatures, implies that the QGP phase is 
strongly interacting, more like a liquid rather than a gas of quarks and gluons, confirming the similar prediction 
from the RHIC experiments.  
For $T<T_c$, the results for all these observables are in agreement with Hadron resonance gas model predictions.

\subsection{Effects of charm quarks on the EoS}
The effects of charm quarks to the pressure in the QGP phase was estimated sometime ago, using next-to leading order 
perturbation theory \cite{charmpt}. It was observed that the contribution of charm quarks become significant 
for temperatures $T>2T_c$. Preliminary data from the LHC already indicates that the charm quarks 
would thermalize quickly as the lighter quarks. It would then affect the EoS and thus the hydrodynamical 
evolution of the fireball formed at LHC energies. Lattice studies are important to quantify the contribution of charm to 
the EoS in the QGP phase. The first lattice studies were done by the RBC \cite{mcheng} as well as the MILC collaboration \cite{milc1}
with quenched charm quarks, i.e., by neglecting quantum fluctuations due to the charm quarks. The quenched charm results for the EoS  
differ from the 2+1 flavour results, already at 1.2 $T_c$. 
Recent results from the Budapest-Wuppertal collaboration with dynamical charm quarks \cite{bwcharm}, however, show that the effects of charm quarks show up only around 300 MeV, more in agreement with the perturbative estimates. 
Both the approaches highlight the fact that the effects of charm quark should be considered for the EoS used as an input for the hydrodynamical evolution of the fireball at LHC energies, which may set in at $T\sim 500$ MeV. It would be also important for the EoS of the Standard Model, important for the cosmological evolution in the early universe~\cite{hind,soldner}. 
\begin{figure}[h!]
\begin{center}
\includegraphics[width=0.27\textwidth]{charmeosmilc.eps}
\includegraphics[width=0.3\textwidth]{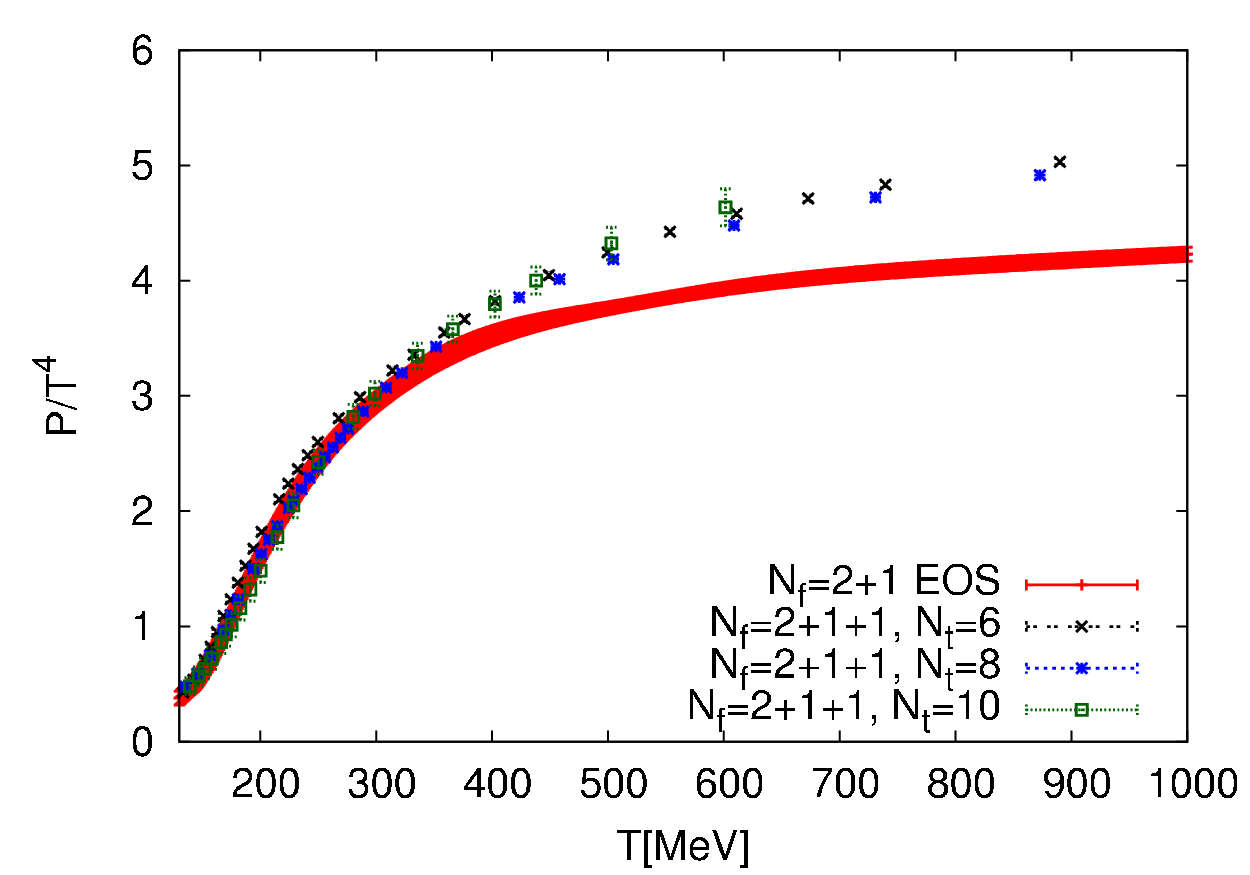}
\caption{In the left panel, the effects of quenched charm quark to the pressure, energy density and trace anomaly are shown as a function of temperature, from \cite{milc1}. The lattice size is $24^3\times 6$. In the right panel, the effects of dynamical charm quarks to the pressure is shown as a function of temperature, from \cite{bwcharm}.}
\label{eoscharm}
\end{center}
\end{figure}

\subsection{The 2 flavour QCD transition and the fate of the $U_A(1)$ anomaly}
The chiral phase transition for $N_f=2$ QCD is still not well understood from lattice studies, as was emphasized at the beginning of this section. 
Though the lattice results for $2+1$ flavours  with different fermion discretizations are in good agreement, the corresponding ones for the two 
light flavour case are still inconclusive.  Two major approaches have been undertaken in the recent years to understand the order of this transition. 
One of them is to check the scaling properties of the order parameter. If the phase transition is indeed a second order one, then the order 
parameter would show O(4) scaling  in the transition region. The second approach is to understand the effects of the $U_A(1)$ anomaly near in 
the phase transition. If the quantum fluctuations responsible for this $U_A(1)$ anomaly decrease significantly with temperature, it would 
result in the degeneracy of the masses of mesons of certain quantum numbers and a characteristic behaviour of 
the density of low lying eigenmodes of the fermion operator. I discuss the major lattice results using both these approaches, 
in the following paragraphs. Most of these approaches are hinting that the two flavour chiral phase transition may be a second order one.

\subsubsection{Scaling analysis in the critical region}
\label{sec:curv} 
\begin{figure}[h!]
\begin{center}
\includegraphics[width=0.3\textwidth]{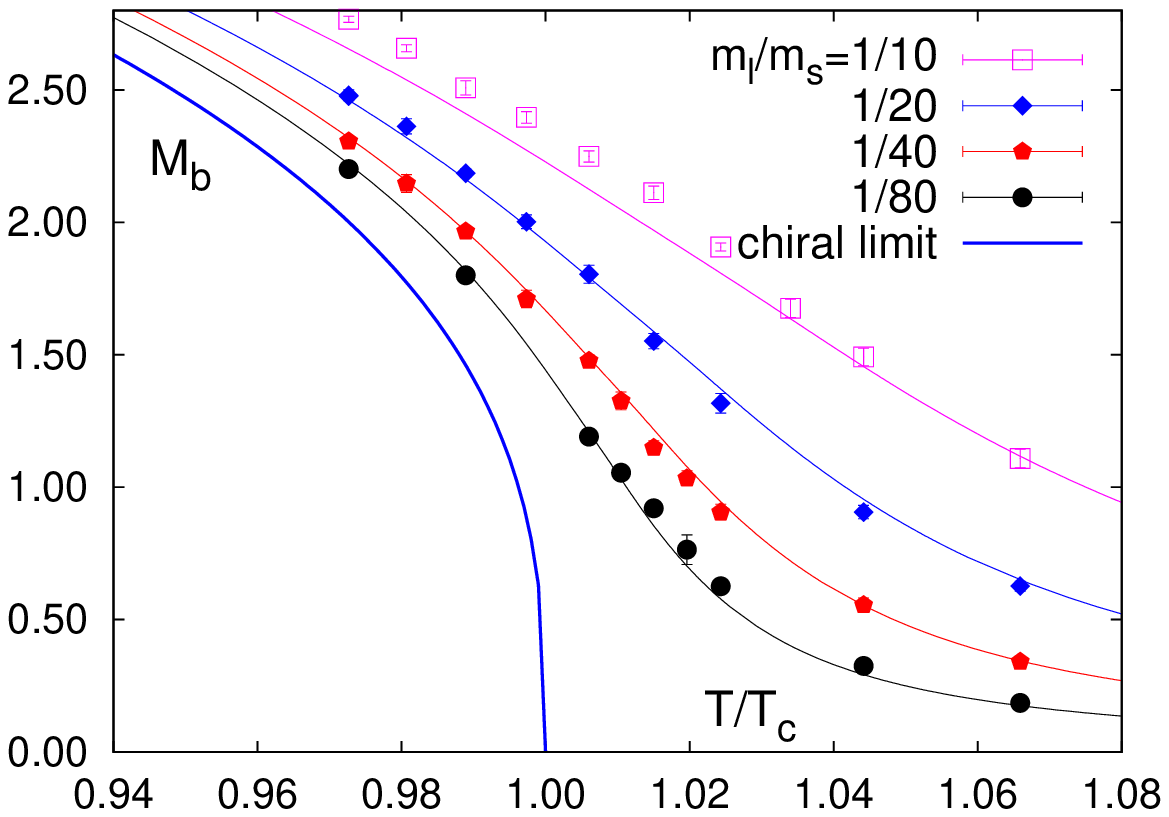}
\includegraphics[width=0.3\textwidth]{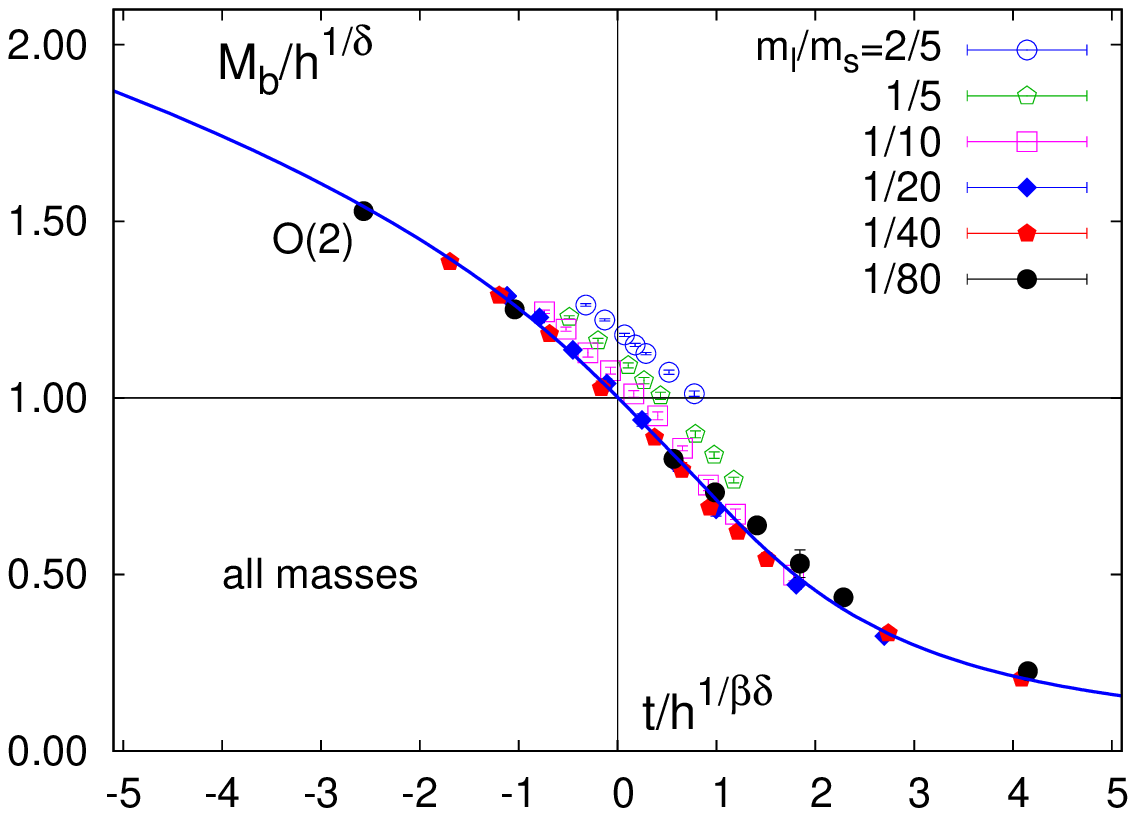}
\caption{The interpolated data for $M_b$ for different light quark masses are compared with the 
corresponding plot for an O(4) spin model in the continuum, denoted by the solid blue line(left panel).
In the right panel, the scaling plots for the renormalized chiral condensate for QCD are shown to match with 
the universal function with O(2) symmetry for $m_l/m_s<1/20$. Both the plots are for p4 staggered quarks, 
from \cite{mageos}.}
\label{meos}
\end{center}
\end{figure}

The order parameter that characterizes the chiral phase transition is the chiral condensate. 
A suitable dimensionless definition of the chiral condensate used in the lattice study 
by the BNL-Bielefeld collaboration \cite{mageos} is,
\begin{equation}
\label{eqn:odparam}
 M_b=m_s\frac{\langle\bar\psi\psi\rangle}{T^4}
\end{equation}
The additive ultraviolet divergences are not explicitly subtracted from the condensate and hence it is the bare
value denoted by subscript b. This additive divergence would be included 
in the regular part and in the transition region, would be much smaller in magnitude than the singular part of $M_b$. In the 
vicinity of the transition region, the order parameter can be written as,
\begin{equation}
\label{eqn:func}
 M_b(T,H)=h^{1/\delta}f_G(t/h^{1/\beta\delta})+f_{reg}(T,H)~,
\end{equation}
where  $f_G$ is the universal scaling function, known from analysis of the $O(N)$ spin-models \cite{on1,on2,on3} with $\beta$ and
$\delta$ being the corresponding critical exponents. The quantities $h$ and $t$ are 
dimensionless parameters that determine the deviations from the critical point and are defined as,
\begin{equation}
 t=\frac{1}{t_0}\frac{T-T_{c,0}}{T_{c,0}}~,~h=\frac{H}{h_0}~,~H=\frac{m_l}{m_s}~,
\end{equation}
 with $T_{c,0}$ being the transition temperature in the chiral regime i.e, for $h\rightarrow0$ and 
$h_0$ and $t_0$ are non universal constants. One of the choice of the regular part of the order 
parameter used in the lattice study is,
\begin{equation}
 f_{reg}=H\left(a_0+a_1 \frac{T-T_{c,0}}{T_{c,0}}+a_2 \left(\frac{T-T_{c,0}}{T_{c,0}}\right)^2\right)~,~
\end{equation}
where one assumes that the regular part is an analytic function of the relevant parameters 
around the transition point. The BNL-Bielefeld collaboration used an improved 
variety of the staggered quarks, called the p4 quarks, to compute the order parameter defined in 
Eq. (\ref{eqn:odparam}) and $\chi_m$, its derivative with respect to $m_l$ for different 
values of the light quark masses, $m_l$. The strange quark mass was fixed at its physical value.
These quantities were fitted to the functional form given in Eq. (\ref{eqn:func}) and its derivative
respectively. The scaling analysis was done for a fixed lattice of size $N^3\times4$, so
the order parameter and its derivatives are expected to have an O(2) scaling in the chiral regime 
since  the fermion discretization only retains a remnant of the continuum O(4) symmetry group.  
From the plots for the order parameter in the left panel of figure \ref{meos}, it is evident that for $m_l/m_s=1/80$ the phase 
transition is indeed a second order one with O(2) critical exponents, though O(4) scaling cannot 
be ruled out completely with the current precision available.  In the scaling regime, 
the variable $M_b/h^{1/\delta}$ should be a universal function of $t/h^{1/\beta\delta}$. 
In the right panel of figure \ref{meos}, the scaled chiral condensate is seen to be 
almost universal for $m_l/m_s<1/20$,  which provides a hint that even for the physical 
quark masses there is a remnant effect of the chiral symmetry. The crossover transition 
for 2+1 flavour QCD should be sensitive to the effects of chiral symmetry and therefore
also to the effects of the $U_A(1)$ anomaly.   
 
\subsubsection{The effects of $U_A(1)$ anomaly}
\begin{figure}[h!]
\begin{center}
\includegraphics[width=0.27\textwidth]{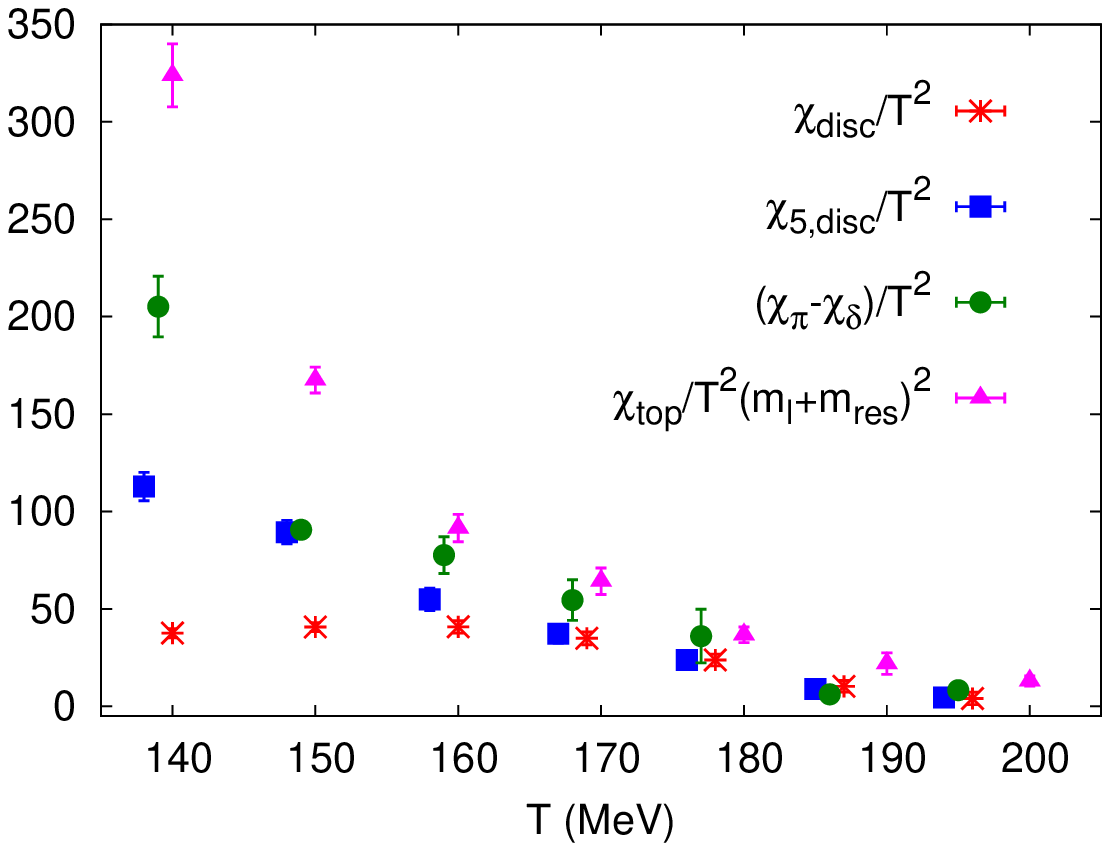}
\includegraphics[width=0.3\textwidth]{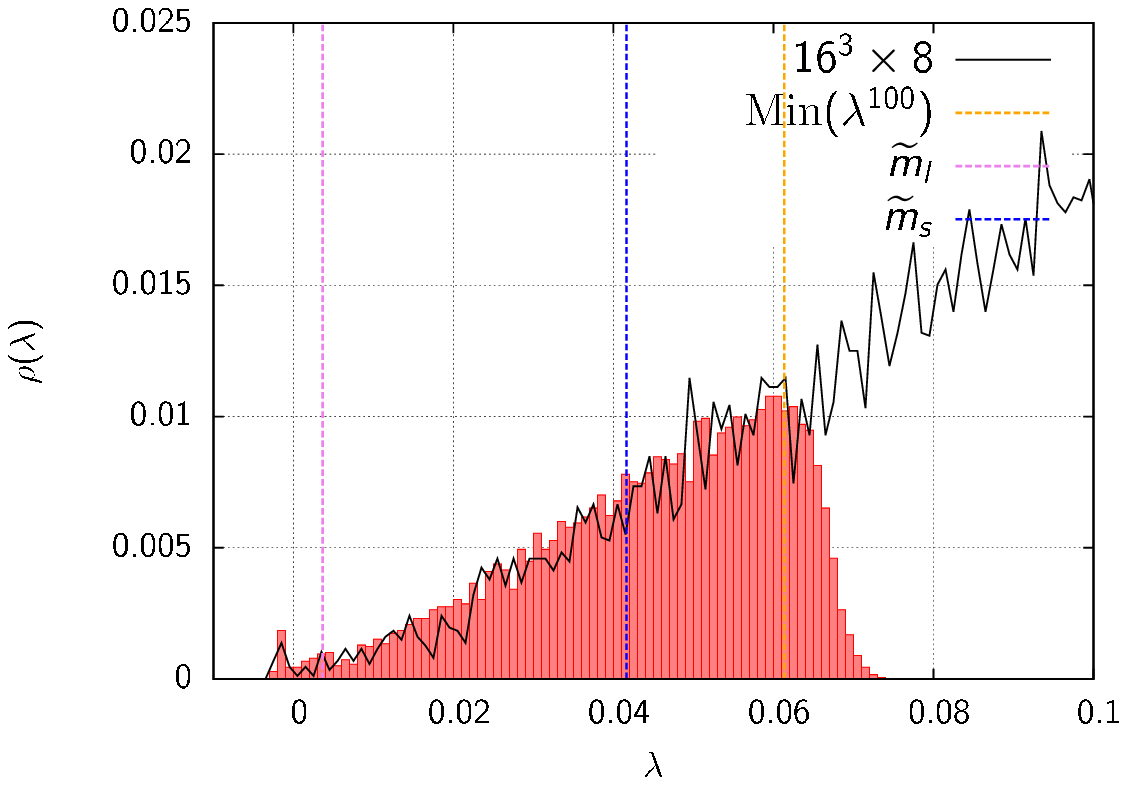}
\caption{The susceptibilities for different meson quantum states constructed with the domain wall fermions are shown as a function 
of temperature in the left panel, from \cite{hotqcddw}. The eigenvalue distribution with domain wall fermions, shown
in the right panel, from \cite{hotqcddwlat}, has a peak in the near zero mode distribution at 177 MeV. The lattice size is 
$16^3\times 8\times N_5$ where $N_5=32$ for $T\geq 160~$MeV and $N_5=48$ otherwise.}
\label{dweigen}
\end{center}
\end{figure}

The QCD partition function breaks $U_A(1)$ symmetry explicitly. However its effects varies with 
temperature since we know that at asymptotically high temperatures, we approach the ideal Fermi gas 
limit where this symmetry is restored. It is important to understand the temperature dependence 
of $U_A(1)$ breaking near the chiral phase transition. If $U_A(1)$ breaking is significantly reduced 
from that at zero temperature, one would then claim that the symmetry is effectively restored.
This would result in the degeneracy of the mass of the isospin triplet pseudoscalar(pion) and scalar
(delta) mesons. The order parameter for such an effective restoration is the quantity defined as,
\begin{equation}
 \chi_\pi-\chi_\delta=\int~d^4 x~ \left[\langle\bar\psi(x)\tau\gamma_5\psi(x)\bar\psi(0)\tau\gamma_5\psi(0)\rangle
-\langle\bar\psi(x)\tau\psi(x)\bar\psi(0)\tau\psi(0)\rangle\right]~,
\end{equation}
and the order parameter for the restoration of the chiral symmetry is the chiral condensate. 
These quantities are also related to the fundamental theory through the density of eigenvalues, 
$\rho(\lambda)$ of the Dirac operator as,  
\begin{eqnarray}
 \nonumber
\langle\bar\psi\psi\rangle&=&\int d~\lambda \rho(\lambda,m)\frac{2 m}{m^2+\lambda^2}\\
\chi_\pi-\chi_\delta&=&\int d~\lambda \rho(\lambda,m)~\frac{4 m^2}{(m^2+\lambda^2)^2}~.
\end{eqnarray}
Different scenarios that could lead to different functional behaviour of $\rho(\lambda)$ were 
discussed in detail in Ref. \cite{hotqcddw}. I summarize the arguments below,
\begin{itemize}
 \item From dilute instanton gas approximation: $\rho(\lambda,m)=c_0m^2\delta(\lambda)~\Rightarrow$
$\langle\bar\psi\psi\rangle\sim m$ and $\chi_\pi-\chi_\delta\sim2$
\item Analyticity of $\rho(\lambda,m)$ as a function of $\lambda$ and $m$ when chiral symmetry 
is restored. To the leading order $\rho(\lambda,m)=c_m m+c_\lambda \lambda+\mathcal{O}(m^2,\lambda^2)$.

If $\rho(\lambda,m)\sim \lambda~\Rightarrow \langle\bar\psi\psi\rangle\sim -2m\ln m~,~
\chi_\pi-\chi_\delta\sim2$.

If $\rho(\lambda,m)\sim m~\Rightarrow \langle\bar\psi\psi\rangle\sim \pi m~,~
\chi_\pi-\chi_\delta\sim\pi$.
\end{itemize}
In fact to understand the effect of anomaly, it is desirable to use fermions with exact chiral 
symmetry on the lattice. The overlap and the domain wall fermions are such candidates, for which the chiral anomaly 
can be defined. Indeed, the overlap fermions satisfies an exact index theorem on the lattice~\cite{hln}.  
A recent study of the eigenvalue spectrum with the domain wall fermions from the HotQCD collaboration 
\cite{hotqcddwlat} seems to favour $\rho(\lambda,m)=c_0m^2\delta(\lambda)+c_1 \lambda$, for
the density of eigenvalues. This would imply that in the chiral limit, the $U_A(1)$ anomaly would still survive when 
the chiral symmetry is restored. This is also consistent with the behaviour of $\chi_\pi-\chi_\delta$ 
as a function of temperature, shown in the left panel of figure \ref{dweigen}. At crossover temperature around 
160 MeV, the $\chi_\pi-\chi_\delta$ is far from zero, implying that the effects of the anomaly may be large 
in the crossover region.

\begin{figure}[h!]
\begin{center}
\includegraphics[width=0.4\textwidth]{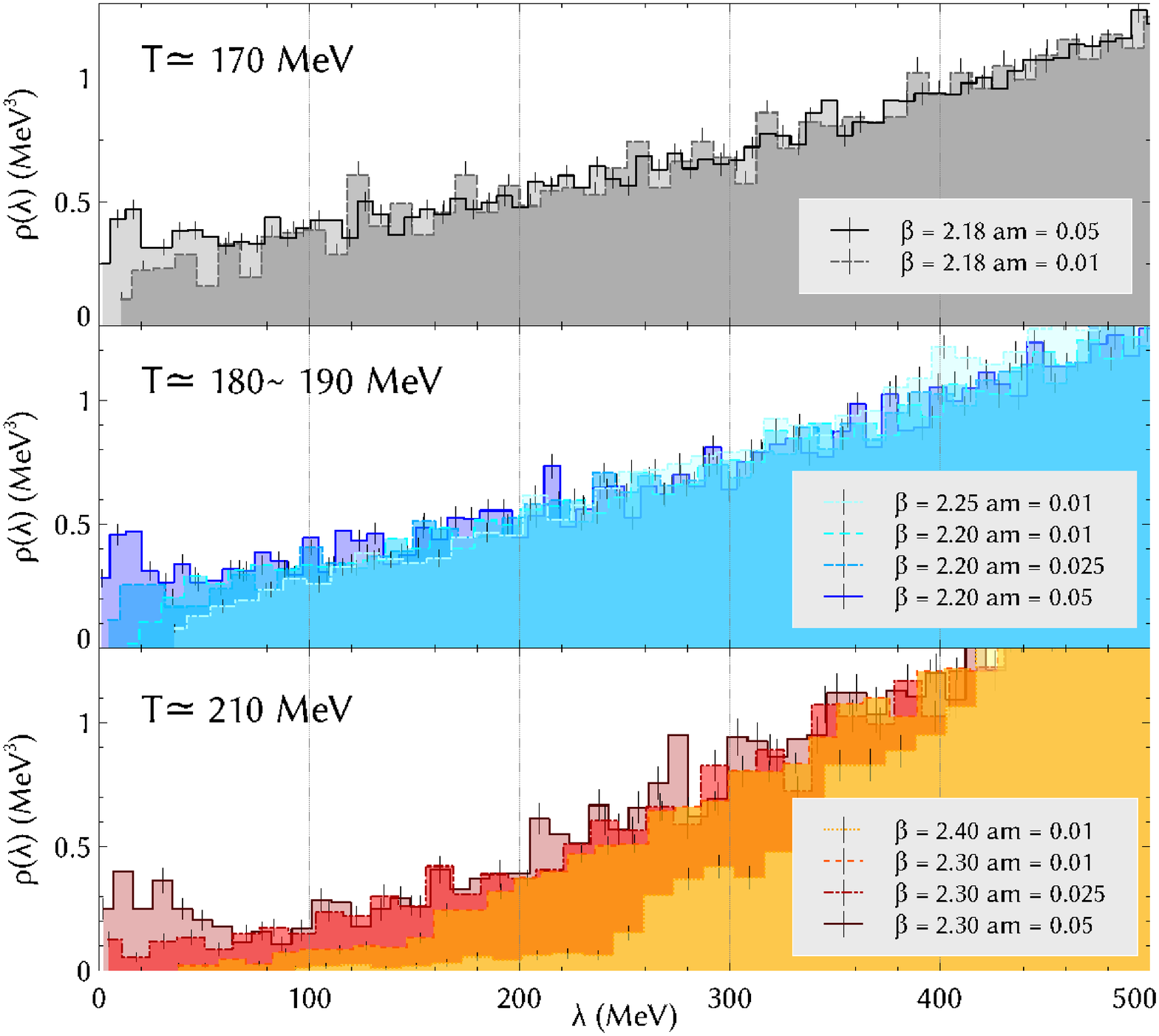}
\includegraphics[width=0.4\textwidth]{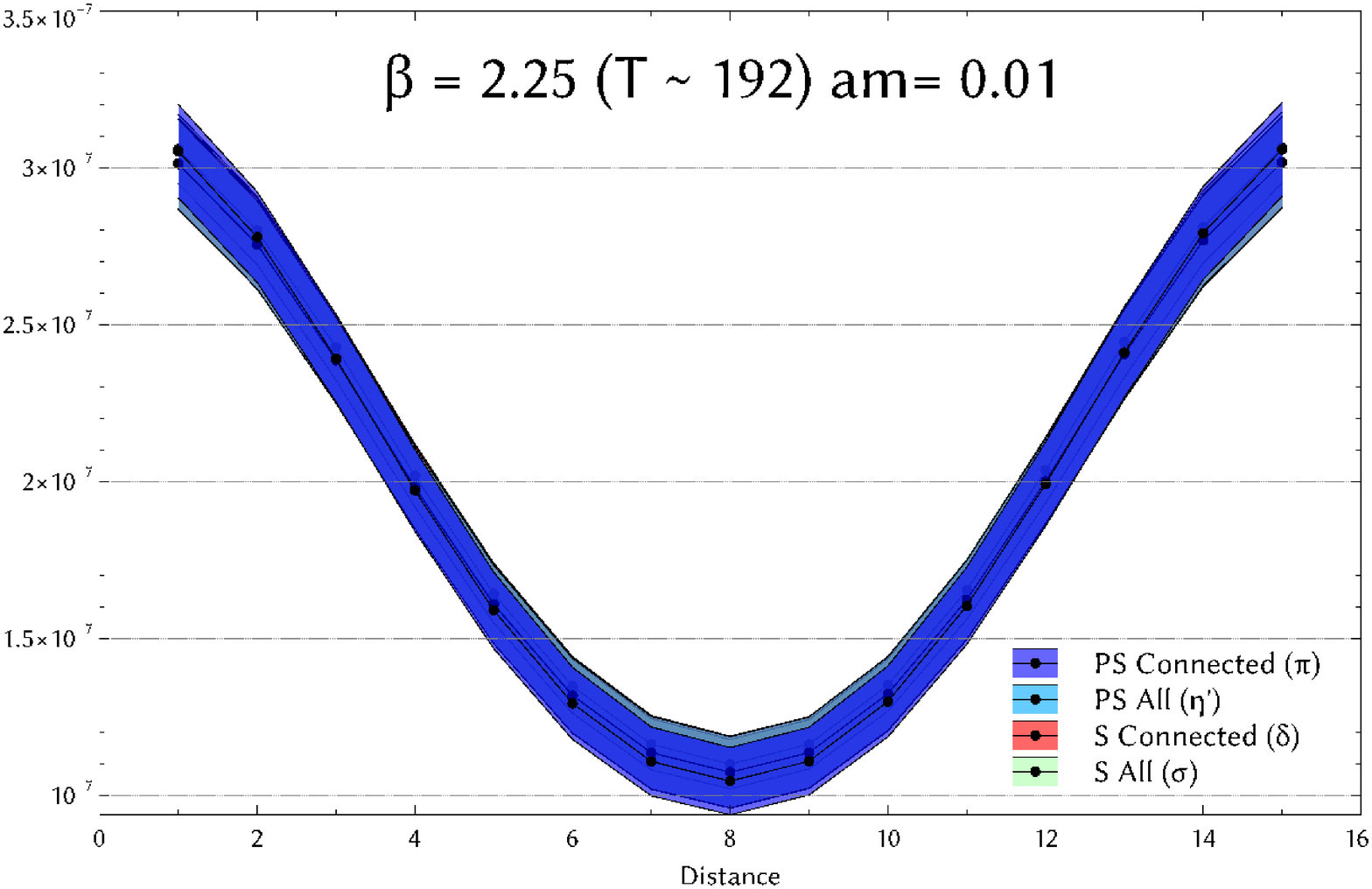}
\caption{In the left panel, the quark mass dependence of eigenvalue distribution for the overlap quarks are compared 
at different temperatures, from \cite{cossu2}. In the right panel, the degeneracy of the scalar and pseudoscalar mesons for 
overlap fermions are shown at a temperature of 192 MeV which is slightly higher than the corresponding pseudo-critical temperature,
from \cite{cossu2}.}
\label{oveigen}
\end{center}
\end{figure}

A recent theoretical study \cite{aoki} with the overlap fermions shows that in the chiral symmetry 
restored phase where $\langle\bar\psi\psi\rangle=0$, the eigenvalue density in the chiral 
limit should behave as,
\begin{equation}
 lim_{m\rightarrow0}\langle\rho(\lambda,m)\rangle=lim_{m\rightarrow0}
\langle\rho(m)\rangle\frac{\lambda^3}{3!}+\mathcal{O}(\lambda^4).
\end{equation}
 which would imply that $\chi_\pi-\chi_\delta\rightarrow0$ as $m\rightarrow0$. 
Moreover it is argued that if an operator is invariant under some symmetry transformation 
then its expectation value becoming zero would not necessarily imply that the symmetry is 
restored, whereas the converse is true \cite{aoki}. This would mean that the observable $\chi_\pi-\chi_\delta$
may not be a good candidate to study the $U_A(1)$ restoration. Rather the equality of the 
correlators of the pion and delta meson could be a more robust observable to indicate the 
restoration of the $U_A(1)$ symmetry. Recent results from the JLQCD collaboration with 
2 flavours of overlap fermions seems to indicate that the $U_A(1)$ may be restored near the chiral symmetry 
restoration temperature, making it a first order transition \cite{cossu2}. Two of their main results are compiled in 
figure \ref{oveigen}. The correlators of the scalar mesons become degenerate at about 196 MeV and at the same 
temperature  a gap opens up in the small eigenvalue region of the eigenvalue spectrum. $T=196~$MeV 
is slightly above the transition temperature which is near about 177 MeV. For $T=177~$ MeV, there is no degeneracy between the 
scalar and the pseudoscalar correlators and the density of zero modes is finite implying that the 
chiral symmetry is broken, which means that the $U_A(1)$ changes rapidly near the phase transition. 
However the lattice size is $16^3\times8$, which is small enough to introduce significant finite volume and cut-off 
effects in the present results.

\begin{figure}[h!]
\begin{center}
\includegraphics[width=0.3\textwidth]{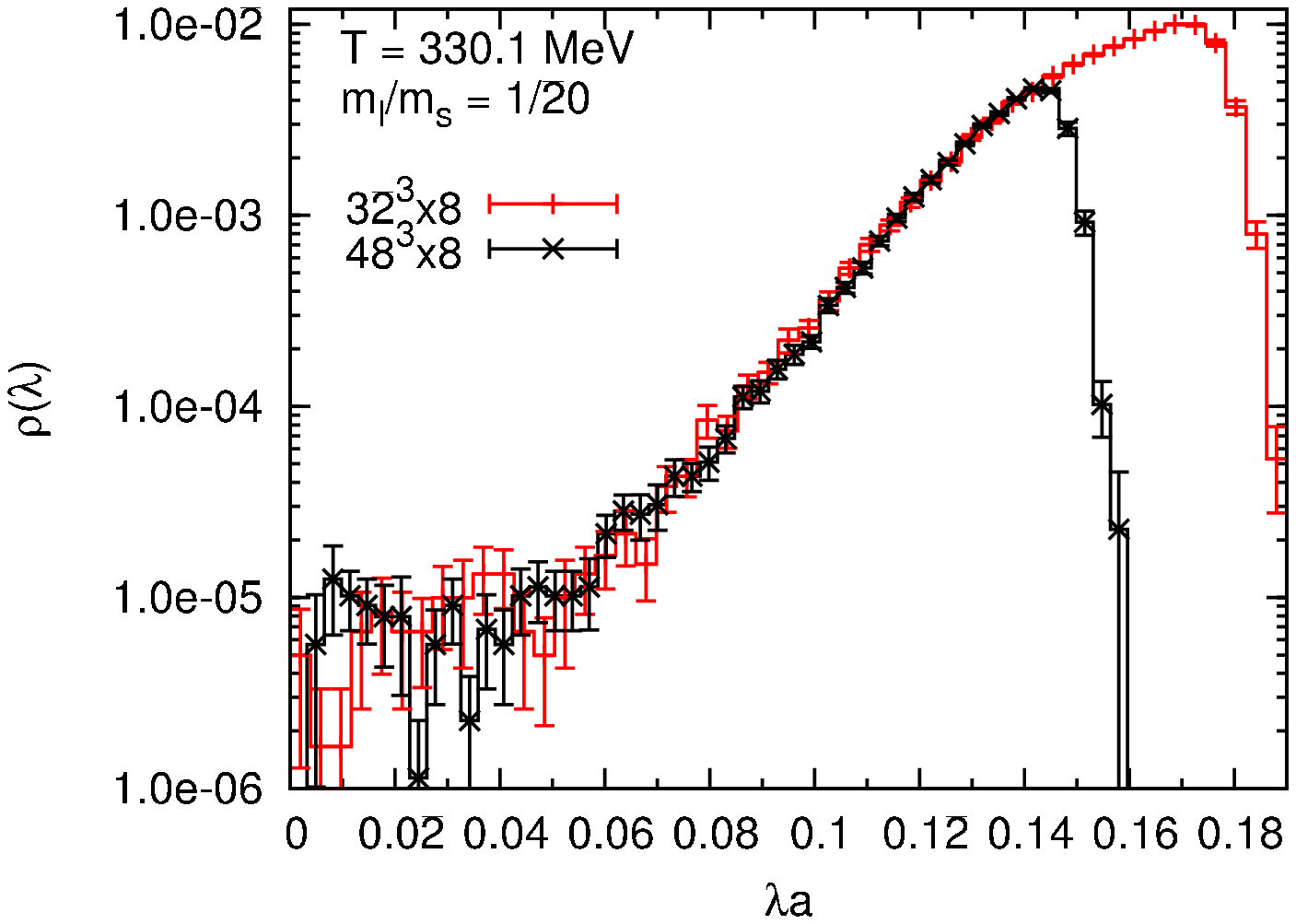}
\includegraphics[width=0.4\textwidth]{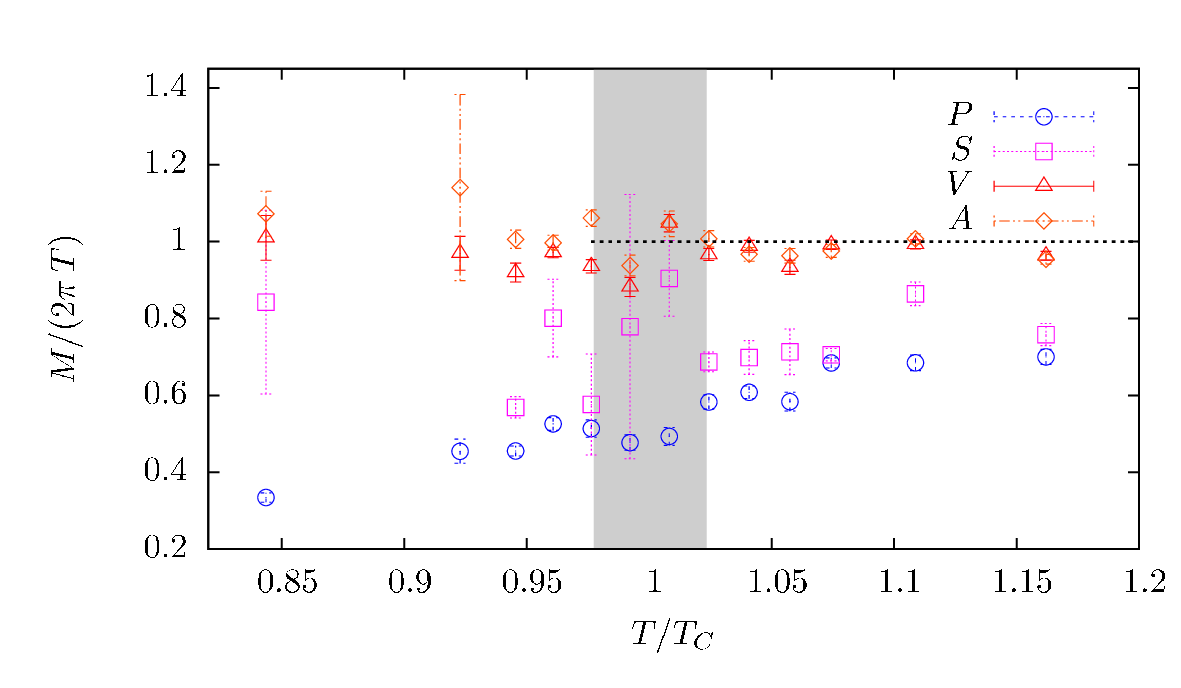}
\caption{The density of eigenvalues at $T=330.1$ MeV for HISQ discretization showing a long tail even with large 
volumes, from \cite{hiroshi}(left panel). In the right panel, the screening masses for scalar, pseudo-scalar, vector and 
axial vector mesons using Wilson fermions are shown as a function of temperature, from \cite{phwil}.}
\label{wilhisqeigen}
\end{center}
\end{figure}

With the chiral fermions, the fate of $U_A(1)$ in the crossover region is still undetermined and more work needs to be done 
for conclusive understanding of this issue. With Wilson and staggered quarks, the anomaly is recovered only in the continuum limit. For 
fine enough lattice spacings, one can however check the behaviour of the low lying eigenmodes and the meson masses for 
different quantum numbers, to understand the effects of the remnant $U_A(1)$ anomaly using these fermions. From the eigenvalue distribution 
of HISQ operator shown in the left panel of figure \ref{wilhisqeigen}, it is evident that the effect of $U_A(1)$ 
still persists at $T=330~$ MeV~\cite{hiroshi}. The long tail in the low lying eigenmodes is not a finite volume artifact since it persists even for 
very large volumes. However, the data is quite noisy and more statistics is required for making a final conclusion. The 
screening masses for the mesons of different quantum numbers were obtained from lattice studies~\cite{phwil} with improved 
Wilson fermions(right panel of figure \ref{wilhisqeigen}). In the transition region, the scalar and pseudoscalar mesons 
are not degenerate and an agreement seen only for temperatures above 1.2 $T_c$. However the input quark masses are quite  
large compared to the physical values and more data is needed to take a final call. At present, the effects of quantum 
anomalies are not yet understood from lattice studies.

\section{Lattice QCD at finite density}
QCD with a finite number of baryons is relevant for the physics of neutron stars and supernovae. It is  
the theoretical setup for the heavy ion physics phenomena occurring at low center of mass energy, $\sqrt{s}$, of the colliding nuclei.
Some of these low $\sqrt s$ collisions are being investigated at the RHIC and to be probed further with the start of 
the heavy ion experiments at FAIR, GSI and NICA, Dubna. In fact, an interesting feature of the QCD 
phase diagram is the critical end-point related to chiral symmetry restoration. The existence of the critical point has 
important consequences on the QCD phase diagram and it is the aim of the extensive Beam Energy Scan(BES) program 
at the RHIC to search for it. 

To explain these experimental results from first principles, we need to extend the lattice QCD formulation to 
include the information of finite baryon density. One of the methods is to work in a grand canonical ensemble. In such 
an ensemble, the partition function is given by
\begin{equation}
 \mathcal{Z}_{QCD}(T,\mu)=Tr(\rm{e}^{\mathcal{H}_{QCD}-\mu N})=\int 
\mathcal {D}U_\mu \prod_{f=1}^{N_f} det D_f(\mu)\rm{e}^{-S_G}
\end{equation}
where the chemical potential $\mu$ is the Lagrange multiplier corresponding to the conserved number density $N$, that commutes with the 
QCD Hamiltonian ${H}_{QCD}$. $N$ can be the baryon number or the net electric charge . The $\mu$ enters into the lattice fermion action as $\exp(\pm\mu a)$ 
factors multiplying the forward and backward temporal links respectively~\cite{hk,kogut}, referred to as the 
Hasenfratz-Karsch method. The naive fermion operator at finite $\mu$ on the lattice would be of the form, 
\begin{equation}
D_f(\mu)_{x,y}=\left[\sum_{i=1}^{3}\frac{1}{2}\gamma_i\left( U_{i}(x)\delta_{y,x+i}-
U^\dagger_{i}(y) \delta_{y,x-i}\right)
+\frac{1}{2}\gamma_4 \left(\rm{e}^{\mu a} U_{4}(x)\delta_{y,x+4}-\rm{e}^{-\mu a} U^\dagger_{\mu}(y)\delta_{y,x-4}
\right)+a m_f \delta_{x,y}\right] .
\end{equation}
This is not a unique way of introducing $\mu$ and it could be also done in several different ways~\cite{gavai}. 
The lattice fermion determinant at finite $\mu$, like in the continuum, is no longer positive definite since 
\begin{equation}
 det D_f^\dagger(\mu)= det D_f(-\mu)\Rightarrow det D_f(\mu)=\vert det D_f(\mu)\vert~\rm{e}^{i\theta}
\end{equation}
and the interpretation of $\int \mathcal{D}U det D_f(\mu)\rm{e}^{-S_G}$ as a probability weight in the 
standard Monte Carlo simulations is no longer well defined.
This is known as the ``sign problem''. One may consider only the real part of the fermion determinant for Monte Carlo 
algorithms and generate configurations, by so called phase quenching. Once the partition function is known in the 
phase quenched limit, one can then use the reweighting techniques to generate the partition function of the full theory
at different values of $\mu$. 
The expectation value of the phase of the determinant, needed for re-weighting, at some finite $\mu$ is given as
\begin{equation}
 \langle \rm{e}^{i\theta}\rangle=\frac{\int 
\mathcal {D}U \prod_{f=1}^{N_f}~\vert det D_f(\mu)\vert~\rm{e}^{i\theta}~\rm{e}^{-S_G}}{\int 
\mathcal {D}U \prod_{f=1}^{N_f} ~\vert det D_f(\mu)\vert~\rm{e}^{-S_G}}=\rm{e}^{\frac{-V~\Delta F}{T}}~.
\end{equation}
 where $\Delta F$ is the difference between the free energy densities of the full and the phase quenched QCD. For 
two degenerate quark flavours, the phase quenched theory is equivalent to a theory with a finite isospin chemical potential 
\cite{son} and $\Delta F$ is the difference of free energies of QCD with finite baryon(quark) chemical 
potential and that at an isospin chemical potential. These two theories are qualitatively quite different and the sign 
problem results in a very small overlap between these two theories. For isospin QCD, 
the charged pions are the lightest excitations and these can undergo a Bose-Einstein condensation for $\mu>m_\pi/2$. 
The difference between the respective free energies in this regime is quite large, leading to a severe sign problem.
This is an algorithmic problem that can arise for any theory which has chiral symmetry breaking. A better understanding of 
the sign-problem has been achieved in the recent years with a knowledge of the regions in the phase diagram with   
severe sign problem and those where it is controllable~\cite{sign1,sign2}. There are several methods followed to circumvent 
this problem on the lattice, some of which are listed below,
\begin{itemize}
 \item Reweighting of the $\mu=0$ partition function~\cite{reweight1, bwreweos,reweight2}
\item Taylor series expansion \cite{tay1,gg1}
\item Canonical ensemble method \cite{cano1,cano2}
\item Imaginary chemical potential approach\cite{imchem1,imchem2,imchem3}
\item Complex Langevin algorithm \cite{complexl1,complexl2,complexl3}
\item Worm algorithms \cite{shailesh,gattringer}
\item QCD on a Lefschetz thimble~\cite{luigi}
\end{itemize}
 The Taylor series method has been widely used in the lattice QCD studies in the recent years, which has lead to interesting
results relevant for the experiments. One such proposal is the determination of the line of chemical freezeout 
for the hadrons in the phase diagram at small baryon density, from first principles lattice study. It was first 
proposed that cumulants of baryon number fluctuations could 
be used for determining the freezeout parameters~\cite{gg5} on the lattice. Last year, another interesting suggestion was made ~\cite{freezeouthisq}, where the experimental data on cumulants of electric charge fluctuations could be used as an input to compute the freezeout curve using lattice data. This and some other results are discussed in the  subsequent subsections. Most of the results are obtained with improved versions of staggered fermions. It has been known that 
the rooting problem may be more severe at finite density~\cite{rootingmu}. It is thus important to explore other fermion formulations 
as well for lattice studies. Wilson fermions have been used but it is important to use chiral fermions, especially for the 
study of the critical point. I outline in the next subsection, the theoretical efforts in the recent years that have led to the 
development of fermion operators at finite density with exact chiral symmetry on the lattice which can be used for future 
lattice studies on the critical point.

\subsection{Chiral fermions at finite density}
 The contribution of the $U_A(1)$ anomaly is believed to affect the order of the chiral phase transition at zero density 
and hence is crucial for the presence or absence of the critical point. If the anomaly is not represented correctly at finite density, it may 
affect the location of the critical point in the phase diagram, if it exists. Overlap fermions 
have exact chiral symmetry on the lattice, in the sense that the overlap action is invariant under suitable chiral transformations
known as the Luscher transformations \cite{luscher}. It can be further shown that the fermion measure in the path integral is not invariant under Luscher transformations, and its change gives the chiral anomaly. The index theorem, relating the anomaly to the difference 
between the fermion zero modes, can be proved for them \cite{hln}. Thus the overlap fermions have the properties analogous to the fermions in the continuum QCD. In the continuum, it is known that the anomaly is not affected in presence of a finite baryon chemical potential. It 
would be desirable to preserve this continuum property with the overlap fermions as well, such that the physical properties 
important for the existence of the critical point are faithfully presented on a finite lattice. Defining an overlap fermion 
action at finite chemical potential is non-trivial as the conserved currents have to be defined with care~\cite{mandula}. 
The first attempt to define an overlap fermion operator at finite density \cite{blochw} was done in the last decade, 
and an index theorem at finite $\mu$ was also derived for them. However these overlap fermions did not have exact chiral symmetry
symmetry on a finite lattice~\cite{bgs}. Moreover, the index theorem for them was $\mu$-dependent, unlike in the continuum.  Recently, 
overlap fermion at finite density has been defined from the first principles~\cite{ns}, which has exact chiral symmetry on the 
lattice \cite{gs1} and preserves the $\mu$-independent anomaly as well.  A suitable domain wall fermion action has been also defined at finite density \cite{gs1}, which was shown to reproduce the overlap action in the appropriate limit. It would be important to 
check the application of these overlap and domain wall fermion operators at finite $\mu$ for future large scale QCD simulations.
 
\subsection{Correlations and fluctuations on the lattice }
The studies of fluctuations of the conserved charges are important to understand the nature of the degrees of freedom in a 
thermalized medium and the interactions among them~\cite{koch,asakawa1}.
The diagonal susceptibility of order $n$, defined as,
\begin{equation}
 \chi^{X}_{n}=\frac{T}{V}\frac{\partial ^{n}ln \mathcal{Z}}{\partial \mu_X^n }~,~X\equiv B,S,Q.
\end{equation}
measures the fluctuations of the conserved quantum number $X$.  In a heavy-ion 
experiment the relevant conserved numbers are the baryon number $B$ and electric charge $Q$. The strangeness $S$, is zero at the 
initial time of collision of heavy nuclei but strange quark excitations are produced at a later time in the QGP and is also believed 
to be a good quantum number. These fluctuations can be computed 
exactly on the lattice at $\mu=0$ from the quark number susceptibilities~\cite{gottleib}. Continuum extrapolated results for the second order susceptibilities of baryon number, strangeness and electric charge exist for both HISQ~\cite{hisqsusc}and stout smeared staggered quarks~\cite{bwsusc}. The fluctuations of baryon number are very well explained by the hadron resonance gas model for $T<160~$ MeV.
However, the fluctuations of the strangeness are usually larger than the HRG values by about 20\% in the freezeout region characterized by 
$160\leq T\leq 170~$ MeV. The electric charge fluctuations, on the other hand, are smaller than the corresponding HRG values by 10\% in the 
same region. The ratio of $\chi_2^Q/\chi_2^B(\mu=0)\simeq 
0.29-0.35$ in the freezeout region. A first principle determination of this ratio is crucial, as it would allow us to relate the net baryon number fluctuations with the net proton number fluctuations, which is an observable in the heavy ion experiments~\cite{hisqsusc}. At high temperatures, these fluctuations slowly approach the corresponding free theory value, with the 
continuum extrapolated data for the baryon number susceptibility showing about 20\% deviation from the free theory value even at 
$2T_c$~\cite{hisqsusc}. The data are in good agreement with resummed perturbation theory estimates at these temperatures
~\cite{sylvain,gm} indicating that the QGP is still fairly strongly interacting even at temperatures around $2T_c$.

To relate to the results of the heavy ion experiments at a lower collision energy, $\sqrt{s}$, one has to compute the fluctuations on the lattice at a finite value of $\mu$. The most widely used lattice method to compute the susceptibilities at a 
finite value of quark chemical potential $\mu$, is through the Taylor expansion of the corresponding quantity at $\mu=0$~,e.g.,
\begin{equation}
\label{eqn:series}
 \chi^{B}_2(\mu)/T^2=\chi^{B}_2(0)/T^2+\frac{\mu^2}{2! T^2}\chi^{B}_4(0)+\frac{\mu^4}{4!  T^4}\chi^{B}_6(0)T^2+...
\end{equation}
The light and strange quark susceptibilities have been 
computed at finite but small densities from Taylor expansion, using asqtad staggered quarks~\cite{milc1} and the ratios of 
baryon number susceptibilities using the unimproved staggered fermions~\cite{gg5} in the region of interest for the 
RHIC experiments. All these ratios agree well with the estimates from the HRG model~\cite{gg5}, the 
results for which are compiled in the right panel of figure \ref{criticalpt}. The ratios of susceptibilities 
serve as a good observable for comparing the lattice and the experimental data since these are free from the unknown
quantities, like the volume of the fireball during freezeout~\cite{sgupta}.

The higher order susceptibilities $\chi_n$, for $n>4$, are important even in the $\mu=0$ regime. In the chiral limit, it 
is expected that the fourth order baryon number susceptibility would have a cusp and the sixth order would diverge
with O(4) scaling at the critical temperature. Even for physical quark masses, $\chi_6^B$ for QCD would show oscillations near 
the pseudo-critical temperature and $\chi_8^B$ would have negative values in the same region~\cite{kr}, quite contrary
to the HRG predictions. Thus the signatures of critical behaviour could be understood 
by the careful study of these quantities already at $\mu\sim0$, which is probed by the experiments at 
LHC ~\cite{kr}. 

\begin{figure}[h!]
\begin{center}
\includegraphics[width=0.3\textwidth]{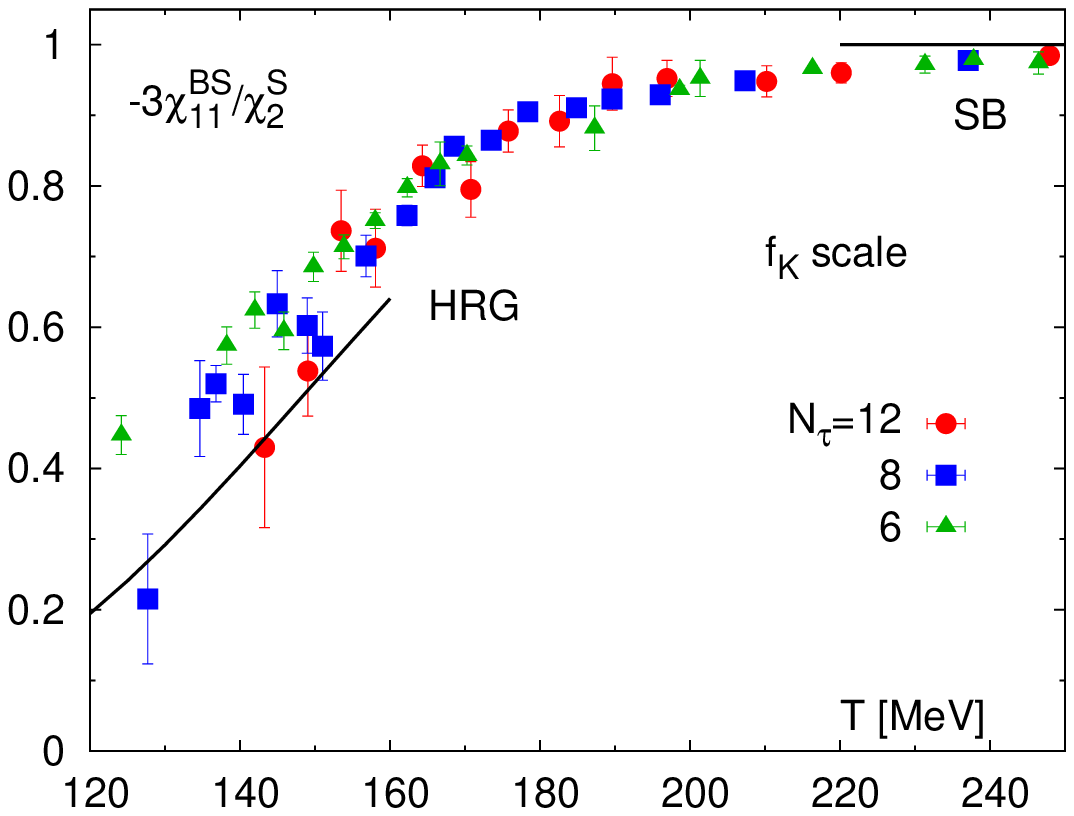}
\includegraphics[width=0.3\textwidth]{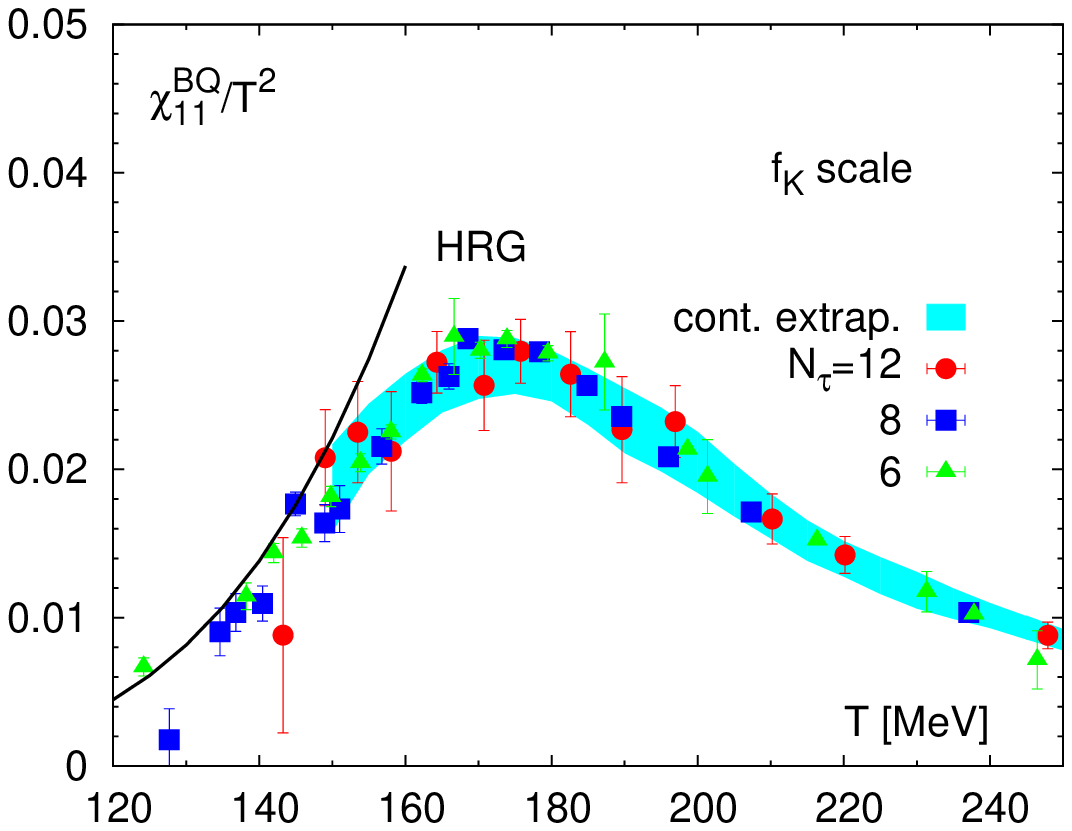}
\caption{The HISQ data for $C_{BS}$(left panel) and $\chi_{11}^{BQ}/T^2$(right panel) 
as a function of temperature, from \cite{hisqsusc}.}
\label{corr}
\end{center}
\end{figure}

Other important quantities  of relevance are the off-diagonal susceptibilities. These defined as
\begin{equation}
 \chi_{ijk}^{BSQ}=\frac{T}{V}\frac{\partial ^{i+j+k}ln \mathcal{Z}}{\partial \mu_B^i\partial \mu_S^j \partial\mu_Q^k}
\end{equation}
are a measure of the correlations between different quantum numbers and hence good observables to 
estimate the effects of interactions in the different phases of the QCD medium. It has been suggested that the
quantity $C_{BS}=-3\chi_{11}^{BS}/\chi^S_2$ is a good observable to characterize the deconfinement in thermal QCD \cite{maz}. 
If the strangeness is carried by quark like excitations, the value of $C_{BS}$ would be identity and would 
be much smaller than unity in the phase where only the baryons and mesons carries the strangeness quantum number. Recent results 
from the HotQCD collaboration using HISQ action~\cite{hisqsusc} show that $C_{BS}$ approaches unity very quickly 
at around 200 MeV implying that almost no strange hadrons survive in the QGP phase above $T_c$.
This is compiled in the left panel of figure \ref{corr}. The HotQCD data is consistent with the corresponding continuum extrapolated data 
with the stout smeared fermions~\cite{bwsusc}. Also $C_{BS}$ is not sensitive to the sea strange quark masses for $T>T_c$ since the 
first partially quenched results~\cite{ggcbs} for this quantity are consistent with the full QCD results. The other 
important observable is the baryon-electric charge correlation. In the confined phase, electric charge 
in the baryon sector is mainly carried by protons and anti-protons, therefore the 
correlation would rise exponentially with temperature if this phase could be described as a non-interacting gas consisting of 
these particles. At high temperatures, however, quark like excitations would be important and their masses being much smaller than the temperature, this correlation would fall to zero. From the behaviour of the continuum extrapolated HISQ 
data for $\chi_{11}^{BQ}$, compiled in the right panel of figure \ref{corr}, it is evident that near the pseudo-critical temperature 
there is a change in the fundamental properties of the degrees of freedom of the medium, with quark like excitations 
dominating at $1.5 T_c$.

\subsection{The freezeout curve from lattice}
\begin{figure}[h!]
\begin{center}
\includegraphics[width=0.3\textwidth]{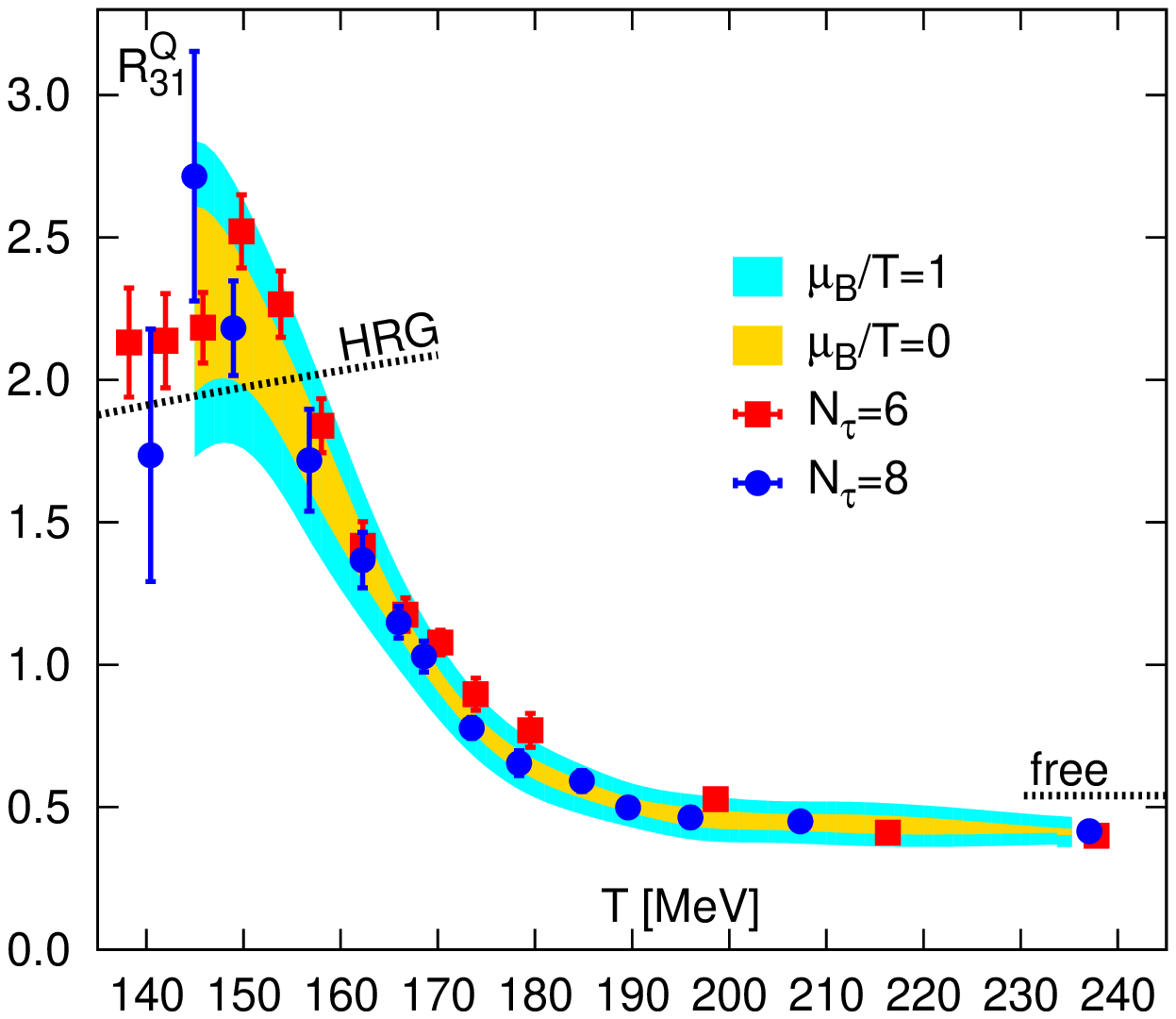}
\includegraphics[width=0.3\textwidth]{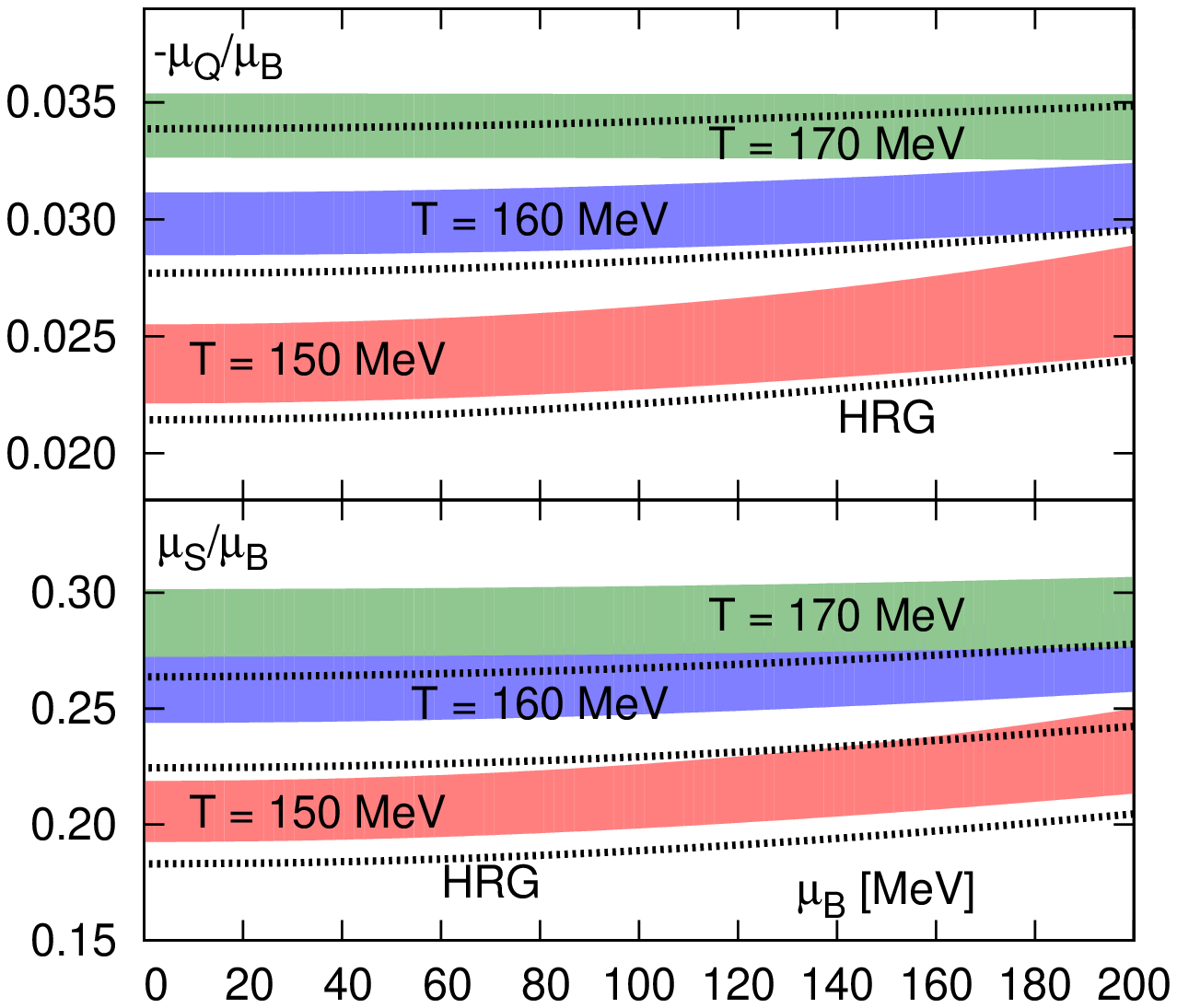}
\caption{In the left panel, the leading term for $R^{Q}_{31}$ shown in the yellow band, is compared to its 
NLO value denoted by the blue band in the 
continuum limit. In the right panel, the ratios of $\mu_Q$ and $\mu_S$ with respect to 
$\mu_B$ are compared with the HRG model predictions at different temperatures. Both figures are from Ref. 
\cite{freezeouthisq}.}
\label{freezeoutmu}
\end{center}
\end{figure}

To relate the results from heavy ion experiments with the lattice data, it is crucial to map
 the center of mass energy of the colliding nuclei in the heavy ion collisions,$\sqrt{s}$, to the corresponding point in the 
$T-\mu_B$ plane of the QCD phase diagram. This is called the freezeout curve. Phenomenologically, the freezeout curve 
is obtained from a particular parameterization of the HRG model obtained through fitting the experimental 
data on hadron abundances~\cite{freezeout}. 
At chemical freezeout, the chemical composition of the baryons gets frozen, meaning that the inelastic collisions between these 
species become less probable under further cooling of the system.
However the systematic uncertainties in determining the hadron yields are not taken into account in the phenomenological 
determination of the freezeout curve. Recent work by the BNL-Bielefeld collaboration shows how lattice techniques 
can provide first principle determination of the freezeout curve through suitable experimental observables~\cite{freezeouthisq} . As emphasized 
in the last subsection, the ratios of susceptibilities are believed to be good observables for comparing the lattice and 
the experimental data.  Two such observables proposed in Ref. \cite{freezeouthisq} are,
\begin{eqnarray}
\nonumber
 R_{12}^{X}\equiv\frac{M_X}{\sigma_X^2}&=&\frac{\mu_B}{T}\left( R_{12}^{X,1}+\frac{\mu_B^2}{T^2}R_{12}^{X,3}+
\mathcal{O}(\mu_B^4)\right)\\
R_{31}^{X}\equiv\frac{S_X\sigma_X^3}{M_X}&=&R_{31}^{X,1}+\frac{\mu_B^2}{T^2}R_{31}^{X,3}+
\mathcal{O}(\mu_B^4)
\end{eqnarray}
where $M_X,~\sigma_X,~S_X$ denotes the mean, variance and the skewness in dimensionless units for the conserved quantum number $X$. These 
observables are chosen because these have are odd and even functions of $\mu_B$, allowing us to independently 
determine $T$ and $\mu_B$ from these two quantities. The quantum number $X$ can either chosen to be the net electric 
charge $Q$ or the net baryon number $B$. In the experiments one can only measure the proton number fluctuations and it 
is not clear whether the proton number fluctuations could be a proxy for the net baryon fluctuation~\cite{asakawa2}.
It was thus suggested, that the ratios of net charge fluctuations would be a better observable to compare with the experiments. Once the 
$R^{Q}_{31}$ is known from experiments, one can determine the freezeout temperature $T_f$ from it by comparing with the 
continuum extrapolated lattice data. Analogously, one can obtain the $\mu_B$ at freezeout from comparison of the 
$R^{Q}_{12}$ data. In the left panel figure \ref{freezeoutmu}, the results for $R^{Q}_{31}$ are shown as a function 
 of temperature. It is evident that the first order correction to the value of the ratio is within 10\% of the 
leading order value for $\mu_B/T<1.3$ and in the freezeout region i.e, $T>140$ MeV. From the leading order 
results of $R^Q_{31}$ one can estimate the freezeout temperature. For $\sqrt{s}$ in the range of 39-200GeV currently 
probed in the Beam Energy Scan(BES) experiment at RHIC, the freezeout temperature from the HRG 
parameterization of the hadron multiplicities is about 165 MeV. At this temperature, the ratio $R^Q_{31}$ calculated from 
the HRG model is quite larger than the lattice estimate which would mean that the freezeout temperature estimated from 
lattice data would differ from the model results by atleast 5\%. Similarly, if $R^Q_{12}$ is known from the 
experiments, $\mu_B$ can be accurately estimated and is expected to be different from the current HRG estimates. This is 
not very surprising because the freezeout of the fluctuations happens due to diffusive processes and is due to a different mechanism from 
the freezeout of hadrons due to decreasing probability of inelastic collisions. Another question that was addressed 
in this work was how relevant are the other  parameters like $\mu_S$ and $\mu_Q$ for the phase diagram and the 
freezeout curve. It was seen that $\mu_S$ and $\mu_Q$ are significantly smaller 
than $\mu_B$ and the ratios of these quantities have a very small $\mu_B$ dependence in the 
entire temperature range of 140-170 MeV relevant for the freezeout studies. It signifies that the relevant axes for the 
phase diagram are indeed $T$ and $\mu_B$ and these two parameters are sufficient for characterizing the freezeout curve.

\subsection{Physics near the critical point}
\begin{figure}[h!]
\begin{center}
\includegraphics[width=0.3\textwidth]{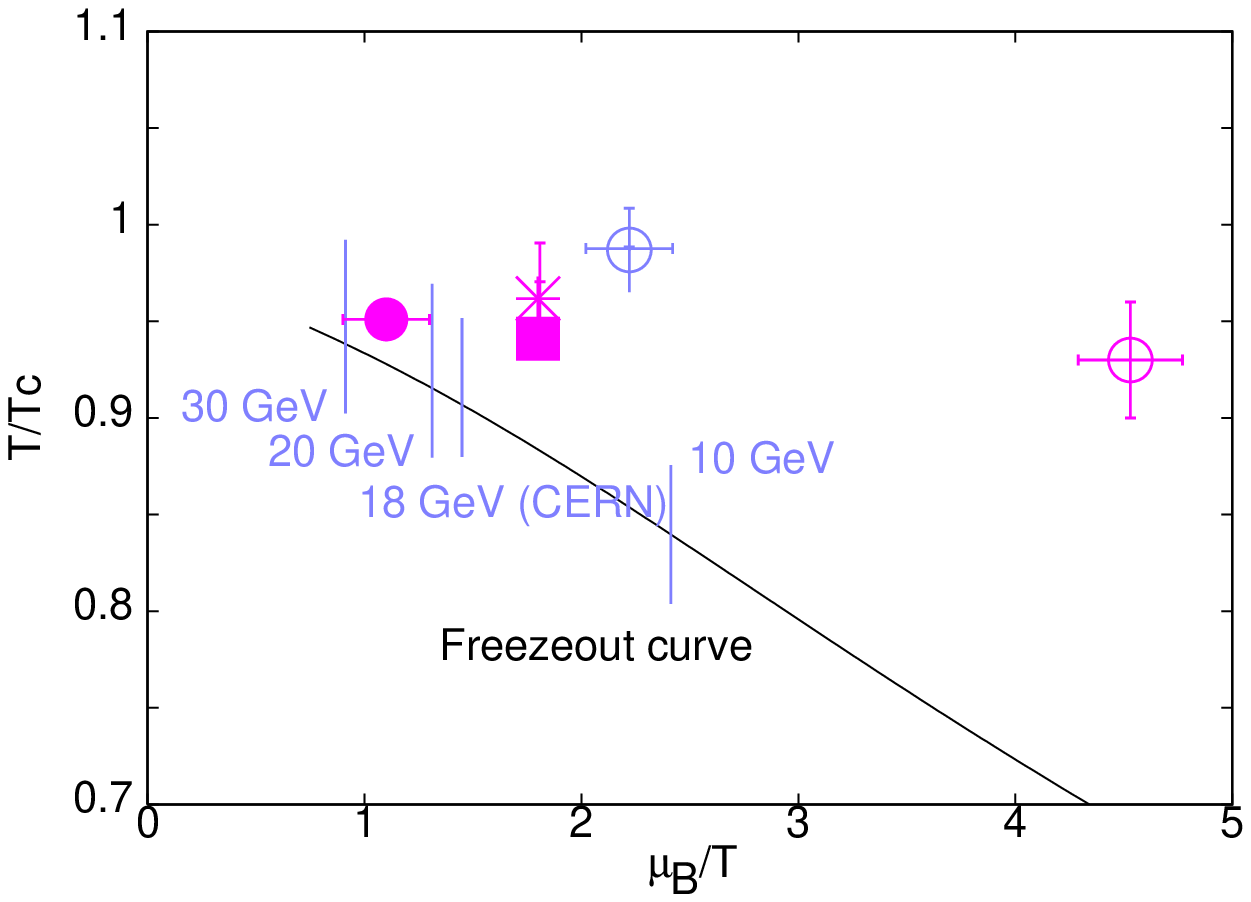}
\includegraphics[width=0.3\textwidth]{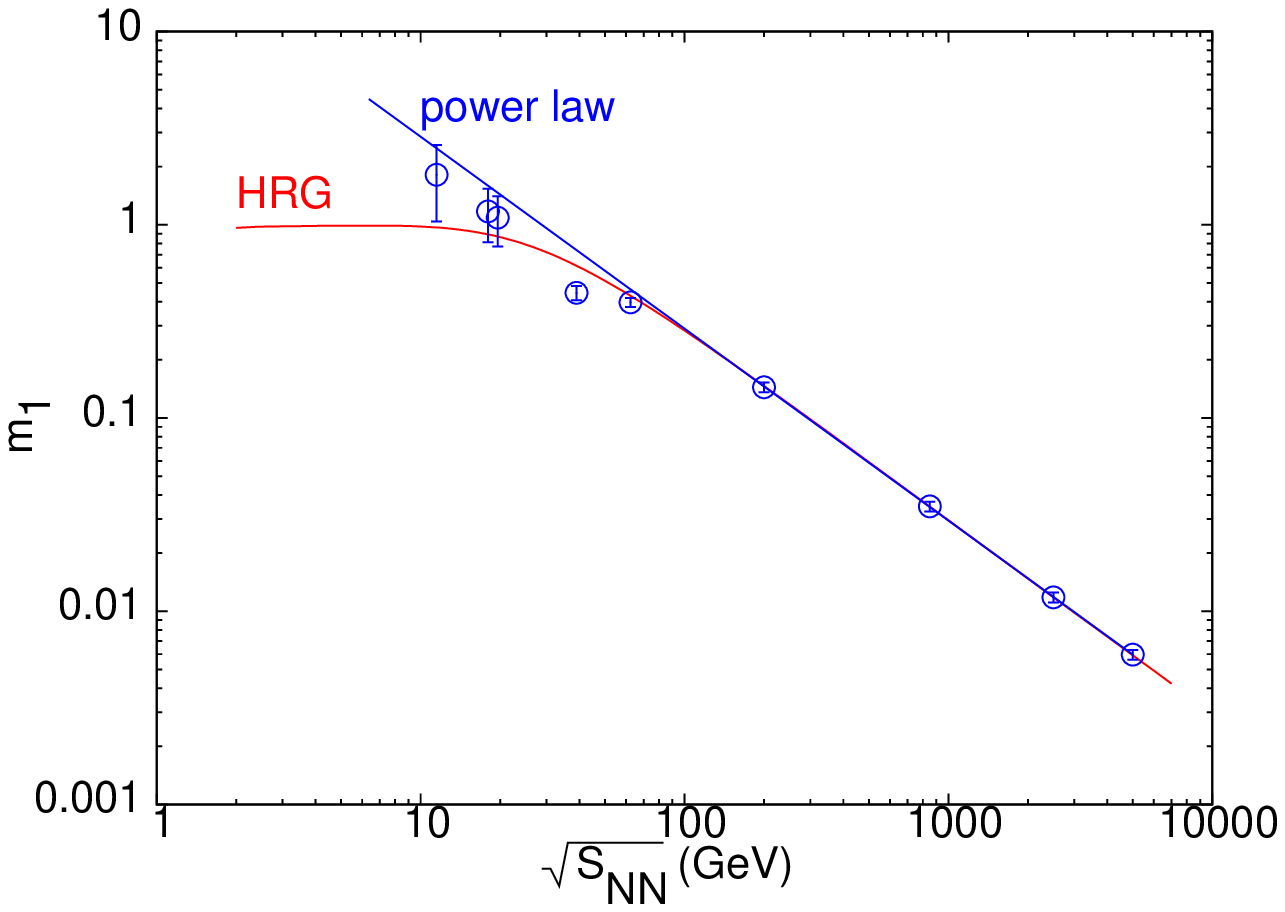}
\caption{The estimates of the critical point from lattice studies are shown in the left panel, from \cite{gg4}. The 
magenta solid circle, box and star denote the $N_\tau=4,6,8$ data respectively for 2 flavours of staggered quarks~\cite{gg2,gg3,gg4} 
and the open circles denote $N_\tau=4$ data for 2+1 flavours obtained with reweighting techniques~\cite{reweight2,reweight3}. 
In the right panel, the ratio of the third and the second order baryon number susceptibility is plotted as a function of $\sqrt s$ relevant 
for the RHIC and LHC experiments and compared with the HRG model data, from ~\cite{gg5}. }
\label{criticalpt}
\end{center}
\end{figure}

It is known from models with the same symmetries as QCD, that the chiral phase transition at $T=0$ and 
finite $\mu$ is a first order one. At zero density and high enough temperatures, QCD undergoes a crossover 
from the hadron to the QGP phase. By continuity, it is expected that the first order line should end 
at a critical end-point in the phase diagram~\cite{crit}. The determination of its existence from first principles 
lattice computation has been quite challenging and the currently available lattice results are summarized in the 
left panel of figure \ref{criticalpt}. These are all obtained using staggered fermions.  The first lattice study 
on the critical point was done using reweighting technique. 
Configurations generated at the critical value of the gauge coupling for $\mu_B=0$ were used to 
determine the partition function at different values of $T$ and $\mu_B$ using two-parameter reweighting~\cite{reweight2}. 
By observing the finite volume behaviour of the  Lee-Yang zeroes of the partition function, it was predicted that for 
2+1 flavour QCD, there is a critical end-point at $T_E=160(4)~$MeV and $\mu_B=725(35)~$MeV. In this study the light quark was 
four times heavier than its physical value. Reducing the light quark mass, shifted the critical end-point to 
$\mu_B=360(40)~$MeV with $T_E=162(2)$ remaining the same~\cite{reweight3}. However, this result is 
for a rather small lattice of size $16^3\times 4$ and is expected to change in the continuum limit 
and with larger volumes. Reweighting becomes more expensive with increasing volume of the lattice, so going to a larger 
lattice seems difficult with this method.

The other results for the critical point were obtained using the Taylor series method. In this method, 
the baryon number susceptibility at finite density is expanded in powers 
of $\mu_B/T$ as a Taylor series as shown in Eq. (\ref{eqn:series}), for each value of temperature. 
The baryon number susceptibility is expected to diverge at the critical end-point~\cite{shuryak1}, so 
the radius of convergence of the series would give the location of the critical end-point~\cite{gg1}. 
However on a finite lattice,  there are no divergences but the different estimates of the radius of 
convergence given as
\begin{equation}
\label{eqn:radiusc}
r_n(\text{n=odd})= \sqrt{\frac{\chi^B_{n+1}}{T^2\chi^B_{n+3}}}~,~r_n(\text{n=even})= 
\left[\frac{\chi^B_{2}}{T^n\chi^B_{n+2}}\right]^{1/n}
\end{equation}
 should all be positive and equal within errors at the critical end-point. Currently, the state of the art 
on the lattice are estimates of baryon number susceptibilities upto $\chi^B_{8}$. This 
gives five different independent estimates of the radius of convergence upto $r_6$, which were shown to be consistent within 
errors for $N_\tau=4,6,8$ at $T_E=0.94(1) T_c$~\cite{gg2,gg3,gg4}. The radius of convergence after finite volume correction 
is $\mu_B/T_E=1.7(1)$ \cite{gg4}, which means $\mu_B=246(15)~$MeV at the critical end-point if we choose $T_c=154~$MeV. The input 
pion mass for this computation is about 1.5 times the physical value and could affect the final coordinates
of the end-point. Moreover the different estimates for the radius of convergence $r_n$ in Eq. (\ref{eqn:radiusc}), agrees 
with each other for asymptotically large 
values of $n$ and one might need to check the consistency of the results with the radii of convergence estimates 
beyond $r_6$. Hints of the critical end-point were also obtained~\cite{critcano} using a different 
fermion discretization and a different methodology as well. Working with the canonical ensemble of improved 
Wilson fermions, the presence of a critical point was reported at $T_E=0.925(5)T_c$ and $\mu_B/T_c=2.60(8)$. 
This is a very preliminary study though, with a small lattice volume and a very heavy pion mass of about 700 MeV.

Though there is growing evidence in support for the existence of the critical end-point, the systematics for all 
these lattice studies are still not under control. It would be desirable to follow a different strategy 
to determine its existence. The alternate method suggested~\cite{dfp1} was to determine the curvature of the surface 
of second order chiral phase transitions as a function of the baryon chemical potential $\mu_B$. If the chiral 
critical surface bends towards larger values of $m_{u,d}$ with increasing baryon chemical potential and for a fixed 
value of the strange quark mass, it would pass through the physical point, ensuring the existence of a critical end-point. 
However if the curvature is of opposite sign, the chiral critical end-point would not exist. For lattice size of 
$8^3\times4$, the critical value of the light quarks was estimated upto $\mathcal{O}(\mu_B^4)$~\cite{dfp2}, 
\begin{equation}
 \frac{m_c(\mu_B)}{m_c(0)}=1-39(8)\left(\frac{\mu_B}{3 \pi T}\right)^2-...~,~
\end{equation}
with the strange quark mass fixed at its physical value. The leading value of the curvature has the same sign even 
for a finer lattice of extent $N_\tau=6$~\cite{dfp3}. These studies show that the region of first order transition 
shrinks for small values of $\mu_B$, which  would mean that the critical point may not exist in this regime of $\mu_B$. 
However for larger values of $\mu_B$, the higher order terms could be important and may bend the chiral critical line 
towards the physical values of quark masses. The finite cut-off effects are still sizeable and 
it is currently premature to make any definite predictions in the continuum limit with this method.

\begin{figure}[h!]
\begin{center}
\includegraphics[width=0.3\textwidth]{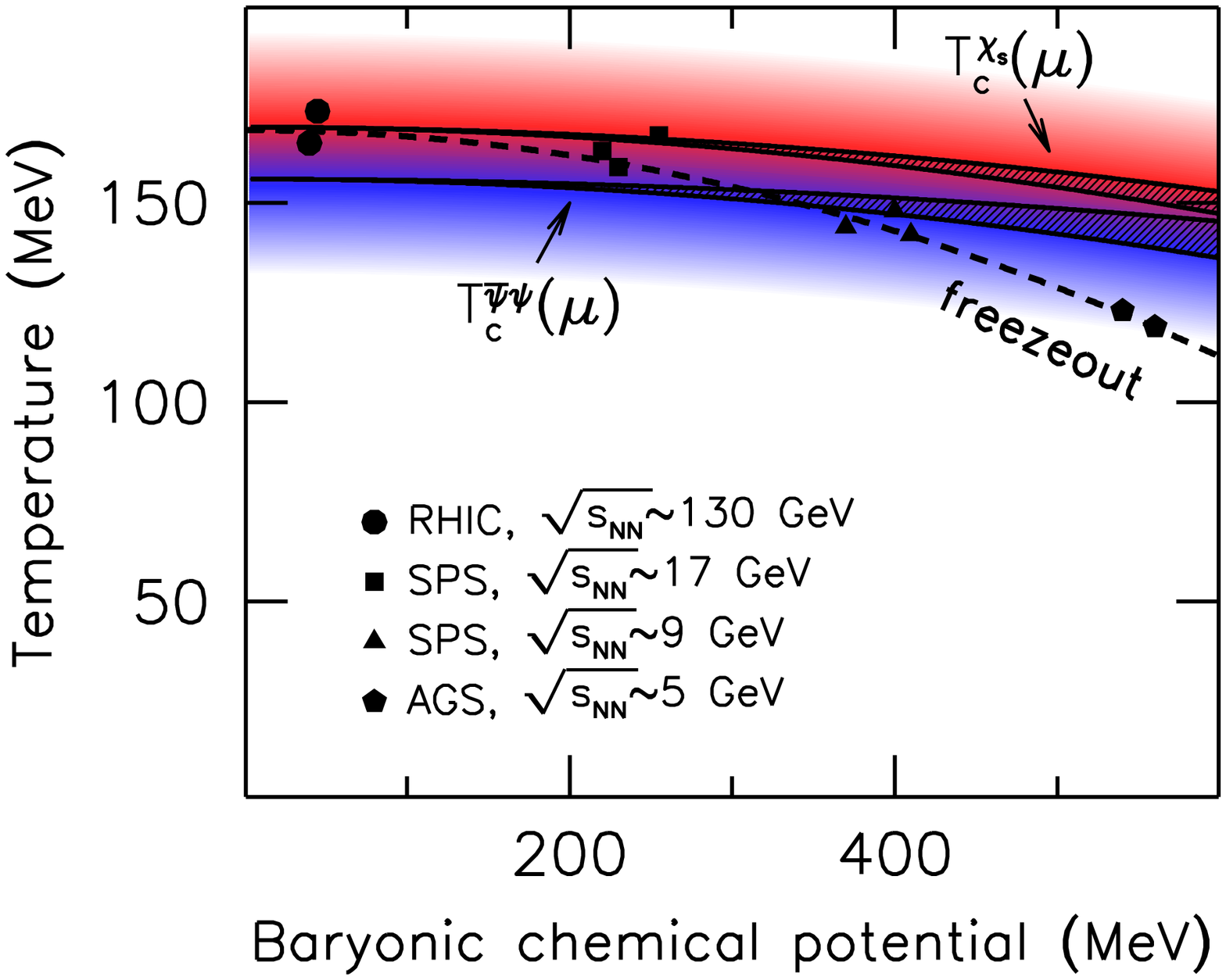}
\includegraphics[width=0.35\textwidth]{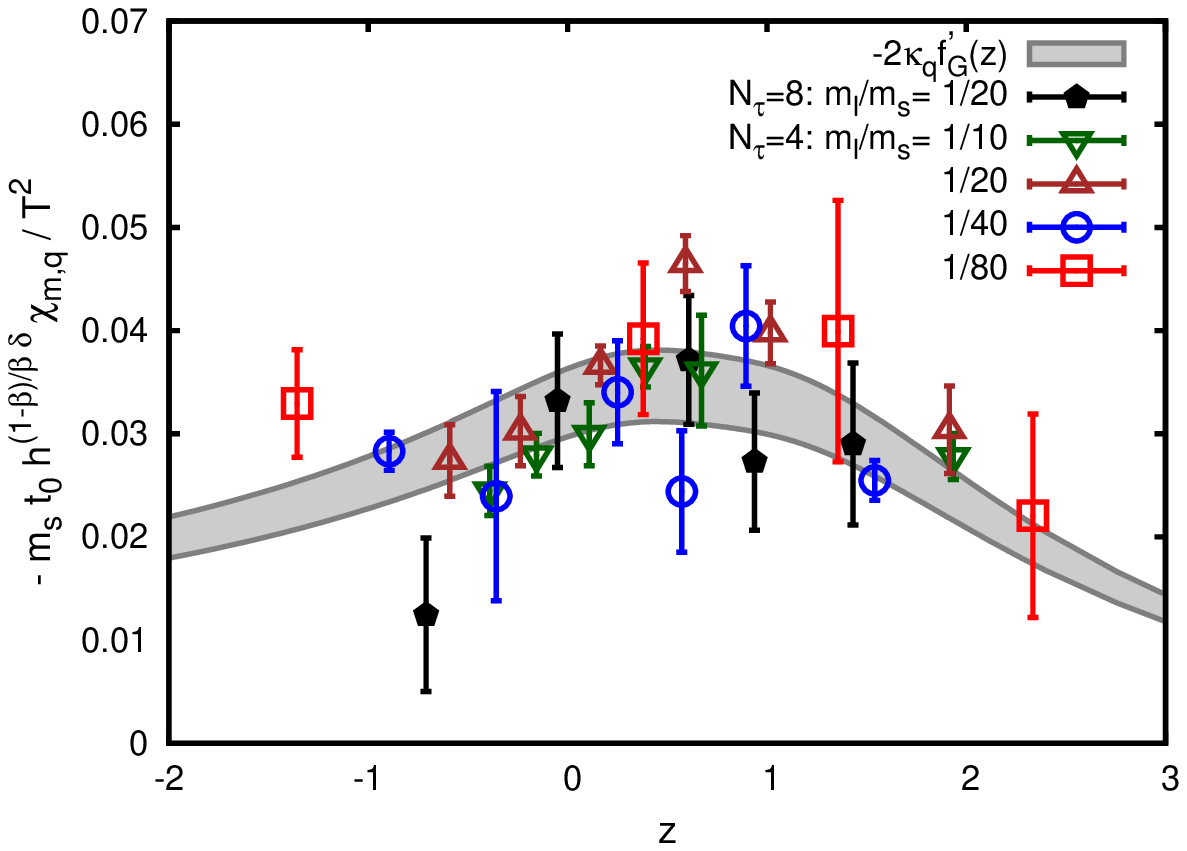}
\caption{In the left panel, the width of the pseudocritical region for chiral condensate is shown as a blue curve and that
for strange quark susceptibility is shown as a red curve, from~\cite{bwfreezeout}. In the right panel, the scaling of the mixed 
susceptibility is shown for different light quark masses and at the physical value of strange quark mass, from~\cite{curvhisq}. }
\label{criticalcv}
\end{center}
\end{figure}

It is equally important to understand the possible experimental signatures of the 
critical point.  The search of the critical end-point is one of the important aims for the extensive 
BES program at RHIC. In a heavy ion experiment,
one measures the number of charged hadrons at the chemical freezeout and its cumulants. During the expansion 
of the fireball, the hot and dense QCD medium would pass through the critical region and cool down eventually 
forming hadrons. If the freezeout and the critical regions are far separated, the system  
would have no memory of the critical fluctuations and the baryon number susceptibility measured from the 
experiments could be consistent with the predictions from thermal HRG models which has no critical behaviour. 
If the freezeout region is within the critical region, the critical fluctuations would be larger than 
the thermal fluctuations. It is thus important to estimate the chiral critical line for QCD from first 
principles.
The curvature of the chiral critical line has been estimated by the BNL-Bielefeld collaboration~\cite{curvhisq}, by extending the 
scaling analysis of the dimensionless chiral condensate $M_b$ outlined in Section \ref{sec:curv} for finite values 
of baryon chemical potential, using Taylor series expansion. The corresponding scaling variables at finite $\mu_B$ are
\begin{equation}
 t=\frac{1}{t_0}\left(\frac{T-T_{c,0}}{T_{c,0}}+\kappa_B\frac{\mu_B}{3T}\right)~,~h=\frac{m_l}{h_0 m_s}. 
\end{equation}
The quantity $M_b$ can be expanded as a Taylor series in $\mu_B/3T$ as,
\begin{equation}
 M_b(\mu)=M_b(0)+\frac{\chi_{m,B}}{2 T}\left(\frac{\mu_B}{3T}\right)^2+\mathcal{O}\left(\frac{\mu_B}{3T}\right)^4
\end{equation}
where $\chi_{m,B}$ is the mixed susceptibility defined as $\chi_{m,B}=\frac{T^2}{m_s}\partial^2  M_b/\partial (\mu_B/3T)^2$ computed
at $\mu_B=0$. In the critical region, it would show a scaling behaviour of the form,
\begin{equation}
\frac{\chi_{m,B}}{T} =\frac{2 \kappa_B T}{t_0 m_s}h^{-(1-\beta)/\beta\delta}f'_G(t/h^{1/\beta\delta}).
\end{equation}
The universality of the scaled $\chi_{m,B}$ data is clearly visible in the right panel of figure \ref{criticalcv}, both for 
p4 staggered quarks on $N_\tau=4$ lattice with mass ratios of light and strange quarks varying from $1/20$ to $1/80$ and 
with HISQ discretization on a $32^3\times8$ lattice with the mass ratio fixed at $1/20$. The fit of the complete lattice 
data set to the scaling relation for $\chi_{m,B}$, gave the value of $\kappa_B=0.00656(66)$. At non-vanishing 
$\mu_B$, the phase transition point is located at $t=0$, which implies that the critical temperature at finite density
can be parameterized as 
\begin{equation}
 \frac{T_c(\mu_B)}{T_c(0)}=1-\kappa_B\left(\frac{\mu_B}{3T}\right)^2+\mathcal{O}\left(\frac{\mu_B}{3T}\right)^4
\Rightarrow T_c(\mu_B)\simeq154(1-0.0066\left(\frac{\mu_B}{3T}\right)^2)~\text{MeV}.
\end{equation}
This estimate of the curvature is about three times larger than the corresponding prediction from the hadron 
resonance gas model. It would be interesting to compare the curvature of the freezeout line computed on the lattice 
with that of the critical line, once the experimental data for the electric charge cumulants are available.

Another complimentary study about the fate of the critical region at finite density was done by the Budapest-Wuppertal group~
\cite{bwfreezeout}. It was suggested that if the critical region shrinks with increasing $\mu_B$, it would imply that 
one slowly converges to the critical end-point. The width of the critical region was measured from two different 
observables, the renormalized chiral condensate and the strange quark number susceptibility. Stout smeared staggered 
quarks were employed and the continuum limit was taken with the $N_\tau=6,8,10$ data. The results are summarized in the left panel of 
figure \ref{criticalcv}. From the plots, it seems that the width of the crossover region does not 
change from its $\mu_B=0$ value significantly for $\mu_B<500~$ MeV, which implies either that the critical end-point does not 
exist at all or is present at a higher value of $\mu_B$. The corresponding curvature measured for the light quark chiral condensate 
is $0.0066(20)$ which is consistent with the result from the BNL-Bielefeld collaboration. The results indicate that the 
chiral pseudo-critical line and the phenomenological freezeout curve would separate apart at larger values of 
$\mu_B$ and would be further away at the critical end-point.

It was noted that the higher order fluctuations are more strongly dependent on the correlation length 
of the system~\cite{stephanov} and would survive even if the chiral and freezeout lines are far apart.
It has been proposed~\cite{kr}, that the signature of the critical point can be detected by monitoring 
the behaviour of the sixth and higher order fluctuations of the electric charge along the freezeout 
curve.

\subsection{The EoS at finite density}
\begin{figure}[h!]
\begin{center}
\includegraphics[width=0.3\textwidth]{Iisentrmilc.eps}
\includegraphics[width=0.3\textwidth]{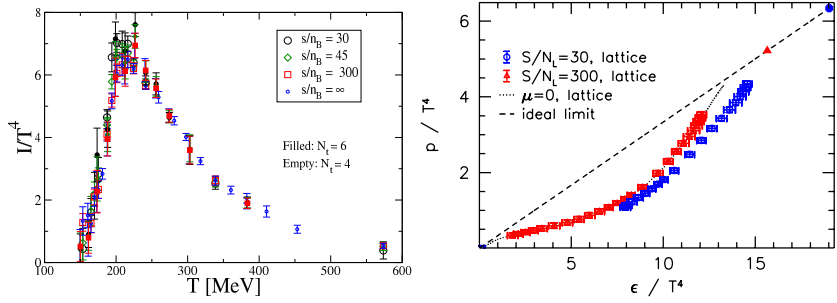}
\caption{The EoS for different isentropes using asqtad quarks are shown in the left panel, from~\cite{milc1}.
In the right panel, the data for the energy density and pressure is compared for different isentropes using stout smeared staggered 
quarks, from~\cite{bweosmu}.}
\label{eosmu}
\end{center}
\end{figure}

The EoS at finite density would be the important input for understanding the hydrodynamical evolution of the 
fireball formed at low values of the collisional energy, at the RHIC and the future experiments at FAIR
 and NICA. It is believed that there is no generation of entropy once the fireball thermalizes~\cite{shuryak2}. 
In that case, as pointed out in Ref.~\cite{christian}, it is important to determine the EoS along lines of constant entropy per net baryon 
number, $S/n_B$ to relate the lattice results with the experiments. The isentrope, determined by a fixed value of $S/n_B$
that characterizes the evolution of the fireball, is $S/n_B\simeq300$ for RHIC experiments, at $\sqrt{s}=200~$GeV. For the future 
experiments at FAIR, the isentropes would be labeled by $S/n_B=30$ nearly as same as the early SPS experiments at CERN where $S/n_B\sim 45$. For two flavour QCD with p4 staggered quarks and with pion mass heavier than its physical value, it was already observed 
that the ratio of pressure and energy density showed little variation as a function of $S/n_B$. The pressure and 
the energy density at finite $\mu$ are usually computed on the lattice as a Taylor series about its value at zero 
baryon density, as,
\begin{equation}
 \frac{P(\mu_l,\mu_s)}{T^4}= \frac{P(0)}{T^4}+\sum_{i,j}\frac{\chi_{ij}(T)}{T^{4-i-j}}\left(\frac{\mu_l}{T}\right)^i
\left(\frac{\mu_s}{T}\right)^j~.
\end{equation}
The formula is valid for two degenerate light quark flavours and a heavier strange quark. The coefficients $\chi_{ij}$ are the quark number susceptibilities at $\mu=0$ and are non-zero for $i+j$=even. The corresponding expression for the trace anomaly is given as,
\begin{equation}
 \frac{I(\mu_l,\mu_s)}{T^4}=-\frac{N_\tau^3}{N^3}\frac{d ln \mathcal{Z}}{d ln a}= \frac{I(0)}{T^4}+\sum_{i,j}b_{ij}(T)\left(\frac{\mu_l}{T}\right)^i
\left(\frac{\mu_s}{T}\right)^j~.
\end{equation}
The $\chi_{ij}$s can be also obtained from the coefficients $b_{ij}$ by integrating the latter along the line of constant physics.
For 2+1 flavours of improved asqtad staggered quarks with physical strange quark mass and $m_l=m_s/10$, the interaction 
measure was computed upto $\mathcal{O}(\mu^6)$ for two different lattice spacings (the left 
panel of figure \ref{eosmu}).  The interaction measure did not change significantly from the earlier results with heavier quarks 
and showed very little sensitivity to the cut-off effects along the isentropes~\cite{milc1}. However, it was observed that the 
light and the strange quark number susceptibilities change significantly from the zero temperature values along the isentropes. 
No peaks were found in the quark number susceptibilities at isentropes $S/n_b=300$, which led to the conclusion that the 
critical point may not be observed at the RHIC~\cite{milc1}. The EoS and the thermodynamic quantities were computed for physical values 
of quark masses by the Budapest-Wuppertal collaboration~\cite{bweosmu}. In this study, they set the values of the light 
quark chemical potentials such that $\mu_l=\mu_B/3$ and the strange quark susceptibility 
is $\mu_s=-2 \mu_l\chi_{11}^{us}/\chi_2^s$ to mimic the experimental conditions 
where the net strangeness is zero. The pressure and the energy density was computed upto $\mathcal{O}(\mu^2)$. The ingredients that 
went into the computations were a) the near continuum values of the interaction measure data from the $N_\tau=10$ lattice and b) the 
spline interpolated values of $\chi_2^s,\chi_{11}^{us}$ for the  range $125<T<400~$MeV obtained using the continuum extrapolated data 
for $\chi_2^s,\chi_{11}^{us}$. 
It was observed, as evident from the right panel of figure \ref{eosmu}, that the finite density effects along the RHIC isentropes are 
negligible consistent with the earlier work. However for isentropes given by $S/n_B=30$, the finite density effects become more 
important. The effect of truncation at $\mathcal{O}(\mu^2)$ was also estimated on a reasonably large $N_\tau=8$ lattice. 
It was observed,
\begin{equation}
\frac{p \textmd{ up to }\mathcal{O}((\mu_B/T)^4)}{p \textmd{ up to }\mathcal{O}((\mu_B/T)^2)} \le 
\begin{cases}
1.1, \hspace*{0.31cm} \textmd{ for $\mu_B/T\le2$}, \\
1.35, \hspace*{0.14cm} \textmd{ for $\mu_B/T\le3$}.
\end{cases}
\end{equation}
 implying that the fourth and higher order terms need to be determined for even modest values of $\mu_B$ in the Taylor series 
method. An independent study about the truncation effects of the Taylor series was performed in Ref. \cite{dfp4}. The derivatives 
of pressure were computed for two flavour QCD with staggered quarks at imaginary chemical potential. These derivatives are related 
to the successive terms of the Taylor coefficients of pressure evaluated at $\mu=0$. By fitting the imaginary $\mu$ data with a polynomial ansatz, these Taylor coefficients were obtained and compared with the exact values. It was observed that for 
$T_c\leq T\leq 1.04 T_c$, at least the 8th order Taylor coefficient is necessary for a good fit. This highlights the necessity 
to evaluate higher order susceptibilities, beyond the currently measured eighth order in the studies of EoS or the critical end-point.
New ideas to extend the Taylor series to higher order susceptibilities are evolving~\cite{gs2,dfp4} and these should be explored 
in full QCD simulations.  

\section{Summary}
As emphasized  in the introduction, I have  tried to compile together some of the important instances to show that 
the lattice results have already  entered into the precision regime with different fermion discretizations  
giving consistent continuum results for the pseudo-critical temperature and fluctuations of different 
quantum numbers.  The continuum result for the EoS would be available in very near future, with consistency 
already observed for different discretizations. The lattice community has opened the door for a very active collaboration 
between the theorists and experimentalists. With the EoS as an input, one can study the phenomenology of the 
hot and dense matter created at the heavy ion colliders. On the hand, there is a proposal of non-perturbative 
determination of the freezeout curve using lattice techniques, once the experimental data on cumulants of the 
charged hadrons are available.

A good understanding of the QCD phase diagram at zero baryon density has been achieved from the lattice studies. 
While the early universe transition from the QGP to the hadron phase is now known to be an analytic crossover 
and not a real phase transition, it is observed that the chiral dynamics will have observable effects in the 
crossover region. One of the remnant effects of the chiral symmetry would be the presence of a critical end-point. 
The search for the still elusive critical endpoint is one of the focus areas of lattice studies, and the 
important developments made so far in this area are reviewed.

While QCD at small baryon density is reasonably well understood with lattice techniques, the physics of 
baryon rich systems cannot be formulated satisfactorily on the lattice due to the infamous sign-problem. 
A lot of conceptual work, in understanding the severity and consequences of the sign problem as well algorithmic 
developments in circumventing this problem is ongoing and is one of the challenging problems in the field of lattice 
thermodynamics.

\section{Acknowledgements}
I would like to thank all the members of the Theoretical physics group at Bielefeld University, and in particular,
Frithjof Karsch, Edwin Laermann, Olaf Kaczmarek and Christian Schmidt for a lot of discussions that have enriched 
my knowledge about QCD thermodynamics and lattice QCD. I express my 
gratitude to Edwin Laermann for a careful reading of the manuscript and his helpful suggestions and Toru Kojo and 
Amaresh Jaiswal for their constructive criticism, that has let to a considerable improvement of this writeup. I also acknowledge
Rajiv Gavai and Rajamani Narayanan for very enjoyable collaboration, in which I learnt many aspects of the subject.


\begin{thebibliography}{150}


\bibitem{lat1}
C. Hoelbling, PoS LATTICE 2010 011 (2010)[arXiv:1102.0410 [hep-lat]].

\bibitem{lat2}
F. Karsch, Z. Phys. C 38  147 (1988). 

  
 \bibitem{hydro}
 P. F. Kolb and U. W. Heinz, 'Quark Gluon Plasma 3', Editors: R.C. Hwa and X.-N. Wang, World Scientific, Singapore [nucl-th/0305084].

\bibitem{maria}
M.-P. Lombardo, PoS LATTICE 2012 016 (2012) [arXiv:1301.7324[hep-lat]].

\bibitem{gaarts}
G. Aarts, PoS LATTICE 2012 017 (2012) [arXiv:1302.3028[hep-lat]].

\bibitem{nn}
S.\ B.\ Nielsen and M.\ Ninomiya,  Nucl.\ Phys.\  B 185, 20 (1981).

\bibitem{wilson}
K.\ G.\ Wilson,  Phys.\ Rev.\  D 10, 2445 (1974).

\bibitem{ks}
J.\ Kogut and L.\ Susskind,  Phys.\ Rev.\  D 11, 395 (1975).



\bibitem{stoutref}
C. Morningstar and M. J. Peardon, Phys. Rev. D 69, 054501 (2004).

\bibitem{hisqref}
E. Follana et al. (HPQCD Collaboration and UKQCD Collaboration), 
Phys. Rev. D 75, 054502 (2007) [arXiv:hep-lat/0610092].

\bibitem{p4ref}
J. Engels, R. Joswig, F. Karsch, E. Laermann, M. L\"{u}tgemeier and B. Petersson,
Phys. Lett. B 396,  210 (1997).



\bibitem{asqtadref}
 K. Orginos and D. Toussaint (MILC), Phys. Rev. D 59
, 014501 (1998) [arXiv:hep-lat/9805009];
J. F. Lagae and D. K. Sinclair, Phys. Rev. D 59, 014511 (1998) [arXiv:hep-lat/9806014]; 
G. P. Lepage, Phys. Rev. D59, 074502 (1999) [arXiv:hep-lat/9809157].




\bibitem{neunar}
R.\ Narayanan and H.\ Neuberger,  Phys.\ Rev.\ Lett.\  71, 3251 (1993); \\
H.\ Neuberger, Phys.\ Lett.\ B 417, 141 (1998).

\bibitem{kaplan}
D.\ B.\ Kaplan, Phys.\ Lett.\ B 288, 342 (1992).

\bibitem{bf1}
 M. Cheng, S. Ejiri, P. Hegde, F. Karsch, O. Kaczmarek, E. Laermann, R. D. Mawhinney and C. Miao
et al., Phys. Rev. D 81 (2010) 054504 [arXiv:0911.2215 [hep-lat]].


\bibitem{bw1}
 S. Borsanyi, G. Endrodi, Z. Fodor, A. Jakovac, S. D. Katz,
S. Krieg, C. Ratti and K. K. Szabo, JHEP 1011
(2010) 077 [arXiv:1007.2580 [hep-lat]].

\bibitem{dwold}
M. Cheng, N. H. Christ, P. Hegde, F. Karsch, M. Li, M. F. Lin, R. D. Mawhinney, D. Renfrew and P. Vranas
Phys. Rev. D 85 054510 (2010).

\bibitem{cossu1}
 G.~Cossu  et al. [ JLQCD Collaboration ], PoS LATTICE 2010, 174 (2010)[arXiv:1011.0257 [hep-lat]].

\bibitem{twqcd}
T-W. Chiu, arXiv:1302.6918[hep-lat].

\bibitem{bwov}
 S. Borsanyi, Y. Delgado, S. D\"{u}rr, Z. Fodor, S. D. Katz, S. Krieg, T. Lippert and 
D. Nogradi et al., Phys. Lett. B 713 (2012) [arXiv:1204.4089 [hep-lat]].


\bibitem{milcscale}
A. Bazavov, C. Bernard, C. DeTar, Steven Gottlieb, U.M. Heller, J.E. Hetrick, J. Laiho, L. Levkova et al., 
Rev. Mod. Phys. 82, 1349 (2010).
\bibitem{karsch1}
G. Boyd, J. Engels, F. Karsch, E. Laermann, C. Legeland, M. L\"{u}tgemeier and B. 
Petersson, Nucl. Phys. B 469,  419 (1996) [hep-lat/9602007].

\bibitem{karsch2}
F. Karsch, Nucl. Phys. Proc. Suppl. 83,  14 (2000) [hep-lat/9909006].

\bibitem{nf3}
 F. Karsch, E. Laermann and C. Schmidt, Phys. Lett. B
520 , 41 (2001), P. de Forcrand and O. Philipsen,
Nucl. Phys. B 673, 170 (2003),  D. Smith and C. Schmidt,
PoS LATTICE 2011, 216 (2011), X. -Y. Jin and R. D. Mawhinney, PoS LATTICE
2011 (2011) 066, M. Cheng,et al., Phys. Rev. D 75, 034506 (2007).

\bibitem{nf2fo}
M. D'Elia, A. Di Giacomo and C. Pica, Phys. Rev. D 72, 114510 (2005) [hep-lat/0503030]; 
C. Bonati, G. Cossu, M. D'Elia, A. Di Giacomo and C. Pica, PoS LATTICE 2008,  204 (2008)[arXiv:0901.3231 [hep-lat]].

\bibitem{mageos}
S. Ejiri, F. Karsch, E. Laermann et al., Phys. Rev. D 80, 094505 (2009) [arXiv:0909.5122 [hep-lat]].

\bibitem{colplot}
S. Gavin, A. Gocksch and R. D. Pisarski, Phys. Rev. D 49 3079 (1994).

\bibitem{milc}
 C. Bernard et al. (MILC Collaboration), Phys. Rev. D 71 , 034504 (2005).

\bibitem{bf2}
M. Cheng, N. Christ, S. Datta, J. van der Heide, C. Jung, et al., Phys. Rev. D 74
, 054507 (2006).

\bibitem{yaoki}
Y. Aoki, G. Endrodi, Z. Fodor, S. Katz, and K. Szabo, Nature 443, 675 (2006).

\bibitem{saito}
H. Saito et al. Phys. Rev. D 84, 054502 (2011).

\bibitem{ding}
 H. -T. Ding,  A. Bazavov, P. Hegde, F. Karsch, S. Mukherjee, P. Petreczky, 
PoS LATTICE 2011  191 (2011) [arXiv:1111.0185 [hep-lat]].

\bibitem{bwnf3}
G. Endrodi, Z. Fodor, S. D. Katz, K. K. Szabo, PoSLATTICE 2007, 182 (2007).

\bibitem{abj}
S.\ L.\ Adler,   Phys.\ Rev.\ 177, 2426 (1969).\\
J.\ Bell and R.\ Jackiw,  Nuovo.\ Cim.\  A 60, 47 (1969).

\bibitem{fujikawa}
K.\ Fujikawa,  Phys.\ Rev. D 21, 2848 (1980).


\bibitem{piswil}
R.\ D.\ Pisarski and F.\ Wilczek,  Phys.\ Rev.\  D 29, 338 (1984).

\bibitem{intmethod}
J. Engels, J. Fingberg, F. Karsch, D. Miller and M. Weber, Phys. Lett. B 252, 625 (1990).

\bibitem{giusti}
 L. Giusti and H. B. Meyer, Phys. Rev. Lett. 106,  131601 (2011) [arXiv:1011.2727 [hep-lat]].

\bibitem{petreczky}
P. Petreczky, HotQCD Collaboration, PoS LATTICE 2012,  069 (2012) [arXiv:1211.1678].

\bibitem{hotqcd1}
A. Bazavov, T. Bhattacharya, M. Cheng, C. DeTar, H.-T. Ding et al. , Phys. Rev. D 85, 054503 (2012) [arXiv:1111.1710 [hep-lat]].


\bibitem{bwcharm}
 S. Borsanyi, G. Endrodi, Z. Fodor, S. D. Katz, S. Krieg, C. Ratti, C. Schroeder 
and K. K. Szabo, PoS LATTICE 2011,  201 (2011) [arXiv:1204.0995 [hep-lat]].

\bibitem{whot}
T. Umeda et al. [WHOT-QCD Collaboration], Phys. Rev. D 85, 094508 (2012)[arXiv:1202.4719[hep-lat]].

\bibitem{tmft}
 F. Burger, M. Kirchner, M. M\"{u}ller-Preussker, E-M. Ilgenfritz, M. P. Lombardo, O. Philipsen, C. Pinke and 
L. Zeidlewicz , PoS LATTICE 2012, 068 (2012) [arXiv:1212.0982[hep-lat]].

\bibitem{fph1}
P. de Forcrand and O. Philipsen, JHEP 0701, 077 (2007) [hep-lat/0607017].

\bibitem{hotqcd2}
 M. Cheng, N. Christ, S. Datta, J. van der Heide, C. Jung, et al., Phys. Rev.
D 77, 014511 (2008).

\bibitem{bwtc}
S. Borsanyi et al. [Wuppertal-Budapest Collaboration], JHEP 1009,  073 (2010)[arXiv:1005.3508
[hep-lat]].

\bibitem{creutz}
 M. Creutz, PoS CONFINEMENT8, 016 (2008).

\bibitem{gsh}
 C. Bernard, M. Golterman, Y. Shamir, and S. R. Sharpe, Phys. Rev. D 77, 114504 (2008).

\bibitem{bwwil}
 S. Borsanyi, S. D\"{u}rr, Z. Fodor, C. Hoelbling, S. D. Katz, S. Krieg, D. Nogradi and B. C. Toth 
et al. , JHEP 1208, 126 (2012) [arXiv:1205.0440 [hep-lat]].

\bibitem{fodor}
Z. Fodor, S.D. Katz and K.K. Szabo, JHEP 0408, 003 (2004).

\bibitem{fixedtop}
S. Aoki, H. Fukaya, S. Hashimoto, and T. Onogi, Phys. Rev. D 76,  054508 (2007).

\bibitem{hotqcddw}
A. Bazavov, T. Bhattacharya, M. I. Buchoff, M. Cheng, N. H. Christ, H.-T. Ding et al. , Phys. Rev. D 86 (2012) 094503 [arXiv:1205.3535 [hep-lat]].

\bibitem{peikert}
 F. Karsch, E. Laermann and A. Peikert, Phys. Lett. B 478,  447 (2000) [hep-lat/0002003].


\bibitem{petreczkymunich}
P. Petreczky, arXiv:1301.6188[hep-lat].

\bibitem{charmpt}
M. Laine and Y. Schroeder, Phys. Rev. D 73, 085009 (2006) [hep-ph/0603048].


\bibitem{mcheng}
M. Cheng, PoS LATTICE 2007,  173 (2007).

\bibitem{milc1}
C. DeTar, L. Levkova, S. Gottlieb, U. M. Heller, J. E. Hetrick,
R. Sugar and D. Toussaint, Phys. Rev. D 81,  114504 (2010) [arXiv:1003.5682 [hep-lat]].

\bibitem{hind}
M. Hindmarsh and O. Philipsen, Phys. Rev. D 71  087302 (2005)[hep-ph/0501232].


\bibitem{soldner}
 M. McGuigan and W. Soldner [arXiv:0810.0265[hep-ph]].


\bibitem{on1}
J. Engels, S. Holtmann, T. Mendes and T. Schulze, Phys. Lett. B 492, 219 (2000).

\bibitem{on2}
 D. Toussaint, Phys. Rev. D 55, 362 (1997).

\bibitem{on3}
J. Engels and T. Mendes, Nucl. Phys. B 572, 289 (2000).

\bibitem{hln} 
P. Hasenfratz, V. Laliena and F. Niedermeyer,  Phys.\ Lett.\  B 427,  125 (1998). 


\bibitem{hotqcddwlat}
Z. Lin PoS LATTICE 2012, 84 (2012).

\bibitem{aoki}
S. Aoki, H. Fukaya, Y. Taniguchi, Phys. Rev. D 86, 114512 (2012).

\bibitem{cossu2}
 G. Cossu, S. Aoki, S. Hashimoto, T. Kaneko, H. Matsufuru, J. -I. Noaki and 
E. Shintani, PoS LATTICE 2011, 188 (2011) [arXiv:1204.4519 [hep-lat]];\\
G. Cossu, S. Aoki, H. Fukaya, S. Hashimoto, T. Kaneko, H. Matsufuru and J. -I. Noaki,
arXiv:1304.6145 [hep-lat].

\bibitem{hiroshi}
H. Ohno, U. M. Heller, F. Karsch and S. Mukherjee, PoS LATTICE 2012,  095 (2012)[arXiv:1211.2591[hep-lat]].

\bibitem{phwil}
B. B. Brandt, A. Francis, H. B. Meyer, H. Wittig and O. Philipsen, PoS LATTICE
2012,  073 (2012)[arXiv:1210.6972 [hep-lat]].

\bibitem{hk}
P.\ Hasenfratz and F.\ Karsch  Phys.\ Lett.\  B 125, 308 (1983).

\bibitem{kogut}
J.\ Kogut et al.,  Nucl.\ Phys.\  B 225, 93 (1983).

\bibitem{gavai}
R.\ V.\ Gavai,  Phys.\ Rev.\ D32, 519 (1985).

\bibitem{son}
D. T. Son and M. A. Stephanov, Phys. Rev. Lett. 86, 592  (2001) [hep-ph/0005225].

\bibitem{sign1}
K. Splittorff, arXiv:hep-lat/0505001;\\
 J. Han and M. A. Stephanov, Phys. Rev.
D 78, 054507 (2008) [arXiv:0805.1939 [hep-lat]].

\bibitem{sign2}
M. P. Lombardo, K. Splittorff and J. J. M. Verbaarschot, Phys. Rev. D 80, 054509 (2009).


\bibitem{reweight1}
I. M. Barbour et al., Nucl. Phys. Proc. Suppl. 60A, 220 (1998) [hep-lat/9705042].

\bibitem{reweight2}
Z. Fodor and S. D. Katz, Phys. Lett. B 534, 87 (2002), [hep-lat/0104001].

\bibitem{bwreweos}
 Z. Fodor, S. D. Katz and K. K. Szabo, Phys. Lett. B 568, 73 (2003);\\
F. Csikor, G. I. Egri, Z. Fodor, S. D. Katz, K. K. Szabo and A. I.Toth, JHEP
0405,  046 (2004).


\bibitem{tay1}
 C. R. Allton, S. Ejiri, S. J. Hands, O. Kaczmarek, F. Karsch, 
E. Laermann and C. Schmidt, Phys. Rev. D 68, 014507 (2003);\\
 C. Allton, M. Doring, S. Ejiri, S. Hands, O. Kaczmarek,
et. al., Phys. Rev. D 71,  054508 (2005) [hep-lat/0501030].

\bibitem{gg1}
R. V. Gavai and S. Gupta, Phys. Rev. D 68, 034506 (2003) [hep-lat/0303013].

\bibitem{cano1}
J. Engels, O. Kaczmarek, F. Karsch, and E. Laermann, Nucl. Phys. B558, 307 (1999) [hep-lat/9903030].


\bibitem{cano2}
K.-F. Liu, Int. J. Mod. Phys. B16, 2017 (2002) [hep-lat/0202026];\\
A. Alexandru, M. Faber, I. Horvath, K.-F. Liu, Phys. Rev. D 72, 114513 (2005);\\
P. de Forcrand and S. Kratochvila, Nucl. Phys. Proc. Suppl. 153, 62 (2006) [hep-lat/0602024].

\bibitem{imchem1}
M. G. Alford, A. Kapustin, and F. Wilczek, Phys. Rev. D 59, 054502 (1999) [hep-lat/9807039].

\bibitem{imchem2}
M.-P. Lombardo, Nucl. Phys. Proc. Suppl. 83, 375 (2000) [hep-lat/9908006];\\
 M. D'Elia and M.-P. Lombardo, Phys. Rev. D 70, 074509 (2004).

\bibitem{imchem3}
P. de Forcrand and O. Philipsen, Nucl. Phys. B642, 290 (2002) [hep-lat/0205016].


\bibitem{complexl1}
G. Parisi, Phys. Lett. B 131, 393 (1983).

\bibitem{complexl2}
F. Karsch and H. W. Wyld, Phys. Rev. Lett. 55,  2242 (1985).\\
J. Ambjorn and S. K. Yang, Phys. Lett. B 165, 140, (1985).

\bibitem{complexl3}
G. Aarts, F. A. James, E. Seiler and I. O. Stamatescu, Phys. Lett. B 687, 154 (2010)  [arXiv:0912.0617
[hep-lat]];\\
G. Aarts, E. Seiler and I. O. Stamatescu, Phys. Rev. D 81, 054508 (2010).

\bibitem{shailesh}
S. Chandrasekharan, Phys. Rev. D 82,  025007 (2010).  

\bibitem{gattringer}
C. Gattringer, Nucl. Phys. B 850, 242 (2011).

\bibitem{luigi}
M. Cristoforetti, F. Di Renzo, and L. Scorzato, Phys. Rev. D 86, 074506 (2012).


\bibitem{gg5}
R.\ V.\ Gavai and S.\ Gupta, Phys.\ Lett.\  B 696, 459 (2011).

\bibitem{freezeouthisq}
A. Bazavov, H.-T. Ding, P. Hegde, O. Kaczmarek, F. Karsch, E. Laermann, Swagato Mukherjee, et al.,
Phys. Rev. Lett. 109, 192302 (2012) [arXiv:1208.1220 [hep-lat]].


\bibitem{rootingmu}
M. Golterman, Y. Shamir, and B. Svetitsky, Phys. Rev. D 74, 071501 (2006) [hep-lat/0602026].


\bibitem{luscher} 
M.~ Luscher,  Phys.\ Lett.\ B  428,  342 (1998). 


\bibitem{mandula}
 J. Mandula, [arXiv:0712.0651 [hep-lat]]. 

\bibitem{blochw}
 J.~C.~R.~Bloch and T.~Wettig,  Phys.\ Rev.\ Lett.\  97, 012003 (2006)   [arXiv:hep-lat/0604020].

\bibitem{bgs}
D. Banerjee, R. V. Gavai and S. Sharma, Phys. Rev. D 78,  014506 (2008) and PoS (LATTICE 2008), 177.

\bibitem{ns}
R.\ Narayanan and S.\ Sharma, JHEP 1110, 151 (2011).

\bibitem{gs1}
R.\ V.\ Gavai and S.\ Sharma,  Phys.\ Lett.\  B 716, 446 (2012).


\bibitem{koch}
S. Jeon and V. Koch, Phys. Rev. Lett. 85, 2076 (2000).

\bibitem{asakawa1}
 M.~Asakawa, U.~W.~Heinz, B.~Muller, Phys.\ Rev.\ Lett.\ 85, 2072 (2000)
  [hep-ph/0003169].

\bibitem{gottleib}
S. A. Gottlieb, W. Liu, D. Toussaint, R. L. Renken, R. L. Sugar, Phys. Rev. Lett. 59, 2247
(1987).

\bibitem{hisqsusc}
A. Bazavov, T. Bhattacharya, C. E. DeTar, H.-T. Ding, Steven Gottlieb, et. al., Phys. Rev. D86, 034509 (2012) [arXiv:1203.0784 [hep-lat]]. 

\bibitem{bwsusc}
S. Borsanyi, Z. Fodor, S. D. Katz, S. Krieg, C. Ratti, et. al., JHEP
1201 (2012) 138, [arXiv:1112.4416[hep-lat]].


\bibitem{sylvain}
J. O. Andersen, S. Mogliacci, N. Su and A. Vuorinen, Phys. Rev. D 87, 074003 (2013) [arXiv:1210.0912 [hep-ph]].

\bibitem{gm}
N. Haque, M. G. Mustafa and M. Strickland, [arXiv:1302.3228].


\bibitem{sgupta}
 S.~Gupta, PoS CPOD, 025 (2009) [arXiv:0909.4630 [nucl-ex]].


\bibitem{kr}
F. Karsch and K. Redlich, Phys. Lett. B 695, 136 (2011).

\bibitem{maz}
V. Koch, A. Majumder and J. Randrup, Phys. Rev. Lett. 95, 182301 (2005).


\bibitem{ggcbs}
R.\ V.\ Gavai and S.\ Gupta,  Phys. Rev. D 73,  014004 (2006).


\bibitem{freezeout}
J. Cleymans, H. Oeschler, K. Redlich and S. Wheaton, Phys . Rev. C 73, 034905 (2006).

\bibitem{asakawa2}
M. Kitazawa and M. Asakawa, Phys. Rev. C 85, 021901 (2012) [arXiv:1107.2755 [nucl-th]].

\bibitem{crit}
M. Asakawa and K. Yazaki, Nucl. Phys. A 504, 668 (1989);\\
J. Berges and K. Rajagopal, Nucl. Phys. B 538, 215 (1999);\\
A. M. Halasz, A. D. Jackson, R. E. Shrock, M. A. Stephanov, and J. J. M. Verbaarschot, Phys. Rev. D 58, 096007 (1998).

\bibitem{reweight3}
Z. Fodor and S. D. Katz, JHEP 0404, 050 (2004). 


\bibitem{shuryak1}
M. Stephanov, K. Rajagopal and E. Shuryak, Phys. Rev. Lett. 81, 4816 (1998).

\bibitem{gg2}
R. V. Gavai and S. Gupta, Phys. Rev. D 71, 114014 (2005) [hep-lat/0412035].


\bibitem{gg3}
 R.~V.~Gavai, S.~Gupta,   Phys.\ Rev.\  D 78, 114503 (2008) [arXiv:0806.2233 [hep-lat]].


\bibitem{gg4}
S. Datta, R. V. Gavai and S. Gupta, Proceedings of Quark Matter 2012 [arXiv:1210.6784[hep-lat]].


\bibitem{critcano}
A. Li, A. Alexandru, K.-F. Liu, Phys. Rev. D 84, 071503 (2011) [arXiv:1103.3045 [hep-ph]].

\bibitem{dfp1} 
Ph.~ de\ Forcrand and O.~ Philipsen  JHEP  0811, 012 (2008).

\bibitem{dfp2} 
J. T. Moscicki et al., Comput. Phys. Commun. 181 (2010) 1715 [arXiv:0911.5682[cs.DC]].

\bibitem{dfp3} 
O. Philipsen, Acta Phys. Polon. Supp. 5,  825 (2012) [arXiv:1111.5370[hep-ph]].

\bibitem{curvhisq}
O. Kaczmarek, F. Karsch, E. Laermann, C. Miao, S. Mukherjee, P. Petreczky, C. Schmidt, W. Soeldner and
 W. Unger, Phys. Rev. D 83, 014504 (2011) [arXiv:1011.3130 [hep-lat]].


\bibitem{bwfreezeout}
 G. Endrodi, Z. Fodor, S. Katz, and K. Szabo, JHEP 1104, 001 (2011) [arXiv:1102.1356[hep-lat]].

\bibitem{stephanov}
M. A. Stephanov, Phys. Rev. Lett. 102, 032301 (2009).

\bibitem{shuryak2}
M. Stephanov, K. Rajagopal and E. Shuryak, Phys. Rev. D 60, 114028 (1999).


\bibitem{christian}
S. Ejiri, F. Karsch, E. Laermann, and C. Schmidt, Phys. Rev. D 73, 054506 (2006) [hep-lat/0512040].

\bibitem{bweosmu}
Sz. Borsanyi, G. Endrodi, Z. Fodor, S. D. Katz, S. Krieg, C. Ratti and K. K. Szabo,
JHEP 1208  053 (2012) [arXiv:1204.6710 [hep-lat]].


\bibitem{dfp4}
T. Takaishi, P. de Forcrand and A. Nakamura, PoS LAT 2009 (2009) 198 [arXiv:1002.0890 [heplat]].

\bibitem{gs2} 
R. V. Gavai and S. Sharma, Phys.\ Rev.\ D 85, 054508 (2012).


\end{thebibliography}
\end{document}